# A Framework for Spontaneous Brillouin Noise: Unveiling Fundamental Limits in Brillouin Metrology

Simeng Jin[1†], Shuai Yao[2†], Zhisheng Yang[1*], Zixuan Du[2], Xiaobin Hong[1], Marcelo A. Soto[3], Jingjing Xie[4], Long Zhang[2], Fan Yang[2*], and Jian Wu[1]

[1] State Key Laboratory of Information Photonics & Optical Communications, Beijing University of Posts and Telecommunications, Beijing, China
[2] Shanghai Institute of Optics and Fine Mechanics, Chinese Academy of Sciences, Shanghai, China
[3] Department of Electronics Engineering, Universidad Técnica Federico Santa María, Valparaíso, Chile
[4] School of Physical Science and Technology & State Key Laboratory of Advanced Medical Materials and Devices, ShanghaiTech University, Shanghai, China

[*] Corresponding authors: zhisheng.yang@bupt.edu.cn; yang@siom.ac.cn.
[†] These authors contributed equally to this work.

**Abstract**
Spontaneous Brillouin scattering (SpBS) provides a non-contact tool for probing the mechanical and thermodynamic properties of materials, enabling important applications such as distributed optical fiber sensing and high-resolution Brillouin microscopy. Achieving metrological precision in these systems relies critically on identifying fundamental noise sources. While a pioneering study three decades ago numerically investigated an intrinsic SpBS noise mechanism, this phenomenon has remained largely unexplored, particularly in the context of Brillouin metrological systems. Here, by revisiting its physical formation process and rethinking its stochastic behaviors, we develop and experimentally validate a comprehensive analytical framework on this long-overlooked noise source. Importantly, we theoretically predict, for the first time, the SpBS noise is a universal and fundamental limit that can dominate over conventional limits such as shot noise in Brillouin metrological systems like imaging, microscopy and sensing. Specifically, we experimentally demonstrate the SpBS-noise-limited regime in Brillouin imaging and sensing scenarios. This framework establishes a critical foundation for understanding and optimizing the performance bounds of current and future Brillouin-based technologies across diverse applications.

**Introduction**
Spontaneous Brillouin scattering (SpBS) is a fundamental inelastic light–matter interaction present in virtually all media[1]. First theorized by Léon Brillouin in 1922, this process arises from the interaction between incident photons and thermally driven density fluctuations - quantized as acoustic phonons - resulting in frequency-shifted scattered photons at Stokes or anti-Stokes wavelengths[2]. The peak frequency shift, linewidth, and intensity of the Brillouin spectrum are inherently linked to the material's physical properties, such as its elastic and thermal characteristics. This fundamental relationship underpins the versatility of SpBS, enabling its application cross non-intrusive long-distance distributed strain and temperature sensing in optical fibers[3–13], photon-phonon interactions in integrated waveguides[14,15], and non-contact, three-dimensional mechanical imaging with high spatial resolution in biological and medical samples via Brillouin microscopy[16–27].

The performance of these techniques critically depends on accurate spectral measurements, which in turn rely on maximizing the system signal-to-noise ratio (SNR). Conventional strategies to enhance SNR focus on reducing the impact of detection noise - primarily thermal and shot noise - pursuing the shot-noise limit broadly regarded as the fundamental bound, where SNR scales with the square root of the signal power.

Yet, in 1990 a seminal work[28] theoretically investigate an intrinsic noise source: intensity fluctuations in the scattered light arising from the stochastic nature of the underlying thermally excited phonons. Beside explaining the physical origin of this noise, this work[28] has also explored its stochastic characteristics from physics point of view via numerical simulation. The relevant conclusion in the deeply stimulated regime was soon experimentally verified (with gain coefficients $G$ =23 and 70)[29]. However, fluctuations in the spontaneous regime ($G \ll 21$)[30] have remained overlooked for over three decades, neither experimentally characterized nor accounted for in SpBS-based applications. The lack of an analytical treatment and a connection to unavoidable practical system parameters has limited our comprehensive understanding of its stochastic nature. More fundamentally, its implications for measurement performance have remained unexplored since the Brillouin scattering effect proposed more than a century ago[2], representing a critical gap in Brillouin metrology.

Here, we address this longstanding gap by establishing an analytical framework that comprehensively characterizes the intrinsic noise in spontaneous Brillouin scattering, based on a pipeline work sequentially revisiting its physical formation process, linking it to the system parameters and analyzing its stochastic properties. Compared to earlier numerical study[28], our framework offers enhanced mechanistic clarity and, initiatively, connects the stochastic behavior of SpBS noise to practical system parameters such as detection bandwidth and sampling configuration. More importantly, our framework reveals that this noise can exceed the conventionally assumed shot-noise limit, imposing a previously unrecognized upper bound on metrological performance. We experimentally validate these predictions, across both coherent and direct detection schemes with $G \approx 0.23$ and $G \approx 3.89$, respectively, reporting the first observation of intrinsic SpBS noise in the spontaneous regime. Finally, we demonstrate the metrological relevance of this noise by re-evaluating the performance limits of several SpBS-based platforms, including Brillouin imaging, microscopy, and distributed fiber sensing. Our findings not only reveal a new fundamental noise floor but also provide a blueprint for rethinking system design in light of these intrinsic constraints.

**Results**

*Revisiting Physical Formation Mechanism of SpBS Noise*
We revisit the physical formation mechanism of SpBS noise by developing a stepwise solution to the classical three-wave coupling equations (**SI Note 1** Eqs. (1)-(3))[1,28,31]. This provides a physically intuitive depiction of how SpBS noise arises from fundamental thermodynamic interactions, as schematically illustrated in **Fig. 1**.

For convenience of intuitively visualizing the relevant physical insight and later on mathematically characterizing the stochastics of the SpBS intensity fluctuations, we firstly discretize the light–matter interaction length into small segments of length $\Delta z$ (**Fig. 1**). In each segment, thermally induced material vibrations are treated as localized Langevin forces, represented by a series of temporally spaced delta functions with random amplitudes $A^f$ and phases $\varphi^f$ at interval $\Delta t$ (**Fig. 1(i)**, **SI Note 1** Eq. (4)). These stochastic forces drive localized acoustic fields $\vec{\rho}(t)$, which can be expressed as a superposition of exponentially decaying sinusoidal waves at a common carrier frequency, with random amplitudes and phases dictated by the Langevin excitation (**Fig. 1(ii)**, **SI Note 1** Eq. (9)).

The resulting SpBS field $\overrightarrow{E_{Sp}}(t)$ is formed by the cumulative interaction of these stochastic acoustic waves with the pump field $\overrightarrow{E_P}$ over the entire interaction length (**Fig. 1**; **SI Note 1** Eq.

(10)). Assuming a constant pump, $\overrightarrow{E_{Sp}}(t)$ becomes proportional to the spatial summation of time-delayed acoustic wave contributions (**Fig. 1(iii)**; **SI Note 1** Eq. (12)). This results in the SpBS power envelope $P_{Sp}(t)$, which intrinsically exhibits stochastic intensity fluctuations (**Fig. 1(iv)**). The SNR of $P_{Sp}(t)$, defined as the ratio of its mean to standard deviation (STD), is analytical derived to be unity (detailed in **SI Note 2**), aligning with early simulation-based observations[28].

This formulation shown by **Fig. 1** provides an intuitive link between microscopic stochasticity and macroscopic noise (defined hereafter as SpBS noise, carried by $P_{Sp}(t)$), as further detailed in **SI** Eqs. (4–36).

*Analyzing the Stochastic Properties of SpBS Noise*
To comprehensively characterize SpBS fluctuations, we extend our analysis to account for the influence of system-level parameters. While prior work based on physical understanding suggested a SNR of unity[28], we show that this condition does not generally hold in practice. Instead, the SNR can vary significantly depending on system bandwidth ($B_m$) and sampling parameters-critical yet previously unaddressed factors that are unavoidable in practical Brillouin metrology.

To interpret the effects of system design more intuitively, we here investigate spectral properties of SpBS fluctuations. The spectrum of the scattered field $\overrightarrow{E_{Sp}}(t)$ with a 3 dB bandwidth $B_{Sp}$[1,28], denoted as $Spec_{es}$ (**Fig. 2a**), is considered to comprise two components (**Fig. 2b**): 1) a deterministic Lorentzian profile, representing conventional Brillouin response, and 2) a stochastic component, reflecting the intrinsic SpBS noise, the focus of this study.

The power spectrum of envelope signal $P_{Sp}(t)$, denoted as $Spec_{ps}$ (**Fig. 2c**), is obtained via autocorrelation of $Spec_{es}$, followed by low-pass filtering (LPF). Its single-sided form (**Fig. 2d**) contains:
1. a realization-dependent alternating-current (AC) component, with a 3dB bandwidth of $B_{Sp}$. According to Wiener-Khinchin theorem[32], its total power corresponds to the noise variance of $P_{Sp}(t)$.
2. a constant direct-current (DC) component, originating from the stochastic spectral component of $Spec_{es}$ (**Fig. 2b**). This DC component, essentially attributed to the SpBS noise (**Fig. 2b**), represents the squared mean (signal power) of $P_{Sp}(t)$[32].

In the idealized full-bandwidth case, the DC and AC powers of $Spec_{ps}$ are equal, reaffirming that SNR=1. However, in practical systems with limited measurement bandwidth $B_m$, only part of the AC noise is captured while the DC term remains unaffected (**Fig. 2e**), leading to a bandwidth-dependent SNR as (**SI Note 2**):

$$SNR_{Sp} = \begin{cases} 1, & B_m \geq B_{Sp} \\ \sqrt{\dfrac{\arctan(1)}{\arctan\left(\dfrac{B_m}{B_{Sp}}\right)}}, & B_m < B_{Sp} \end{cases} \qquad (1)$$

This relationship implies that different Brillouin-based systems, operating across varying bandwidth regimes, will exhibit distinct but predictable SNR behaviors when constrained by SpBS noise.

Furthermore, sampling interval and total number of samples also strongly impact the observed SNR. **Figure 2f** presents a theoretical 2D SNR map (see **Methods** for the data generation) as a function of these sampling parameters, with contour lines illustrating how undersampling can artificially elevate the SNR by distorting the relative DC/AC contributions. Representative spectral decompositions under three sampling conditions (**Fig. 2g**) illustrate these effects.

Together, this analysis refines the condition under which SNR = 1 holds, showing that it is only achieved when both $B_m \geq B_{Sp}$ and sufficient sampling conditions are met. These results offer practical design guidance for selecting system parameters to accurately capture the intrinsic SpBS noise, while also establishing a quantitative foundation for predicting and optimizing SNR in diverse Brillouin metrology platforms.

*Experimental Validation of the Framework via two experimental setups*
To date, experimental evidence of spontaneous Brillouin scattering (SpBS) power fluctuations, and particularly their stochastic nature, has been lacking. Here, we provide the first experimental observation and quantitative characterization of SpBS noise. Our framework is validated under controlled conditions using two foundational detection schemes: coherent detection and direct detection.

**Coherent detection setup.** Coherent detection is widely adopted for its optimal balance between simplicity and SNR. We implemented a fiber-based coherent detection scheme (**Fig. 3a**), incorporating a 400 m polarization-maintaining fiber (PMF) to mitigate polarization fading (**Methods**). A continuous-wave (CW) laser with 1.8 mW pump power was launched into the fiber, and the output spectrum was obtained via fast Fourier transform (FFT) analysis (**Fig. 3b**). The non-averaged spectra reveal pronounced amplitude fluctuations near the Brillouin peak (~220 MHz), which decay with the spectral envelope and vanish when the pump is turned off (**Fig. 3c**), strongly implicating SpBS noise as their origin.

We developed a quantitative SNR model for the Brillouin peak power $r_{Co}^{FFT}$, by leveraging our framework to incorporate all relevant noise sources, as (**SI Note 3**):

$$\text{SNR}\{r_{Co}^{FFT}\} = \frac{\mathcal{R}_p^2 P_{Lo} \overline{P_{Sp}}}{\sqrt{\frac{1}{SNR_{Sp}^2}\mathcal{R}_p^4 P_{Lo}^2 \overline{P_{Sp}}^2 + \frac{\pi B_{Sp}}{2}\mathcal{R}_p^2 P_{Lo}\overline{P_{Sp}}\sigma_e^2 + \left(\frac{\pi B_{Sp}}{4}\right)^2 \sigma_e^4}} \leq SNR_{Sp} \qquad (2)$$

where $\mathcal{R}_p$ is the photodiode responsivity; $P_{Lo}$ is the local oscillator (OLO) power; $\overline{P_{Sp}}$ is the mean SpBS optical power entering the photodiode; $\sigma_e^2$ is and power spectral density (PSD) of the detection noise; $B_{Sp}$ is the Brillouin linewidth. Compared to Eq. (1), the SNR in this case is lower than or equal to $SNR_{Sp}$, due to the contribution of three types of noise:
1) **Signal-dependent noise (SpBS noise):** The STD scales linearly with $\overline{P_{Sp}}$ and dominates at high powers.
2) **Cross-term noise:** Arises from interaction between signal and detection noise; scales with $\sqrt{\overline{P_{Sp}}}$.
3) **Detection noise:** Includes shot and thermal noise, largely independent of $\overline{P_{Sp}}$ due to the strong OLO.

At high signal powers, the first term dominates, and the SNR asymptotically approaches the theoretical SpBS noise limit $SNR_{Sp} = 1$, consistent with FFT-based measurements where $B_m \geq B_{Sp}$ always holds (**SI Note 3**).

Experimental measurements of the Brillouin signal amplitude, noise STD, and SNR over a range of pump powers (-25.5 dBm to 3.5 dBm, **Fig. 3d-f**) show excellent agreement with theoretical predictions, validating the saturation behavior predicted by Eq. (2). As the highest pump power here corresponds to a gain coefficient $G \approx 0.23$, which remains well below the stimulated scattering threshold $G = 21$, these results constitute the first experimental confirmation of SpBS fluctuations.

To further isolate SpBS noise, we replaced the scattered signal with a CW laser of equivalent power. This eliminates the signal-dependent noise term while preserving other noise components. The resulting SNR (pink asterisks in **Fig. 3d–g**) exceeds unity and deviates from the SpBS-limited case (blue), confirming that SpBS noise indeed defines the saturation floor.

The importance of sampling conditions is further evidenced in a 2D experimental SNR map (**Fig. 3h**). Compared to the idealized model (**Fig. 2f**), a slight leftward shift is observed, attributable to additional detection noise. This highlights the importance of system optimization, as emphasized in our framework but previously overlooked.

Finally, replacing FFT analysis with an electrical envelope detector (**Fig. 3a(ii)**) allows direct time-domain observation of SpBS power fluctuations. This approach sets the measurement bandwidth $B_m$ by the detector bandwidth, thus making Eq. (1) directly applicable. For two representative bandwidths (50 MHz and 2 MHz), the observed SNR (**Fig. 3i,j**) matches the theoretical curves, verifying the bandwidth-dependent SNR behavior predicted uniquely by our framework.

**Direct detection setup.** To further validate the stochastic nature of SpBS noise, we turn to direct detection measurement on a 10-m long fiber sample using a virtually imaged phased array (VIPA)-spectrometer and a camera (**Fig. 4a**; **Methods**). Since the Brillouin frequency shift of the fiber (~22 GHz at 780 nm pump) exceeds one free spectral range (FSR) of the VIPA (~15.3 GHz), we employ a reflective grating to provide orthogonal dispersion, enabling full spectral separation of the Brillouin signal in combination with VIPA dispersion (**SI Fig. 5**). Compared to coherent detection, this direct detection scheme offers linear superposition of noise sources and clearer interpretability, enabling a direct expression of the SNR of the camera response $r_{Di}$, which includes contributions from SpBS noise, shot noise, and scientific complementary metal-oxide-semiconductor (sCMOS) camera read noise (background noise):

$$\text{SNR}\{r_{Di}\} = \frac{\mathcal{R}_c \overline{P_{Sp}}}{\sqrt{\frac{1}{\text{SNR}_{Sp}^2}\mathcal{R}_c^2 \overline{P_{Sp}}^2 + 2q\mathcal{R}_c\overline{P_{Sp}}B_m + \sigma_{re}^2}} \leq \text{SNR}_{Sp} \qquad (3)$$

where $\mathcal{R}_c$ is camera responsivity; $q = 1.6 \times 10^{-19}$ is the elementary charge. $\sigma_{re}^2$ represents the constant variance of the camera read noise, which, unlike thermal noise, remains largely insensitive to bandwidth variations under the low-noise mode of typical sCMOS used in this study[33,34]. The model predicts square-root scaling of SNR at low powers (shot-noise-limited), and saturation at high powers (SpBS-noise-limited), governed by Eq. (1).

Measurements for bandwidths $B_m$ = 50 kHz (maximum bandwidth of the sCMOS camera used) and 5 kHz (**Fig. 4b, c**), both within the $B_m < B_{Sp}$ regime, show strong agreement with Eq. (3). The maximum gain coefficient $G \approx 3.89$ remains well within the spontaneous scattering regime. Replacing SpBS light with CW light again yields higher SNR values (pink asterisks), consistent with the model excluding SpBS noise (red solid lines). These results confirm that the observed SNR saturation in direct detection is governed by SpBS noise, in full agreement with our framework.

***Implementation of our Brillouin noise framework on Brillouin metrological applications***
To underscore the technological relevance of our framework, we apply it to evaluate the performance limits of key Brillouin-based metrological platforms, incorporating both spontaneous Brillouin scattering (SpBS) noise and conventional noise sources. We focus on three applications: Brillouin imaging, Brillouin microscopy, and distributed optical fiber sensing.

***Brillouin imaging and microscopy.*** We first consider Brillouin imaging, implemented using a free-space optics setup (**Fig. 5a**) that builds on the configuration detailed in **Fig. 4a**, with two modifications. First, a high numerical aperture (NA=0.7) objective is employed to focus light onto the cleaved end face of an optical fiber (SMF-28e, length = 1 km), which is mounted on a 3D piezo stage for precise scanning (see **SI Fig. 6** for the fiber mounting details). Second, a double-pass Fabry-Pérot interferometer is integrated to suppress amplified spontaneous emission noise (**Methods; SI Fig. 7**). The SNR model derived from our framework for this imaging approach closely resembles Eq. (3), incorporating imaging-specific parameters such as $\overline{P_{Sp}}$ and $B_m$ (**Methods**). This model predicts a saturation behavior in SNR with increasing pump power (**Fig. 5b**, yellow curve), which is in good agreement with experimental data (yellow dots). The SNR dependency on $B_m$ is also demonstrated to match the anticipation of our SNR model (**SI Fig. 8**). More illustratively, **Fig. 5c** presents spatial maps of Brillouin shift and linewidth at pump powers from 1 dBm to 10 dBm in 3 dB increments. The precision of both parameters plateaus between 7 dBm and 10 dBm, confirming the predicted SNR saturation regime where SpBS noise dominates. Conversely, below 7 dBm, the precision deteriorates rapidly with decreasing power - consistent with the shot-noise-limited regime observed in our direct detection experiments (**Fig. 4b**).

We further extend our framework to Brillouin microscopy, using the same setup (**Fig. 5a**) to investigate polydimethylsiloxane (PDMS) beads as a phantom and HeLa cells as a biological sample. While the underlying noise model remains the same, the effective interaction lengths in both PDMS and cell samples are reduced to the micrometer scale. Combined with limited optical power delivery, this results in significantly weaker scattered signals compared to the kilometer-long fiber sample used in Brillouin imaging. Consequently, the system does not enter the SpBS-limited saturation regime, and its performance remains shot-noise-limited across the entire range of pump powers tested. This behavior is evidenced by the square-root scaling of Brillouin shift and linewidth precision with increasing pump power (**Fig. 5d,e**), and is similarly observed in measurements on a reference sample of double-distilled water (**Fig. 5b**, blue curve). These results align with previous reports of shot-noise-limited Brillouin microscopy[25,26,35]. Critically, however, our work introduces the first unified analytical framework that explicitly incorporates SpBS noise in Brillouin microscopy. This framework further elucidates the ultimate SNR limit that will arise when significantly higher optical powers are employed in future implementations.

***Brillouin distributed optical fiber sensing.*** We next apply our framework to Brillouin optical time-domain reflectometry (BOTDR), a widely adopted technique in structural health monitoring due to its single-ended operation, long sensing range, and large dynamic range[3,4,6,8,13]. We evaluate two implementations: using polarization-maintaining fiber (PMF), and standard single-mode fiber (SMF), each representing distinct noise regimes.

*BOTDR with polarization-maintaining fiber.* For PMF-based BOTDR, which eliminates polarization-induced fluctuations, the SNR of the single-pulse response $r_{Sg}^{PM}(z)$ at position $z$ is derived as (**SI Note 4**):

$$\text{SNR}\{r_{Sg}^{PM}(z)\} = \frac{\mathcal{R}_p{}^2 P_{Lo} \overline{P_{Sp}(z)}}{\sqrt{\frac{1}{2}\left(1 + \frac{1}{SNR_{Sp}{}^2}\right)\mathcal{R}_p{}^4 P_{Lo}{}^2 \overline{P_{Sp}(z)}^2 + 2\mathcal{R}_p{}^2 P_{Lo}\overline{P_{Sp}(z)}\sigma_e{}^2 B_{BPF} + \sigma_e{}^4 B_{BPF}{}^2}}$$

$$\leq \frac{1}{\sqrt{\frac{1}{2}\left(1 + \frac{1}{SNR_{Sp}{}^2}\right)}} \qquad (4)$$

where $\overline{P_{Sp}(z)}$ is the mean SpBS power at position $z$, proportional to the pump pulse duration and peak power, and $B_{BPF}$ is the bandwidth of BPF used before envelope extraction process (**Methods**). For meter-scale spatial resolutions, increasing the pump power initially improves the SNR, but the benefit saturates as SpBS noise dominates, particularly near the fiber input or in shorter fiber spans. Since optimized BOTDR systems require $B_m \approx B_{Sp}$, corresponding to $SNR_{Sp} = 1$, substituting which into Eq. (4) result in a saturated SNR (without averaging) of 1.

Experimental validation (**Fig. 6a**; **Methods**) uses a 400 m PMF and pump powers from 10 to 31 dBm in 3 dB intervals. Measured SNRs (**Fig. 6b**) show clear saturation trends consistent with theoretical predictions for various spatial resolutions (1 m, 2 m, 6 m, and 10 m). Note that, larger spatial resolutions, which yield higher $\overline{P_{Sp}(z)}$, require lower pulse peak powers to reach SNR saturation.

Due to the negligible attenuation in the 400 m PMF, the SNR remains nearly uniform along the fiber, allowing us to perform more accurate analysis by averaging the SNR over the fiber length (**Fig. 6c**), showing that the measured data (circle markers) closely match theoretical predictions (solid lines). These results directly challenge the conventional view that PMF-based BOTDR is inherently shot-noise-limited [8,36–41].

Furthermore, the SNR dependence on both spatial resolution and peak power (**Fig. 6d**) provides actionable guidance for optimizing BOTDR system performance, balancing sensitivity with power efficiency.

*BOTDR with standard single-mode fiber.* SMFs are more practical for deployment, but exhibiting intrinsic birefringence leading to polarization fluctuations [8,36–40,42–47]. Extending our framework to include a polarization scrambling factor $k_{Pol}$, the SNR of the SMF-based BOTDR becomes (**SI Note 5**):

$$\text{SNR}\{r_{Sg}^{SMF}(z)\} = \frac{\frac{1}{2}\mathcal{R}_p{}^2 P_{Lo}\overline{P_{Sp}(z)}}{\sqrt{\frac{3SNR_{Sp}{}^2(k_{Pol}{}^2 + 1) + k_{Pol}{}^2 + 3}{24 SNR_{Sp}{}^2}\mathcal{R}_p{}^4 P_{Lo}{}^2 \overline{P_{Sp}(z)}^2 + \mathcal{R}_p{}^2 P_{Lo}\overline{P_{Sp}(z)}\sigma_e{}^2 B_{BPF} + \sigma_e{}^4 B_{BPF}{}^2}}$$

$$\leq \sqrt{\frac{6 SNR_{Sp}{}^2}{3SNR_{Sp}{}^2(k_{Pol}{}^2 + 1) + k_{Pol}{}^2 + 3}} \qquad (5)$$

where $k_{Pol} \in [0,1]$ quantifies the self-polarization-scrambling effect caused by the fiber birefringence. It reflects the degree to which polarization fluctuations are averaged over the pump pulse duration (**SI Note 6**): longer pulse yields lower $k_{Pol}$. This framework uncovers a new theoretical SNR upper bound (between 0.77 and 1 depending on the spatial resolution), even in the presence of polarization noise - contradicting earlier assumptions that such systems are purely polarization-noise-limited[45]. Experimentally (**Fig. 6e,f**), SMF-based BOTDR exhibits lower SNR and slower scaling than PMF, fully consistent with our model. Importantly, these results represent the first experimental demonstration of an SpBS - imposed upper SNR limit in SMF-based BOTDR.

**Discussion**

In summary, we have developed and experimentally validated a comprehensive theoretical framework to describe intensity fluctuations arising from spontaneous Brillouin scattering (SpBS). This work advances the understanding of Brillouin noise through four key contributions:

1. **Intuitive physical mechanism.** We describe the formation of SpBS noise through a stepwise analytical model. This decomposition - visualized in **Fig. 1** - provides an intuitive and analytically tractable explanation of the stochastic generation of Brillouin intensity fluctuations.

2. **Complement stochastic behavior.** Building on prior reports of a unity SNR for SpBS signal from an idealized physical standpoint[28], we extend the analysis to account for practical system design and signal processing factors. Our results reveal that the SNR is critically dependent on the phonon lifetime ($\tau$), measurement bandwidth ($B_m$), as well as sampling parameters. We show that the unity SNR is achievable only when $B_m > B_{Sp}$ and proper sampling parameters are used. This refined understanding provides actionable criteria for optimizing experimental configurations in Brillouin-based systems.

3. **Experimental validation.** Through both coherent and direct detection experiments, with gain coefficients $G \approx 0.23$ and $G \approx 3.89$, respectively, we report the first observation of SpBS intensity fluctuations in the spontaneous regime ($G \ll 21$)[30]. This addresses a long-standing experimental gap, more than a century after the theoretical prediction of Brillouin scattering[2]. Our SNR predictions (Eqs. (2)-(5)), derived from the framework, match observed behaviors across various Brillouin metrological platforms, including Brillouin imaging, microscopy, and distributed fiber sensing.

4. **Revised noise limit.** Our framework establishes SpBS noise as a fundamental noise floor in Brillouin-based metrology. Once a system enters the SpBS-noise-limited regime, this intrinsic noise source can dominate over conventional limits such as shot noise. We experimentally confirm this regime in both imaging (**Fig. 5**) and distributed sensing (**Fig. 6**), showing that the SpBS noise sets a new sensitivity baseline. Importantly, this regime reveals an optimal pump power threshold, beyond which increasing optical power yields no further SNR improvement. Instead, performance must be enhanced through increased measurement time.

Beyond the standard implementations discussed, our framework generalizes to more advanced Brillouin sensing systems. For example, we anticipate that the polarization diversity coherent detection approach, employed to mitigate polarization noise in SMF-based BOTDR[46], offers limited SNR improvement when SpBS noise is dominant (experimentally demonstrated in **SI Note 7**). Additionally, our analysis and experimental demonstration (**SI Note 8**) reveal that, while pulse coding technique was originally designed to boost SNR[41,48], its effectiveness in BOTDR is constrained by the combined effects of SpBS and polarization noises. Nevertheless, it is worth noting that, in Brillouin-integrated sensing and communication systems, where background noise is higher, such coding strategies regain utility[47]. These results underscore the importance of our framework and the full understanding of SpBS noise.

Although shot noise remains the dominant noise source in Brillouin microscopy - and we did not observe a significant SpBS noise contribution in this case due to the short interaction length and limited pump power permitted for biological samples - our framework highlights the importance of considering SpBS noise in high-power conditions. For example, in applications such as Brillouin spectroscopy[49,50] and Brillouin Lidar systems[51,52], where megawatt-scale peak-power pulsed lasers are employed, our model predicts a critical contribution from SpBS noise, which has thus far been largely overlooked.

**FIGURES**

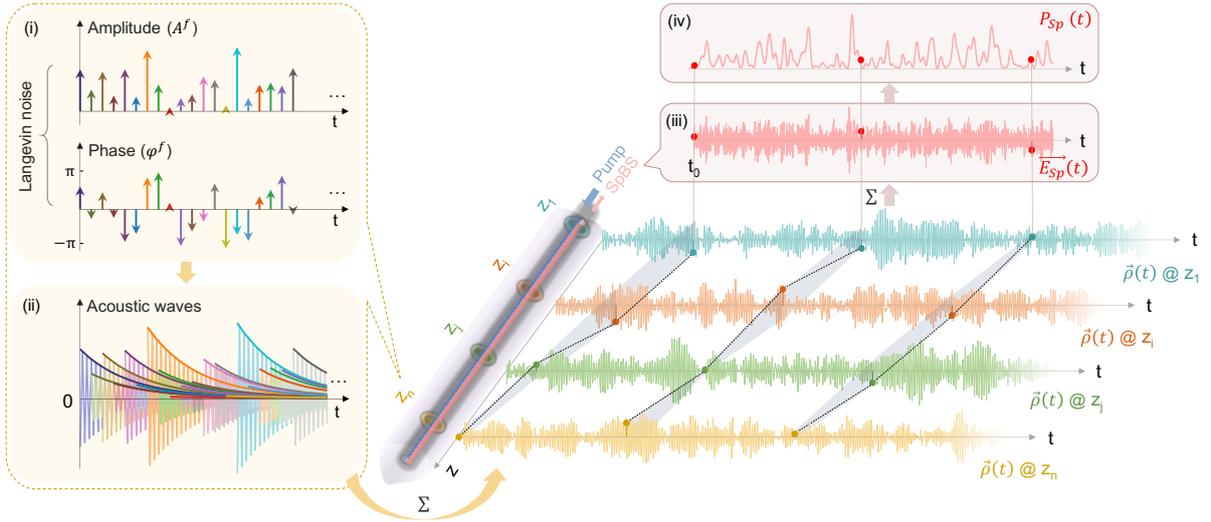

**Figure 1 | Schematic description on physical formation process of spontaneous scattering (SpBS) fluctuations.** The light purple irregular cylinder marks the region where light interacts with matter, while the blue and red beams represent the incident pump light and the backward-propagating SpBS light, respectively. Four specific positions ($z_1$, $z_i$, $z_j$, and $z_n$) within the interaction region are highlighted in different colors as examples. Insets **(i)-(ii)** exemplify the local excitation of acoustic waves by Langevin noise at $z_n$. Inset **(i)**: the amplitude ($A^f$) and corresponding phase ($\varphi^f \in [-\pi, \pi]$) of Langevin noise in the time domain. Inset **(ii)**: the acoustic waves driven by each Langevin noise element, where different light-colored curves denote individual acoustic waves, and the corresponding dark-colored curves depict the corresponding acoustic wave envelopes. The acoustic waves ($\vec{\rho}(t)$) at the four example positions are color-matched to their respective locations. Inset **(iii)** shows the SpBS wave $\overrightarrow{E_{Sp}}(t)$ results from the cumulative effect of these acoustic waves observed at the same moment. Inset **(iv)** shows the time-domain power envelope of SpBS, $P_{Sp}(t)$, extracted from $\overrightarrow{E_{Sp}}(t)$ in inset **(iii)**.

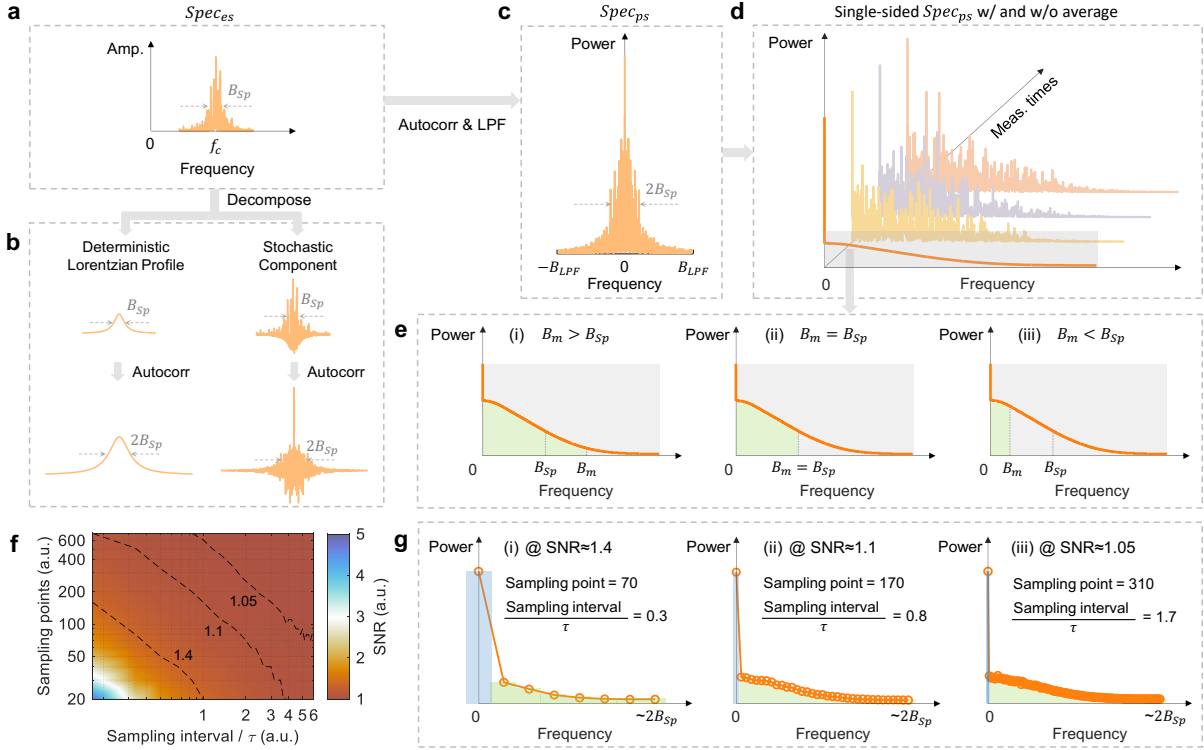

**Figure 2 | Spectral analysis on the stochastic behaviors of SpBS fluctuations. a,** Spectrum of the scattered field $\overrightarrow{E_{Sp}}(t)$, denoted as $Spec_{es}$, centered at $f_c$ with a full width at half maximum (FWHM) of $B_{Sp}$. The property of $Spec_{es}$ is consistent across detection methods, differing only in central frequency $f_c$ (optical or radio frequency) depending on whether direct or coherent detection is used. It can be decomposed into a deterministic Lorentzian profile and a stochastic component, illustrated in **b**, along with their respective autocorrelation results. **c,** Double-sided spectrum ($Spec_{ps}$) of $P_{Sp}(t)$, obtained via the autocorrelation of $Spec_{es}$, followed by a low-pass filtering (LPF) process. **d,** Single-sided form of $Spec_{ps}$, in the cases with and without averaging. Light yellow, light purple, and light pink curves correspond to three unaveraged power envelope spectra, while dark orange curve denotes the power envelope spectrum after averaging. The averaged spectral segment within the gray box is used for further analysis in **e**. **e.** Effect of different system bandwidths $B_m$ on the SpBS power envelope spectrum with a FWHM of $B_{Sp}$. Insets **(i)**, **(ii)**, and **(iii)** depict the cases of $B_m > B_{Sp}$, $B_m = B_{Sp}$, and $B_m < B_{Sp}$, respectively. Green shading indicates the captured part of the AC component, representing the preserved SpBS noise after filtering. **f,** Theoretical 2D SNR map as a function of sampling interval and total number of samples, with acoustic lifetime $\tau = 6$ ns. Dashed contour lines (black dashed lines) on the map indicate SNR levels of 1.4, 1.1, and 1.05. **g.** Power envelope spectra of SpBS under different sampling conditions. Inset **(i)-(iii)** showcase the DC (blue shading) and AC (green shading) contributions within spectral range $[0, 2B_{Sp}]$, under three sampling conditions associated with the contour lines in **f**, representing SNR values of ~1.4, ~1.1, and ~1.05, respectively.

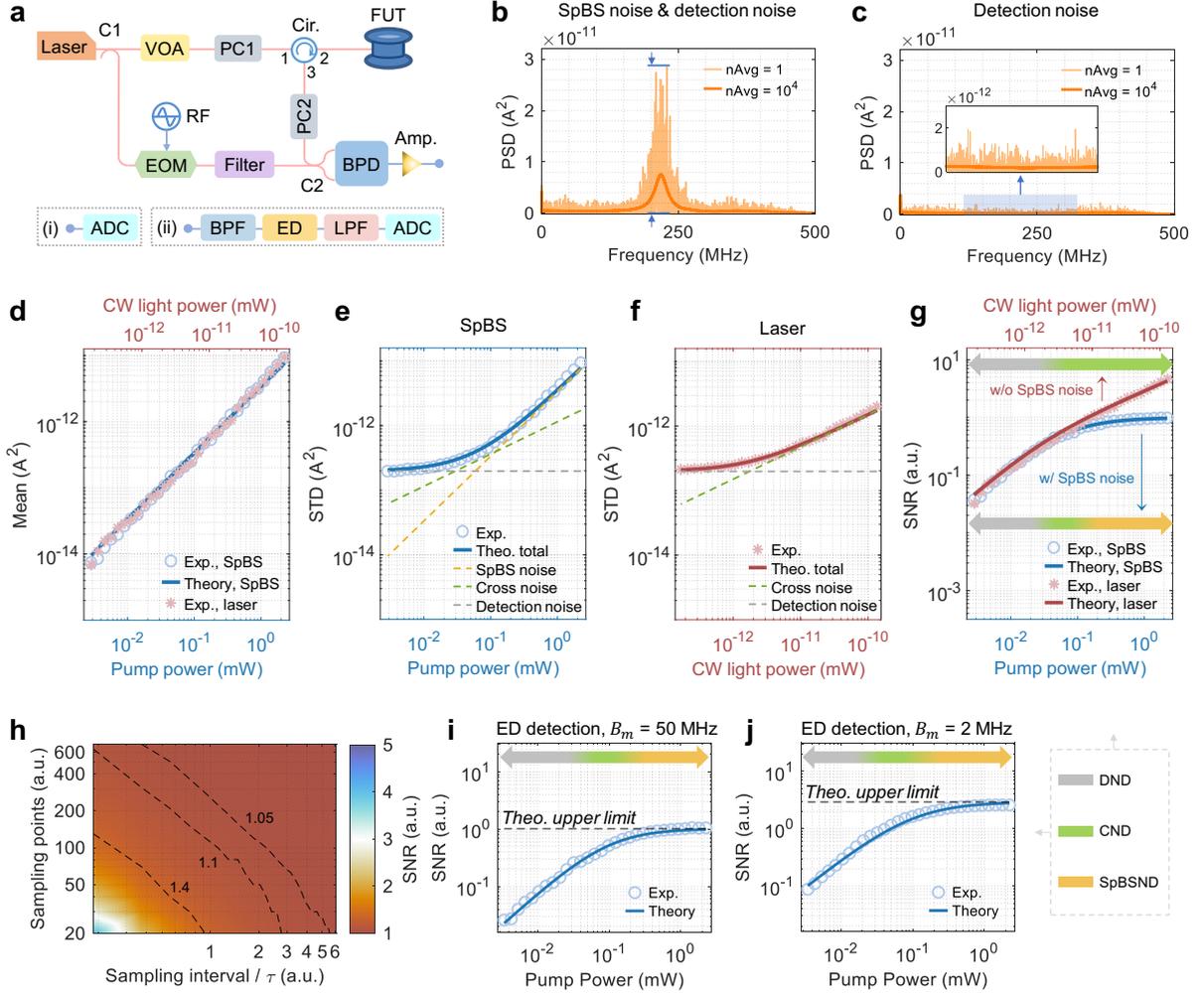

**Figure 3 | Experimental validation of the proposed framework via optical fiber-based coherent detection. a,** Experimental setup for **(i)** fast Fourier transform (FFT)-based and **(ii)** envelope detection (ED)-based coherent detection using a 400 m polarization maintaining fiber (PMF). C: PM optical coupler; VOA: variable optical attenuator; PC: polarization controller; Cir.: circulator; FUT: fiber under test; RF: radio frequency; EOM: electro-optic modulator; BPD: balanced photodetector; Amp.: electric amplifier; ADC: analogue to digital converter; BPF: band pass filter; ED: envelope detector; LPF: low pass filter. **b,** PSD of the acquired beating signal at CW pump power of 1.8 mW, with 10,000 averages (dark orange curve) and no average (light orange curve). The two short blue arrows indicate the peak-to-peak noise fluctuation level at the central position of Brillouin spectra. **c,** PSD of noise floor, obtained in the absence of incident CW pump light. **Inset**: zoom in of the noise floor around the Brillouin spectrum location in **b**. **d,** Experimental (light blue circles) and theoretical (dark blue solid line) signal mean versus pump powers ranging from -25.5 dBm to 3.5 dBm in 1 dB increment; Pink asterisks denote results obtained using a CW laser light of the same power as SpBS light. **e,** With SpBS signal, the noise STDs obtained experimentally (light blue circles) and theoretically (dark blue solid line). The latter is calculated by taking the square root on the square sum of all theoretical noise STDs (yellow dashed line: SpBS noise STD; green dashed line: cross-term noise STD; grey dashed line: detection noise STD). **f,** Replacing SpBS light by CW laser light of the same power, noise STDs obtained experimentally (pink asterisks) and theoretically (red

solid line). **g,** SNRs obtained under the conditions of the same pump power range and CW light power range as in **d**, **e**, and **f**. The data include the experimentally measured SNR affected by SpBS noise (light blue circles) and its theoretical prediction (dark blue solid line), as well as the experimentally measured SNR (pink asterisks) and theoretical prediction (dark red solid line) with direct laser measurement. A color-gradient arrow composed of grey, green, and yellow reflects the relative contributions of detection noise, cross-term noise, and SpBS noise at different pump power levels. DND: detection noise domination; CND: cross noise domination; SpBSND: SpBS noise domination. **h,** Experimentally obtained 2D SNR map as a function of sampling interval and total number of sampling points. The three SNR contour lines correspond to SNR values of 1.4, 1.1, and 1.05. **i,** SNR as a function of CW pump power (same range as in **d**) for a system bandwidth $B_m$=50 MHz, obtained from ED-based experimental measurements (light blue circles) and corresponding theoretical calculations (dark blue solid lines). The black dashed line indicates the SpBS-limited SNR upper bounds. A color-gradient arrow, transitioning from grey to green to yellow, illustrates the evolving contributions of detection noise, cross-term noise, and SpBS noise with increasing pump power, similar with **g**. **j,** SNR as a function of CW pump power $B_m$ = 2 MHz (same range as in **d**), displayed in the same format as in **i**.

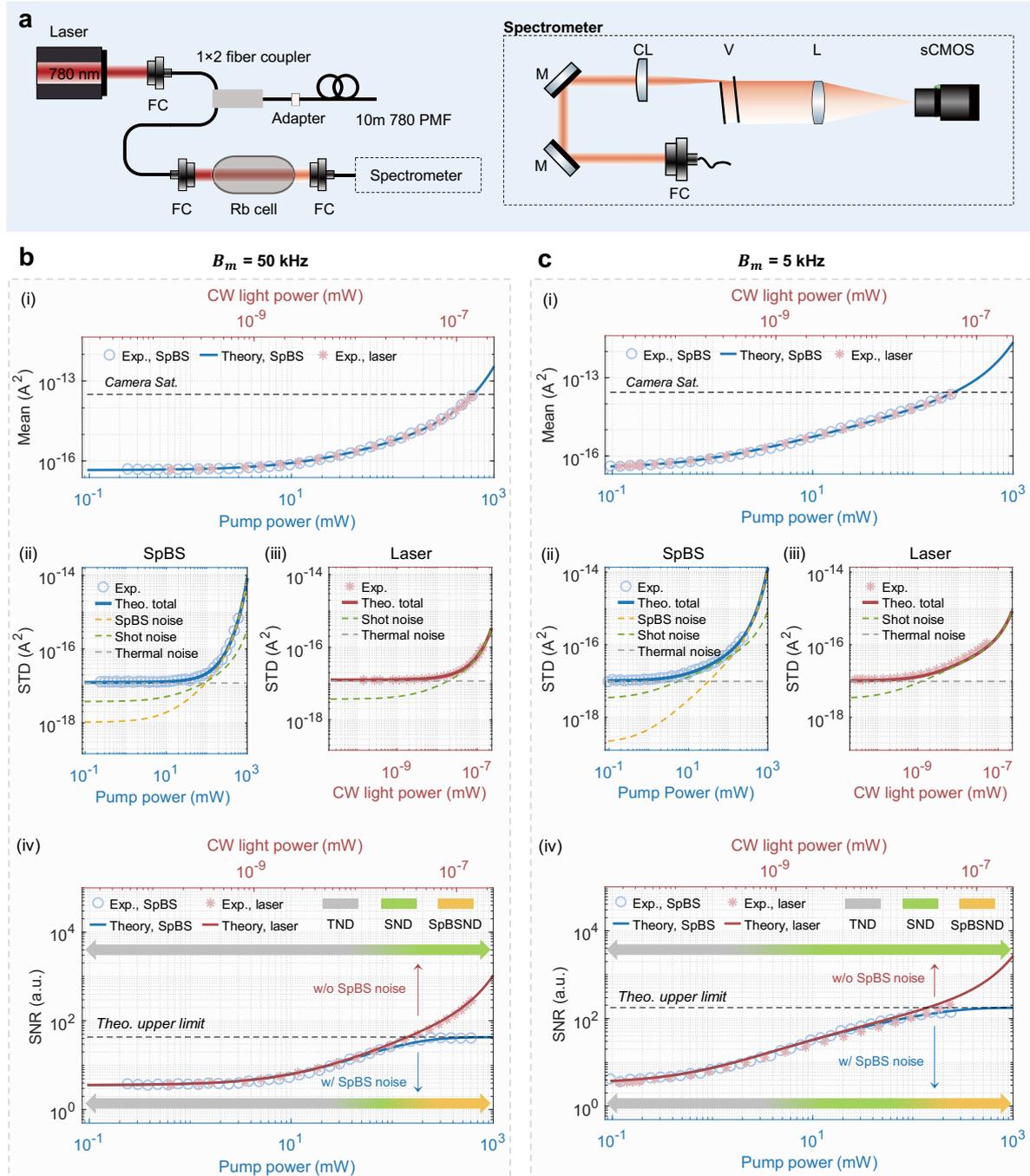

**Figure 4 | Experimental validation of the proposed framework via direct detection. a,** Experimental setup. Rb cell: Rubidium cell; FC: fiber coupler / collimator; CL: cylindrical lens; M: mirror; V: Virtually Imaged Phased Array (VIPA); L: lens. **b,** Experimental (blue circles: using SpBS light, pink asterisks: using CW laser light) and theoretical (blue solid lines: using SpBS light, red solid lines: using CW laser light) results with system bandwidth $B_m =$ 50 kHz, presented in four subplots: **(i)** signal mean, **(ii)** noise STD with SpBS signal, **(iii)** noise STD with CW laser light replacing SpBS signal, and **(iv)** SNR. The pump powers range from −6.2 dBm to 27.8 dBm in 1 dB step. The maximum pump power is limited by the camera saturation (black dashed line in **(i)**). The theoretical total noise STD is obtained by taking the square root on the square sum of all theoretical noise STDs (yellow dashed line: SpBS noise STD; green dashed line: shot noise STD; grey dashed line: thermal noise STD). In **(iv)**, the

SNR upper limit constrained by SpBS noise is denoted as the black dashed line. Color-gradient arrow bars, transitioning from grey to yellow, illustrate the evolving domination of thermal noise, shot noise, and SpBS noise with increasing pump power. TND: thermal noise domination; SND: shot noise domination; SpBSND: SpBS noise domination. **c,** Experimental and theoretical results with system bandwidth $B_m = 5$ kHz, following the same illustration approach as in **b**. The pump powers used in experiments (in the presence of the SpBS signal) range from -10.2 dBm to 23.8 dBm in 1 dB increments.

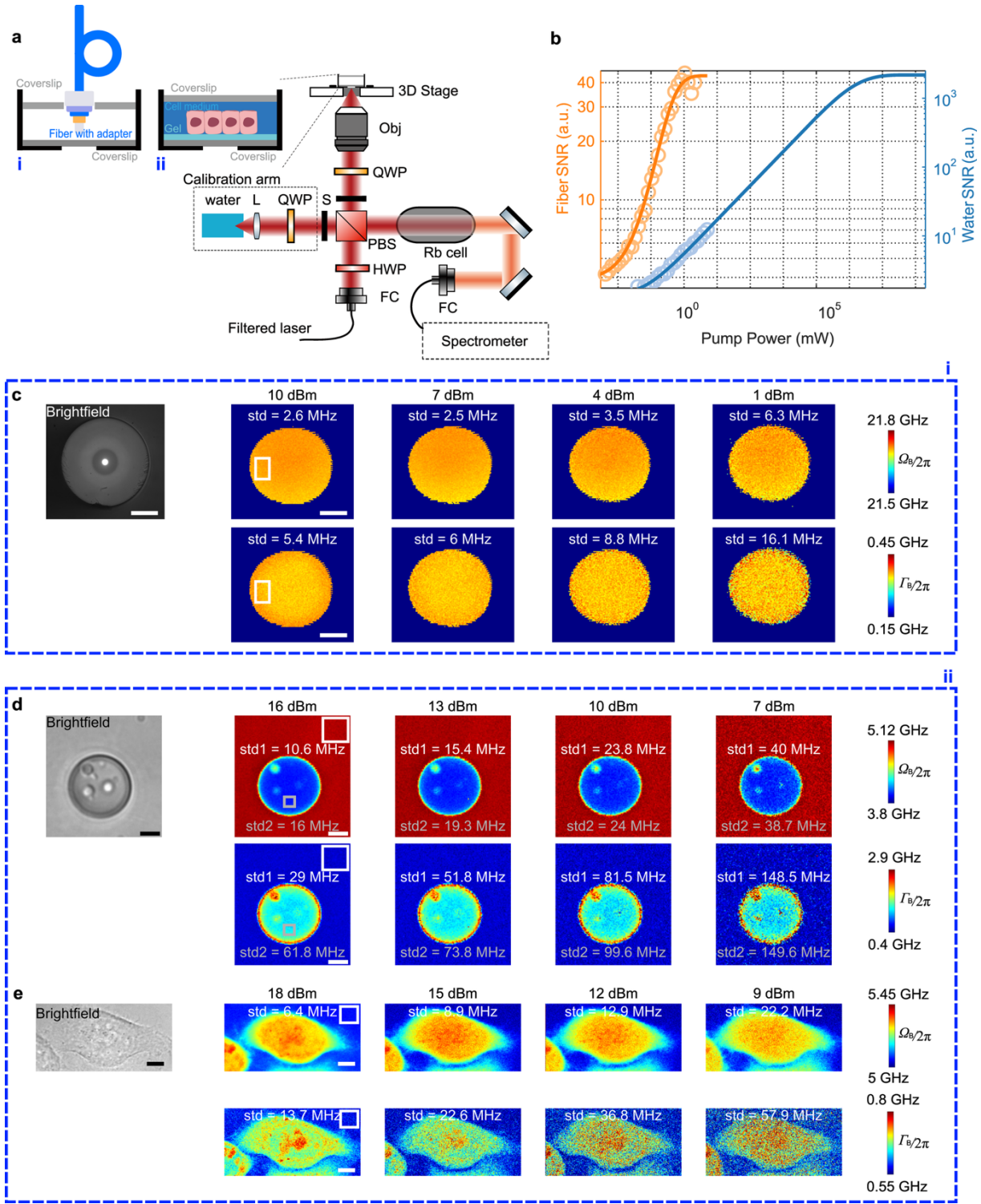

**Figure 5 | Implementation of the proposed framework on Brillouin imaging and microscopy. a,** Experimental setup. Obj: objective; QWP: quarter-wave plate; S: optical shutter; PBS: polarizing beam splitter; HWP: half-wave plate. **b,** SNR as a function of pump power for two samples: the fiber sample for Brillouin imaging (theoretical curve) and the other one is double distilled water sample for Brillouin microscopy. For the fiber and water cases, the pump power varies from -20.8 dBm to 13.2 dBm and from -4 dBm to 17 dBm, respectively, in 1 dB step, with the system bandwidth $B_m$ set to 50 kHz and 50 Hz, respectively. The preparation method of the water sample is consistent with that described for sample ii in panel

**a**. **c,** Brillouin imaging of a fiber sample with brightfield, Brillouin shift and Brillouin linewidth images under different pump powers. The scale bars in the brightfield and Brillouin images are 40 μm and 2 μm, respectively. $\Omega_B/2\pi$ on the colorbar denotes the Brillouin shift (see **SI Fig. 7** for a detailed analysis on the Brillouin frequency shift of the fiber sample), while $\Gamma_B/2\pi$ represents the Brillouin linewidth. White and gray text labels overlaid on the images indicate the fitting precision (for both shift and linewidth) corresponding to the white and gray square regions of interest (ROI, selected based on a region with strongest Brillouin amplitude as illustrated in **SI Fig. 8**), respectively. **d,** Brillouin microscopy of a phantom bead in agarose with brightfield, Brillouin shift and Brillouin linewidth images under different pump powers. The scale bars in the brightfield and Brillouin images are both 5 μm. The white square ROI corresponds to a region within the agarose, while the gray square ROI marks a part area of an embedded phantom bead. **e,** Brillouin microscopy of a HeLa cell with brightfield, Brillouin shift and Brillouin linewidth images under different pump powers. The scale bars in the brightfield and Brillouin images are both 10 μm. The white square ROI identifies a region within the culture medium surrounding a HeLa cell.

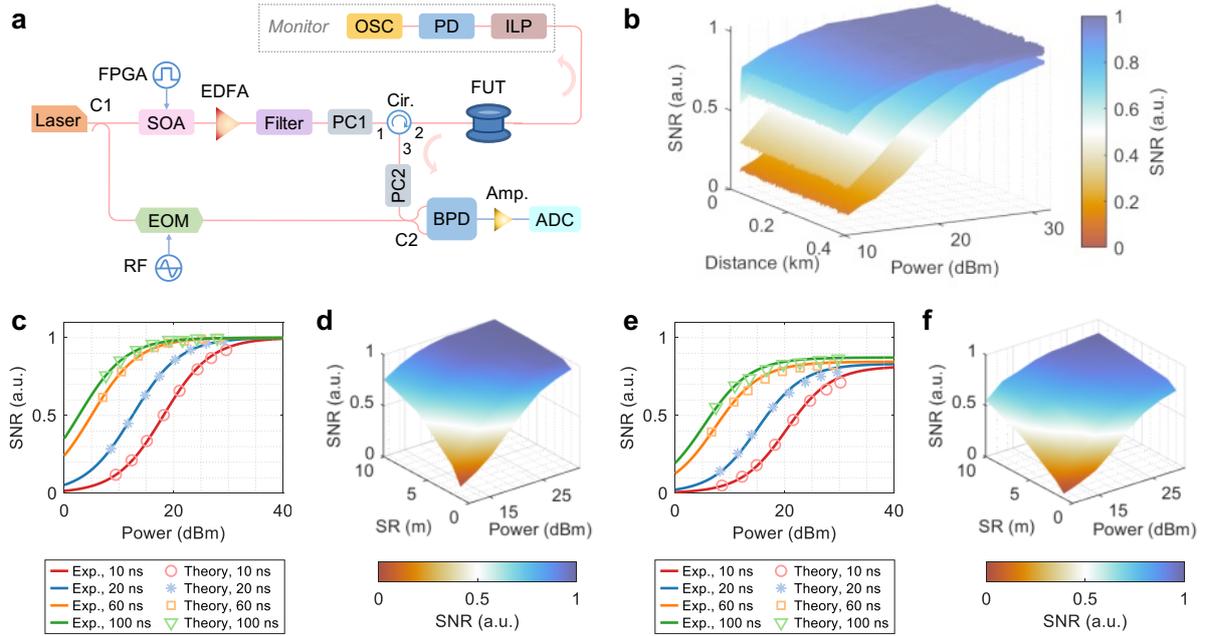

**Figure 6 | Single-pulse BOTDR experimental configuration and results. a,** Experimental configuration of single-pulse BOTDR based on PMF. FPGA: field-programmable gate array; SOA: semiconductor optical amplifier; EDFA: erbium-doped fiber amplifier; EOM: electro-optic modulator; ILP: in-line polarizer; PD: photodetector; OSC: oscilloscope. **b,** Measured SNR surfaces representing SNR distribution along a 400 m-long PM fiber for spatial resolutions of 1 m, 2 m, 6 m, and 10 m, respectively. The different values of the SNR are represented by the different colors on the colorbar to the right. **c,** Correspondence between the theoretically calculated and experimentally measured SNR values for a 400 m PMF, covering spatial resolutions of 1 m, 2 m, 6 m and 10 m. **d,** Theoretical SNR behavior of a 400 m PMF under the combined influence of varying spatial resolutions and pump pulse powers. **e,** Correspondence between the theoretically calculated and experimentally measured SNR values for a 1.9 km standard SMF, following the same parameter settings and adjustment logic as in **c**. **f,** Theoretical SNR behavior of a 1.9 km standard SMF under the combined influence of various spatial resolutions and various pump pulse powers.

**Methods**
**Data generation for the 2D SNR map shown by Fig. 2f.**
The 2D SNR map shown in **Fig. 2f** is constructed from a large ensemble of realizations of $P_{Sp}(n)$, the discrete-time representation of the power envelope of the spontaneous Brillouin field $\vec{E_{Sp}}(t)$, under varying sampling conditions. As established theoretically, the stochastic properties of $\vec{\rho}(t)$ and $\vec{E_{Sp}}(t)$ are equivalent, owing to the similar structure of their Fourier expanded expressions (see **SI** Eqs. (16) and (23)). To simplify data generation, we produce the discrete-time envelope of $\vec{\rho}(t)$, denoted as $P_\rho(n)$, which we treat as stochastically equivalent to $P_{Sp}(n)$.

In detail, $P_\rho(n)$ is obtained by squaring the discrete-time form of **SI Note 1** Eq. (8) that rigorously models $\vec{\rho}(t)$, followed by low-pass filtering of the result. This process is repeated 500 times, each with a fixed carrier frequency $\omega_\rho$ but with randomly generated amplitude $A^f$ and phase $\varphi^f$, respectively. The resulting dataset consists of 500 independent realizations of $P_\rho(n)$. The key parameters include: phonon energy decay time $\tau$ = 6 ns, delay time $\Delta t$ = 0.5 ns, carrier frequency of $\omega_\rho/2\pi$ = 220 MHz (chosen arbitrarily as it does not impact stochastic properties of $P_\rho(n)$), sampling rate of 1 GSa/s (to ensure dense sampling), low-pass filtering bandwidth of 30 MHz, and 35,000 samples per realization.

The abovementioned 500 realizations of $P_\rho(n)$ form a $500 \times 35000$ matrix. The SNR is computed for each row under varying sampling intervals and number of sampling points. Specifically, sampling interval/$\tau$ is set from 0 to 6, in 1/6 increments, and the number of sampling points ranges from 20 to 700 points, in 10-point increments. For each combination of sampling interval and sampling number, we compute the SNR for all 500 realizations and average the results to obtain a single pixel value on the 2D SNR map shown in **Fig. 2f**.

**Experimental setup and data processing**
**Coherent detection**. The experimental configuration of coherent detection at 1550 nm is shown in **Fig. 3a**. A CW light from a distributed feedback laser operating at 1550 nm is split into pump and OLO branches using a PM optical coupler (C1). In the pump branch (upper arm), the CW light serves directly as the pump. A PM VOA controls the optical power of pump light before it is launched into a 400 m-long PM FUT via a non-PM circulator. A PC (PC1) preceding the circulator is adjusted to maximize the incident pump power, ensuring proper alignment of the incident polarization. The pump powers range from -25.5 dBm to 3.5 dBm in 1 dB step. In the OLO branch (lower arm), the CW light is modulated by an EOM driven by an RF signal, to produce a carrier-suppressed dual-sideband OLO wave. A narrowband optical filter selects the lower OLO sideband that interacts with the Stokes SpBS light. The optical power of the selected single-sideband OLO is set to approximately 3 dBm. Another PC (PC2) aligns the polarization direction of the SpBS light with that of the OLO. At the receiver stage, the OLO and SpBS light are mixed via a 50/50 PM coupler (C2) and then detected by a BPD with a 400 MHz bandwidth. The resulting electrical beating signal after BPD is amplified by a low-noise electrical amplifier (Mini-Circuits ZFL-500LNB+) to suppress the ADC quantization noise, and then directly acquired at 1 GSa/s by the ADC (**Fig. 3a(i)**).

The beating signal is digitally post-processed to extract the Brillouin spectrum. Specifically, beat signals with a sampling duration of 10.24 μs (i.e., the number of sampling points $N_F$ = 10,240) are continuously acquired over 10,000 realizations. Each of beat signal undergoes an FFT, followed by modulus extraction, squaring, and normalization by $N_F$, yielding the normalized Brillouin power spectral density (PSD). To maintain consistency with the units in Eq. (2), the obtained PSD is divided by the square of the squared BPD conversion gain ($k_{VA}^2$) and by the gain factor of the electrical amplifier ($k_{amp}$). The result under a high pump power (2.5 dBm, 1.8 mW) is illustrated in **Fig. 3b**. The single-sided spectral range spans 500 MHz, determined by the sampling rate of 1 GSa/s. The light orange curve represents the Brillouin spectrum from a single realization, while the dark orange curve corresponds to the average of 10,000 spectra. The Brillouin peak is centered at approximately 220 MHz, consistent with the frequency offset between the OLO and SpBS signals. The noise spectrum shown in **Fig. 3c** is obtained by applying the same procedure in the absence of the pump light as that in **Fig. 3b**. For each pump power, the same post-processing procedure is applied as previously described, followed by subtraction of the noise baseline (**Fig. 3c**) to obtain a bias-free Brillouin spectrum. The mean peak amplitude across the 10,000 bias-free spectra is taken as the signal, the STD of the peak amplitude as the noise level, and their ratio as the SNR. These are respectively represented by the light blue circles in **Fig. 3d**, **e**, and **g**. As a control experiment, CW laser output (without Brillouin scattering) with power matched to that of the SpBS light is used instead, yielding the pink asterisks in **Fig. 3d**, **f**, and **g**. The parameters used to obtain the theoretical curves in **Fig. 3d-g** based on Eq. (2) are as follows: the responsivity of the photodiode $\mathcal{R}_p$ = 0.95 A/W; the BPD conversion gain $k_{VA}$ = 5000 V/A; the gain factor of the electrical amplifier $k_{amp}$ = 316; the mean SpBS optical power (including Stokes and anti-Stokes components) $\overline{P_{Sp}} = \beta_1(e^{\beta_2 P_p} - 1)$, where $\beta_1$ = 0.25, $\beta_2$ = 101.86, and $P_p$ is the pump power; the dual-sideband OLO power $P_{Lo}$ = 3 dBm; $SNR_{Sp}$ = 1 (**SI Note 3**); the STD of squared detection noise $\pi B_{Sp} \sigma_e^2 / 4 = 2\times 10^{-13}$ A$^2$.

Under a pump power of 2.5 dBm, a 30 μs-long beat signal is sampled at 1 GSa/s sampling rate with 500 realizations. A corresponding noise signal is recorded under identical conditions without pump light. Both the beat signal and noise undergo the same digital envelope detection as the simulated case (**Fig. 2f**). The true experimental envelope signal is derived by subtracting the noise envelope baseline from the directly obtained envelope (i.e., LPF output). The resulting envelope signal is evaluated for SNR with the same sampling intervals and sampling points as the simulated case. Each SNR value is obtained from 500 repeated measurements, yielding the results in **Fig. 3h**. The phonon energy decay time $\tau$ in the experiment is estimated to be 5.4 ns.

Additionally, an experimental envelope detection scheme shown by **Fig. 3a(ii)** is implemented to physically examine the impact of system bandwidth on SNR. In this approach, the amplified beating signal passes through a physical BPF (188 MHz centre frequency, 50 MHz bandwidth), selecting the desired spectral components of both signal and noise. Then a commercial logarithmic ED (ANALOG DEVICES, AD8318) is utilized to directly extract the temporal envelope signal. To explore the effects of system bandwidth, physical LPFs with two bandwidths of 50 MHz and 2 MHz are employed before ADC conversion. For each LPF configuration, the pump power is swept from 3.5 dBm to -25.2 dBm in 1-dB steps. At each pump power level, 10,000 realizations of the envelope signal (10 μs duration, sampled at 1 GSa/s) are recorded. Noise measurements (without pump light) are subsequently performed under identical sampling conditions. The unbiased envelope signal at each power level is obtained by subtracting the noise floor from the directly measured envelope signal. The mean values and fluctuation STD of the 10,000 unbiased envelope signals are extracted as signal mean and noise STD, respectively. Their ratio defines the SNR, as presented in **Fig. 3i** and **j**.

**Direct detection.** The experimental setup for direct detection of SpBS signal is shown in **Fig. 4a**. A frequency-locked, CW tapered amplifier laser operating at 780 nm is injected into one of the two output ports of a 1×2 50:50 PMF coupler. The 10 m-long PMF is connected to the input port of the coupler, and the backscattered SpBS signal is collected from the second output port of the coupler and directed through a custom-built Rubidium vapor gas-cell filter, which provides ~50 dB extinction to suppress residual elastic scattering, and finally entering the VIPA-based spectrometer.

The VIPA-based spectrometer firstly includes a fiber collimator to collimate the SpBS light. The light is line-focused by a cylindrical lens. Different frequency components of the input light emerge at different angles and are spatially dispersed. The spectrally resolved signal is then imaged onto different position of a scientific complementary metal oxide semiconductor (sCMOS) camera. The Brillouin frequency shift and linewidth are extracted based on the known angular dispersion characteristics of the VIPA.

An example image acquired by the sCMOS camera is shown in **SI Fig. 4a**, where the two resolved peaks (from top to bottom) correspond to the anti-Stokes and Stokes Brillouin signals of the 10 m-long fiber sample. A gray value intensity profile along the longitudinal direction of the selected region in the image is shown in **SI Fig. 4b**. Lorentzian fitting is applied to each of the two Brillouin peaks to extract their spectral characteristics. The extracted Brillouin spectrum is corrected by subtracting the offset introduced by the camera's background setting. To ensure consistency with the units in Eq. (3), the resulting grayscale spectrum is divided by the conversion gain coefficient of camera $k_{IG}$. We analyze the STD of the Stokes peak intensity as a function of pump power to quantify the noise behavior and signal stability under varying excitation conditions.

The theoretical curves in **Fig. 4b** and **c** are obtained based on below parameters: the camera responsivity $\mathcal{R}_c = 0.5$ A/W, the camera conversion gain $k_{IG} = 2\times10^{18}$, the Brillouin linewidth $B_{Sp} = 120$ MHz, the noise-induced bias grey value is 1500, and the constant STD of camera background noise $\sigma_{re} = 25/k_{IG}$. In addition to these common parameters, in **Fig. 4b,** we use $\beta_1 = 1.25\times10^3$, $\beta_2 = 6.47$, and measurement bandwidth $B_m = 50$ kHz, while in **Fig. 4c** we use $\beta_1 = 2.24\times10^4$, $\beta_2 = 5.6$, and $B_m = 5$ kHz.

**Brillouin imaging and microscopy setup with VIPA system.** The experimental setup for Brillouin imaging and microscopy at 780 nm is shown in **Fig. 5a**. A frequency-locked, CW tapered amplifier laser at 780 nm is spectrally filtered using an ASE-suppressed module (as described in **SI Fig. 7**) and delivered to the main optical path via a PM fiber. The output laser is collimated using a fiber collimator and focused onto the sample mounted on a high-precision XYZ piezoelectric stage using a 60×, 0.7 NA objective lens. The backscattered Brillouin signal is collected along the same optical path, filtered by a Rubidium vapor cell (to suppress elastic scattering), and coupled into the VIPA-based spectrometer for spectral analysis.

A separate calibration arm, incorporating the standard water sample, is used to correct for thermal drift of the VIPA dispersion. The Brillouin shift of water at room temperature is well known to be 5.07 GHz at 780 nm and serves as a reference for calibration. A computer-controlled shutter alternates between the main imaging arm and the calibration arm, enabling scheduled corrections throughout the imaging process to ensure measurement accuracy.

For fiber sample scanning in **Fig. 5c**, the XYZ three-axis input ports of the piezoelectric stage are independently controlled by three analog output channels from a data acquisition module (NI USB-6343 BNC), using a custom-built LabVIEW program. The data acquisition module also generates its internal clock signal of the created output voltage task, which serves as an external trigger for the sCMOS camera to ensure synchronized image acquisition. Due to the strong scattering signal from the fiber, the camera is operated in low gain mode. The scanning area is set to 18 μm × 18 μm with XY step size of 0.1 μm, and an exposure time of 20 μs per pixel. To accommodate the limited scanning speed of the piezo stage, a dwell time of 10 ms is implemented via the LabVIEW program. Additionally, to ensure positional stability after each scan step, a 5 ms trigger delay is configured in Micro-Manager for the sCMOS camera.

For phantom bead and HeLa cell sample scanning in **Fig. 5d** and **e**, the sCMOS camera provides the trigger signal to the data acquisition module, enabling synchronized operation of sample scanning (via piezo control voltages) and image acquisition. The camera is operated in high gain mode to compensate for the weaker scattering signal from these samples. For the phantom bead sample, the scan area is 28 μm × 28 μm with XY step size of 0.25 μm and an exposure time of 20 ms per pixel. For the HeLa cell sample, the scanning area is 60 μm × 30 μm with the same step size and exposure time.

In addition to the parameters shared with **Fig. 4b** and **c**, the theoretical fiber curve in **Fig. 5b** is calculated using $k_{IG} = 2 \times 10^{18}$, $\beta_1 = 1.39 \times 10^3$, $\beta_2 = 178$, and $B_m = 50$ kHz; while for the water curve in **Fig. 5b**, the parameters are $k_{IG} = 1 \times 10^{20}$, $\beta_1 = 6.63 \times 10^8$, $\beta_2 = 1.5 \times 10^{-4}$, and $B_m = 50$ Hz.

**Single-pulse BOTDR based on PMF.** The experimental setup for single-pulse BOTDR based on PMF is shown in **Fig. 6a**. A CW light from a distributed feedback laser operating at 1550 nm is split into pump pulse and OLO branches by a PM coupler (C1). Unlike the fundamental coherent detection at 1550 nm, the pump pulse branch (upper arm) employs a high-extinction-ratio SOA driven by an FPGA to intensity-modulate the light into a single optical pulse. Before being launched into a 400 m-long PM fiber under test (FUT), the optical pulse successively passes through an EDFA (for peak power adjustment from 10 dBm to 31 dBm in 3 dB increment), a narrowband optical filter (for suppression of the broadband amplified spontaneous emission (ASE) noise), and a PC (PC1, to align the incident polarization with the principal axis of the PM FUT). The duration of pump pulse is set to 10 ns, 20 ns, 60 ns, and 100 ns, corresponding to the spatial resolutions of 1 m, 2 m, 6 m, and 10 m, respectively. After propagating through the PM FUT, the pump pulse light subsequently passes through an in-line polarizer before being detected by a PD with a 400 MHz bandwidth. The PD output is routed to a commercial OSC for monitoring the pulse shape and power. PC1 is adjusted to maximize the monitored pulse peak power, confirming proper alignment of the incident polarization. The OLO branch (lower arm) follows the same configuration as in **Fig. 3a**, except that both sidebands are retained. The optical power of dual-sideband OLO is set to approximately 3 dBm. PC2 aligns the polarization of the SpBS light with that of the OLO. The configuration used at the receiver stage is the same as that in **Fig. 3a(i)**. The uncorrelated Stokes and anti-Stokes signals beat with the lower and upper sidebands of OLO, respectively, resulting in an electrical beating signal in the distance domain merged with zero-mean additive noise (detection noise). For each combination of pulse power and duration, 10,000 realizations are recorded within a 6.144 μs sampling window. Instead of a physical ED, digital post-processing is employed to extract the intensity envelope signal (similar with the method used for **Fig. 3h**). Specifically, the acquired 10,000 beating signals under every parameter set are first filtered by a digital BPF, with a bandwidth approaching the FWHM of the Brillouin spectrum. The filtered signal is then squared, followed by low-pass filtering with a bandwidth on the same order of magnitude as the FWHM of the Brillouin spectrum, to yield the intensity envelope signal. After correcting for the noise-induced bias by applying the identical processing procedure to the ADC output recorded without the pump pulse, the mean values and STD across 10,000 measurements are computed per pulse parameter configuration. The resulting SNR surface along the fiber is shown in **Fig. 6b**.

The parameters required in the theoretical calculations of **Fig. 6** include: dual-sideband OLO power $P_{Lo}$ = 3 dBm, backscattering coefficient $k_{Sp}$ which is medium-dependent and the common value in silica is about -95 dB/m, effective group index of the propagating mode $n_{eff}$ = 1.44 which is medium-dependent, mode group velocity in vacuum $c = 3 \times 10^8$ m/s, pump pulse duration $D_p$ = 10, 20, 60, 100 ns. The pump pulse power $P_p$, ranging from 10 dBm to 31 dBm with an interval of 3 dB, are merely theoretical reference powers and may deviate from the actual values due to the influence of nonlinear effects (e.g., stimulated Brillouin scattering) and the limited performance of the devices (e.g., EDFA).

**Single-pulse BOTDR based on SMF.** The single-pulse BOTDR experiment based on SMF adopts a standard polarization-scrambled coherent detection scheme [45]. The FUT used is a 1.9 km-long SMF. The OLO power, pump pulse power and duration, as well as the detection components (BPD, amplifier, ADC), are identical to those in the PMF-based BOTDR setup. The electrical beating signal is recorded with a duration of 25.088 μs (25,088 samples at 1 GSa/s), repeated 10,000 times for each pulse configuration. Envelope extraction is performed digitally using the same processing and filtering parameters as in the PMF case.

**Sample preparations**
**Water sample preparation.** The water sample is prepared by sandwiching 10 μL double-distilled water between two #1.5 coverslips, separated by a 120 μm thick imaging spacer (Grace Bio-Labs, SecureSeal).

**PDMS beads sample preparation.** To prepare polydimethylsiloxane (PDMS) beads embedded in agarose gel, 5 μL 1% (w/v) low-melting-point agarose solution in water is first deposited onto a #1.5 coverslip pre-mounted with a 120 μm thick imaging spacer. After gelation, PDMS beads are applied onto the gel surface, followed immediately by the addition of 7 μL of the same agarose solution. A second #1.5 coverslip is placed on top to seal the sample. This three-layer configuration effectively reduces pump light reflection.

**Fiber sample preparation for Brillouin imaging.** The fiber sample (see **SI Fig. 6** for details) is assembled by attaching a fiber adaptor (Thorlabs, SM1FCA) and a #1.5 coverslip with one imaging spacer with thickness of 120 μm. For preventing the high reflection of pump light from the interface of fiber end and air, index-matching refractive index liquid (Cargille, Series AA, 1.414) is filled in the interspace. A 1 km SMF fiber is then connected to the fiber adaptor slowly to avoid the formation of air bubbles. The output power from the distal end of the fiber is monitored for checking the focus point location and determining the scan range and center.

**Cells.** HeLa cell (CBP60232, Cobioer) are cultured according to the American Type Culture Collection (ATCC) guidelines. Cells are maintained in DMEM (Gibco, 11960044) supplemented with 10% (v/v) FBS (Gibco, 30044333) and 100 U/mL penicillin-streptomycin (Gibco, 15140122). For imaging, cells are seeded at a density of 5000 cells/cm$^2$ onto polyacrylamide (PAA)-coated imaging dishes (MATRIGEN, SV3510-EC-12) and allowed to adhere overnight.

**References**

1. Boyd, R. W. *Nonlinear Optics*. (Academic Press, Burlington, MA, 2008).
2. Brillouin, L. Diffusion de la lumière et des rayons X par un corps transparent homogène - Influence de l'agitation thermique. *Ann. Phys.* **9**, 88–122 (1922).
3. Kurashima, T., Horiguchi, T., Izumita, H., Furukawa, S. & Koyamada, Y. Brillouin Optical-Fiber Time Domain Reflectometry. *IEICE Trans. Commun.* **E76-B**, 382–390 (1993).
4. Shimizu, K., Horiguchi, T., Koyamada, Y. & Kurashima, T. Coherent self-heterodyne detection of spontaneously Brillouin-scattered light waves in a single-mode fiber. *Opt. Lett.* **18**, 185 (1993).
5. Mizuno, Y., Zou, W., He, Z. & Hotate, K. Proposal of Brillouin optical correlation-domain reflectometry (BOCDR). *Opt. Express* **16**, 12148–12153 (2008).
6. Motil, A., Bergman, A. & Tur, M. [INVITED] State of the art of Brillouin fiber-optic distributed sensing. *Opt. Laser Technol.* **78**, 81–103 (2016).
7. Mizuno, Y., Hayashi, N., Fukuda, H., Song, K. Y. & Nakamura, K. Ultrahigh-speed distributed Brillouin reflectometry. *Light Sci. Appl.* **5**, e16184–e16184 (2016).
8. Hartog, A. H. An Introduction to Distributed Optical Fibre Sensors. (2017).
9. Yang, F., Gyger, F. & Thévenaz, L. Intense Brillouin amplification in gas using hollow-core waveguides. *Nat. Photonics* **14**, 700–708 (2020).
10. Merklein, M., Kabakova, I. V., Zarifi, A. & Eggleton, B. J. 100 years of Brillouin scattering: Historical and future perspectives. *Appl. Phys. Rev.* **9**, 041306 (2022).
11. Huang, L., Fan, X., He, H., Yan, L. & He, Z. Single-end hybrid Rayleigh Brillouin and Raman distributed fibre-optic sensing system. *Light Adv. Manuf.* **4**, 171–180 (2023).
12. Youn, J. H., Kim, J. H. & Song, K. Y. Brillouin Optical Correlation Domain Analysis at MHz Sampling Rates Using Lock-in-Free Orthogonally Polarized Probe Sidebands. *J. Light. Technol.* **42**, 6312–6317 (2024).
13. Romanet, M., Rochat, É., Beugnot, J.-C. & Huy, K. P. Extended-range and faster photon-counting Brillouin optical time domain reflectometer. *Optica* **12**, 564–569 (2025).
14. Eggleton, B. J., Poulton, C. G., Rakich, P. T., Steel, M. J. & Bahl, G. Brillouin integrated photonics. *Nat. Photonics* **13**, 664–677 (2019).
15. Gyger, F. *et al.* Observation of Stimulated Brillouin Scattering in Silicon Nitride Integrated Waveguides. *Phys. Rev. Lett.* **124**, 013902 (2020).
16. Bevilacqua, C. & Prevedel, R. Full-field Brillouin microscopy based on an imaging Fourier-transform spectrometer. *Nat. Photonics* **19**, 494–501 (2025).
17. Scarcelli, G., Besner, S., Pineda, R., Kalout, P. & Yun, S. H. In Vivo Biomechanical Mapping of Normal and Keratoconus Corneas. *JAMA Ophthalmol.* **133**, 480–482 (2015).
18. Scarcelli, G. & Yun, S. H. Confocal Brillouin microscopy for three-dimensional mechanical imaging. *Nat. Photonics* **2**, 39–43 (2008).
19. Scarcelli, G. *et al.* Noncontact three-dimensional mapping of intracellular hydromechanical properties by Brillouin microscopy. *Nat. Methods* **12**, 1132–1134 (2015).
20. Prevedel, R., Diz-Muñoz, A., Ruocco, G. & Antonacci, G. Brillouin microscopy: an emerging tool for mechanobiology. *Nat. Methods* **16**, 969–977 (2019).
21. Kabakova, I. *et al.* Brillouin microscopy. *Nat. Rev. Methods Primer* **4**, 1–20 (2024).
22. Elsayad, K. *et al.* Mapping the subcellular mechanical properties of live cells in tissues with fluorescence emission–Brillouin imaging. *Sci. Signal.* **9**, rs5–rs5 (2016).
23. Margueritat, J. *et al.* High-Frequency Mechanical Properties of Tumors Measured by Brillouin Light Scattering. *Phys. Rev. Lett.* **122**, 018101 (2019).

24. Bailey, M. *et al.* Viscoelastic properties of biopolymer hydrogels determined by Brillouin spectroscopy: A probe of tissue micromechanics. *Sci. Adv.* **6**, eabc1937 (2020).
25. Bevilacqua, C. *et al.* High-resolution line-scan Brillouin microscopy for live imaging of mechanical properties during embryo development. *Nat. Methods* **20**, 755–760 (2023).
26. Zhang, J., Nikolic, M., Tanner, K. & Scarcelli, G. Rapid biomechanical imaging at low irradiation level via dual line-scanning Brillouin microscopy. *Nat. Methods* **20**, 677–681 (2023).
27. Keshmiri, H. *et al.* Brillouin light scattering anisotropy microscopy for imaging the viscoelastic anisotropy in living cells. *Nat. Photonics* 1–10 (2024) doi:10.1038/s41566-023-01368-w.
28. Boyd, R. W., Rzązewski, K. & Narum, P. Noise initiation of stimulated Brillouin scattering. *Phys. Rev. A* **42**, 5514–5521 (1990).
29. Gaeta, A. L. & Boyd, R. W. Stochastic dynamics of stimulated Brillouin scattering in an optical fiber. *Phys. Rev. A* **44**, 3205–3209 (1991).
30. Smith, R. G. Optical Power Handling Capacity of Low Loss Optical Fibers as Determined by Stimulated Raman and Brillouin Scattering. *Appl. Opt.* **11**, 2489–2494 (1972).
31. Beugnot, J.-C., Tur, M., Mafang, S. F. & Thévenaz, L. Distributed Brillouin sensing with sub-meter spatial resolution: modeling and processing. *Opt. Express* **19**, 7381 (2011).
32. Proakis, J. G. & Manolakis, D. G. DIGITAL SIGNAL PROCESSING. 1033.
33. Maximise the Sensitivity and Decrease the Noise in Scientific Cameras. *Oxford Instruments* https://andor.oxinst.com/learning/view/article/sensitivity-and-noise-of-ccd-emccd-and-scmos-sensors.
34. Understanding Read Noise in sCMOS Cameras. *Oxford Instruments* https://andor.oxinst.com/learning/view/article/understanding-read-noise-in-scmos-cameras.
35. Zhang, J. & Scarcelli, G. Mapping mechanical properties of biological materials via an add-on Brillouin module to confocal microscopes. *Nat. Protoc.* 1–25 (2021) doi:10.1038/s41596-020-00457-2.
36. Li, C., Lu, Y., Zhang, X. & Wang, F. SNR enhancement in Brillouin optical time domain reflectometer using multi-wavelength coherent detection. *Electron. Lett.* **48**, 1139–1141 (2012).
37. Lu, Y., Yao, Y., Zhao, X., Wang, F. & Zhang, X. Influence of non-perfect extinction ratio of electro-optic modulator on signal-to-noise ratio of BOTDR. *Opt. Commun.* **297**, 48–54 (2013).
38. Lalam, N., Ng, W. P., Dai, X., Wu, Q. & Fu, Y. Q. Performance Improvement of Brillouin Ring Laser Based BOTDR System Employing a Wavelength Diversity Technique. *J. Light. Technol.* **36**, 1084–1090 (2018).
39. Lalam, N., Ng, W. P., Dai, X., Wu, Q. & Fu, Y. Q. Performance analysis of Brillouin optical time domain reflectometry (BOTDR) employing wavelength diversity and passive depolarizer techniques. *Meas. Sci. Technol.* **29**, 025101 (2018).
40. Bai, Q. *et al.* Enhancing the SNR of BOTDR by Gain-Switched Modulation. *IEEE Photonics Technol. Lett.* **31**, 283–286 (2019).
41. Soto, M. A., Bolognini, G. & Di Pasquale, F. Enhanced Simultaneous Distributed Strain and Temperature Fiber Sensor Employing Spontaneous Brillouin Scattering and Optical Pulse Coding. *IEEE Photonics Technol. Lett.* **21**, 450–452 (2009).
42. van Deventer, M. O. & Boot, A. J. Polarization properties of stimulated Brillouin scattering in single-mode fibers. *J. Light. Technol.* **12**, 585–590 (1994).
43. Zadok, A., Zilka, E., Eyal, A., Thévenaz, L. & Tur, M. Vector analysis of stimulated Brillouin scattering amplification in standard single-mode fibers. *Opt. Express* **16**, 21692 (2008).

44. Gao, X. *et al.* Impact of optical noises on unipolar-coded Brillouin optical time-domain analyzers. *Opt. Express* **29**, 22146–22158 (2021).
45. Jin, S., Yang, Z., Hong, X. & Wu, J. Analytical Signal-to-Noise Ratio Model on Frequency-Scanned Brillouin Optical Time-Domain Reflectometry. *J. Light. Technol.* **42**, 5786–5796 (2024).
46. Jostmeier, T., Marx, B., Buntebarth, C., Rath, A. & Hill, W. Long-Distance BOTDR Interrogator with Polarization- Diverse Coherent Detection and Power Evaluation. in *Optical Fiber Sensors Conference 2020 Special Edition* T3.21 (Optica Publishing Group, Washington, DC, 2021). doi:10.1364/OFS.2020.T3.21.
47. Jin, S. *et al.* Single-Channel Integrated Sensing and Communication Based on Spontaneous Brillouin Scattering. in *ECOC 2024; 50th European Conference on Optical Communication* 1679–1682 (2024).
48. Sun, X. *et al.* Genetic-optimised aperiodic code for distributed optical fibre sensors. *Nat. Commun.* **11**, 5774 (2020).
49. Pan, X. Coherent Rayleigh-Brillouin scattering in molecular gases. *Phys. Rev. A* **69**, (2004).
50. Vieitez, M. O. *et al.* Coherent and spontaneous Rayleigh-Brillouin scattering in atomic and molecular gases and gas mixtures. *Phys. Rev. A* **82**, 043836 (2010).
51. Wang, Y. *et al.* Brillouin scattering spectrum for liquid detection and applications in oceanography. *Opto-Electron. Adv.* **6**, 220016–10 (2023).
52. Zhou, Y. *et al.* Shipborne oceanic high-spectral-resolution lidar for accurate estimation of seawater depth-resolved optical properties. *Light Sci. Appl.* **11**, 261 (2022).

## ACKNOWLEDGMENTS

We would like to thank Prof. Luc Thévenaz for insightful discussions that helped deepen our understanding of the fundamental aspects of SpBS noise, and for his valuable suggestions on the preparation of this manuscript. Z.Y. acknowledges the supports from National Natural Science Foundation of China (NSFC, 62375023; 62275028) and the Fundamental Research Funds for the Central Universities (2023RC51). F.Y. acknowledges the Excellent Young Scientists Fund from NSFC. F.Y. and L.Z. acknowledge the supports from Strategic Priority Research Program of the Chinese Academy of Sciences (XDB0650000).

## AUTHOR CONTRIBUTIONS

Z.Y. and F.Y. conceived the project. Z.Y. initialized the idea that SpBS noise can serve as the ultimate performance-limiting factor in Brillouin metrological systems. Z.Y. and S.J. developed the theory on the physical formation and stochastic properties of SpBS noise, and established the SNR models for the experimental systems presented, with support from F.Y. and M.A.S. S.J., Z.Y., and X.H. designed and carried out the coherent detection and distributed sensing experiments, for which S.J. processed and analyzed the resulting data under the guidance of Z.Y. F.Y., S.Y., and Z.D. designed and performed the direct detection, imaging, and microscopy experiments, for which S.Y., Z.D., and S.J. processed the experimental data under the supervision of F.Y. and Z.Y. J.X. provided the biological samples. S.J., Z.Y., S.Y., and F.Y. wrote the manuscript with input from all authors. J.W., X.H., L.Z. and F.Y. supervised the project.

## COMPETING INTERESTS

All authors declare no competing interests.

## DATA AVAILABILITY

The dataset and code will be made publicly available at the point of publication.

# SUPPLEMENTARY INFORMATION

## A Framework for Spontaneous Brillouin Noise: Unveiling Fundamental Limits in Brillouin Metrology

Simeng Jin[1†], Shuai Yao[2†], Zhisheng Yang[1*], Zixuan Du[2], Xiaobin Hong[1], Marcelo A. Soto[3], Jingjing Xie[4], Long Zhang[2], Fan Yang[2*], and Jian Wu[1]

[1] State Key Laboratory of Information Photonics & Optical Communications, Beijing University of Posts and Telecommunications, Beijing, China
[2] Shanghai Institute of Optics and Fine Mechanics, Chinese Academy of Sciences, Shanghai, China
[3] Department of Electronics Engineering, Universidad Técnica Federico Santa María, Valparaíso, Chile
[4] School of Physical Science and Technology & State Key Laboratory of Advanced Medical Materials and Devices, ShanghaiTech University, Shanghai, China

[*] **Corresponding authors**: zhisheng.yang@bupt.edu.cn; yang@siom.ac.cn.
[†] These authors contributed equally to this work.

**This PDF file includes:**

Supplementary Figures 1-13
Supplementary Notes 1-8
Supplementary Tables I-IV

**SUPPLEMENTARY FIGURES**

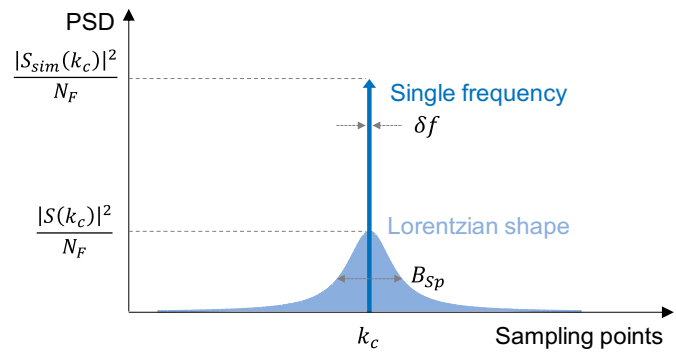

SI Figure 1: Schematic diagram of single-frequency and Lorentzian spectra.

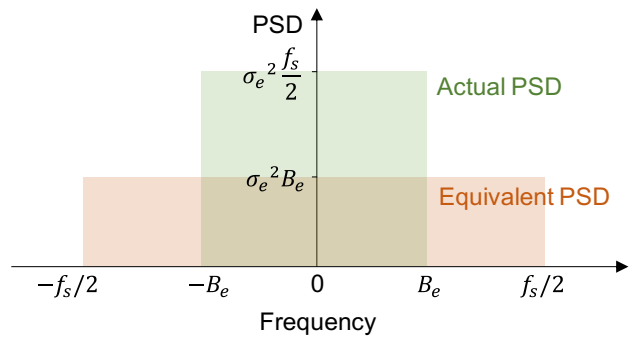

SI Figure 2: Schematic diagram illustrating the equivalent and actual noise power spectral density (PSD).

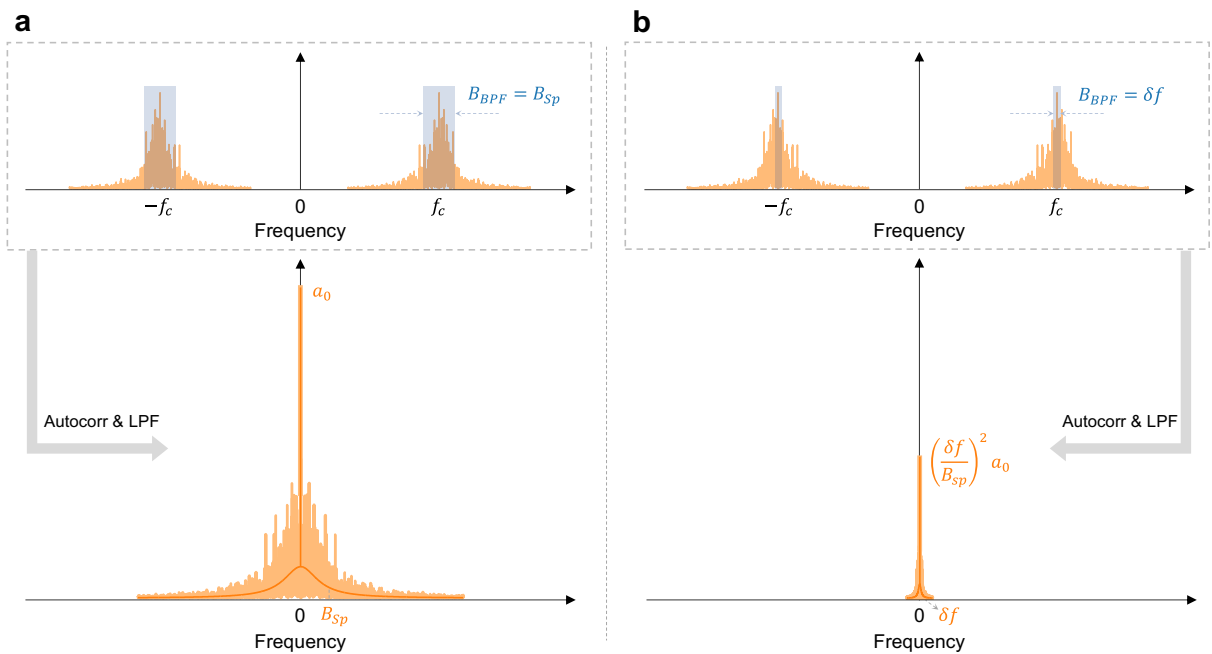

SI Figure 3: Spectral interpretation on SNR behavior of FFT-based coherent detection. **a,** Spectrum of $\overrightarrow{E_{Sp}}(t)$ with a BPF of bandwidth $B_{BPF} = B_{Sp}$, and its corresponding power envelope spectrum. **b,** Spectrum of $\overrightarrow{E_{Sp}}(t)$ with a BPF of bandwidth $B_{BPF} = \delta f$, and its corresponding power envelope spectrum.

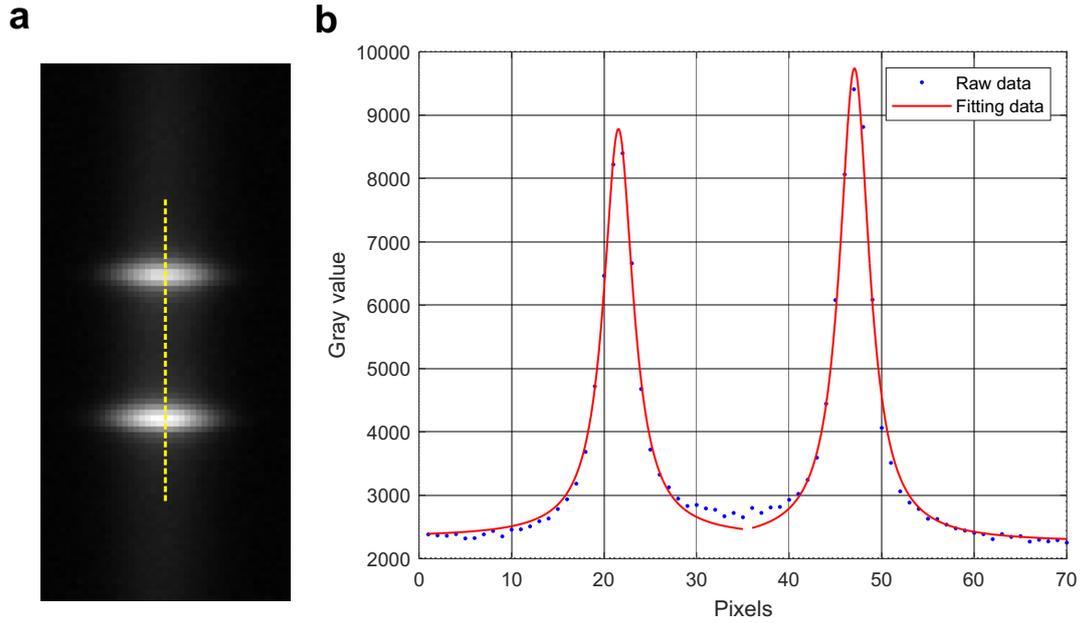

**SI Figure 4: Raw data and Lorentzian fit of the Brillouin signal from a 10 m fiber acquired via direct detection. a**, sCMOS image of the Brillouin spectrum from a 10 m-long polarization-maintaining fiber. Two distinct peaks are visible, corresponding (from top to bottom) to the anti-Stokes and Stokes Brillouin signals. Note that due to the flipped image orientation in our detection setup, the shorter-wavelength Stokes signal appears at the bottom of the image. **b**, Intensity profile (gray value plot) extracted along the longitudinal yellow dashed line in panel a, along with corresponding Lorentzian fits to the anti-Stokes and Stokes peaks.

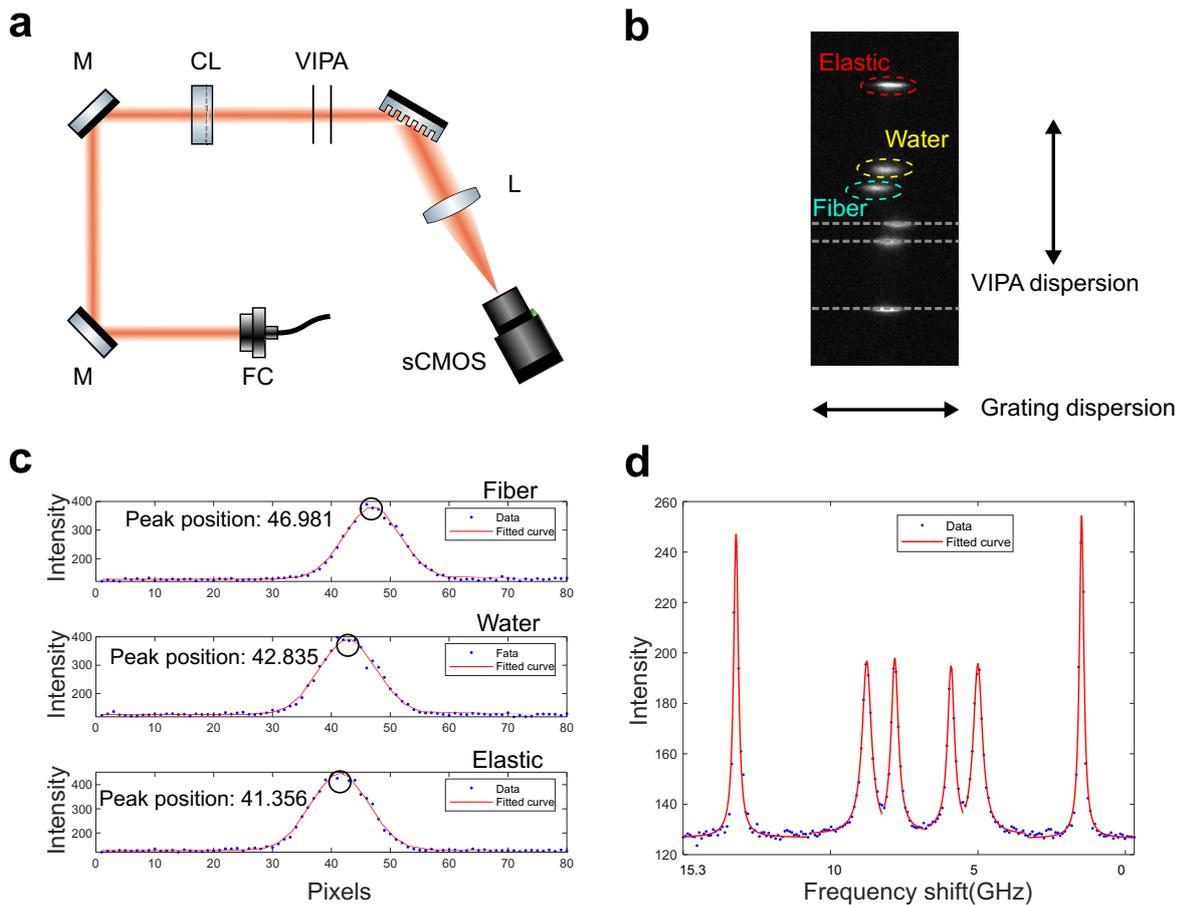

**SI Figure 5: Quantitative measurement of fiber frequency shift using VIPA with grating spectrometer. a**, Schematic of the VIPA spectrometer setup incorporating a 2D dispersive grating. FC: fiber collimator; M: mirror; CL: cylindrical lens; VIPA: Virtually Imaged Phased Array; RHG: reflective holographic grating; L: lens; sCMOS: scientific Complementary Metal–Oxide–Semiconductor camera. **b**, Image acquired from sCMOS showing elastic scattering and Brillouin signals from water and fiber, measured using the VIPA with grating spectrometer. **c**, Horizontal gray-value profiles extracted from the image in b along three gray dashed lines, corresponding (from top to bottom) to the anti-Stokes Brillouin signal of the fiber, the anti-Stokes signal of water, and the elastic scattering peak. Based on the known Brillouin frequency shift of water (5.07 GHz at 780 nm pump laser) and the linear dispersion of the grating, the relative horizontal displacement of the fiber and water signals with respect to the elastic peak (Δfiber / Δwater ≈ 3.8) suggests that the fiber's Brillouin shift lies between approximately 19.3 GHz and 35 GHz. **d**, Simultaneous plots of the three signal groups (raw and fitted data) along the VIPA dispersion axis. Using the known dispersion characteristics of the VIPA, the 15.3 GHz free spectral range (FSR) of the VIPA, and the 5.07 GHz Brillouin shift of water under 780nm laser excitation, the Brillouin frequency shift of the fiber is accurately determined to be 21.7 GHz.

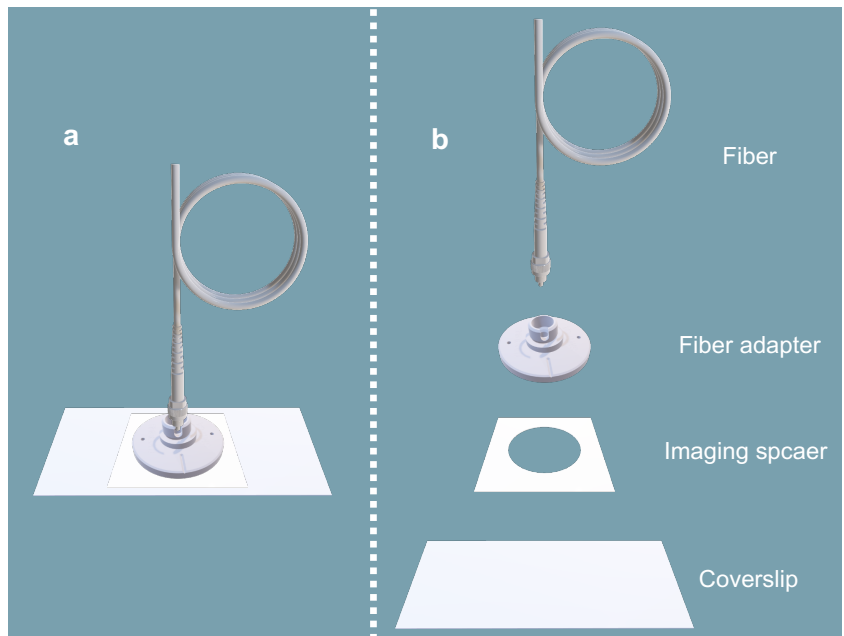

**SI Figure 6: Schematic of the fiber sample used for Brillouin imaging. a**, Overview schematic illustrating the full composition and structure of the fiber sample. **b**, Enlarged view highlighting the internal subdivisions and specific regions of interest within the fiber sample.

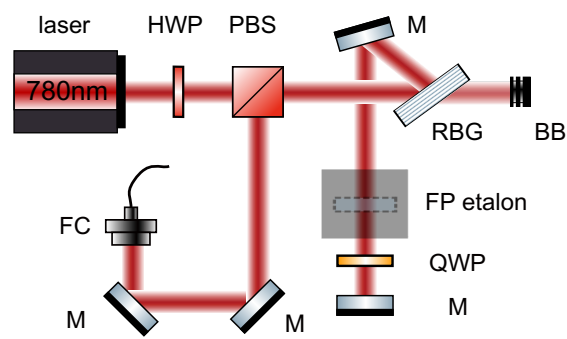

**SI Figure 7: Setup of the ASE-filtered module with temperature-controlled FP etalon.** The setup comprises a double-pass Reflective Bragg grating (RBG) and a temperature-controlled Fabry–Pérot (FP) etalon to form a narrowband filter (<250 MHz), effectively suppressing amplified spontaneous emission (ASE) noise from the 780 nm pump laser. HWP: half-wave plate; QWP: quarter-wave plate; PBS: polarizing beam splitter; M: mirror; RBG: Reflective Bragg grating; BB: beam block; FC: fiber coupler.

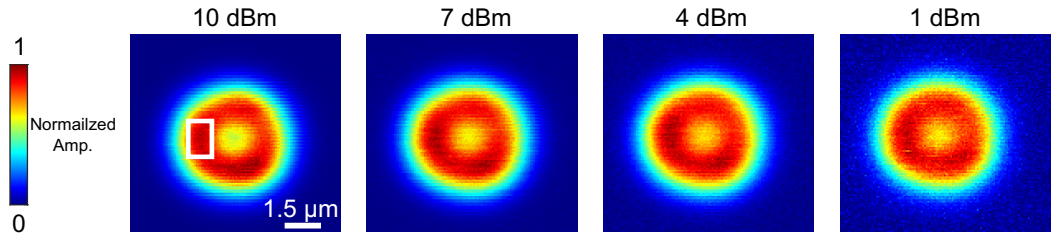

**SI Figure 8: Brillouin signal amplitude of a fiber sample using the VIPA spectrometer.** Brillouin signal amplitude images of a 1 km, 1550 nm fiber sample acquired under varying pump powers in dBm. White square in the images marks the region with the strongest signal from the fiber sample.

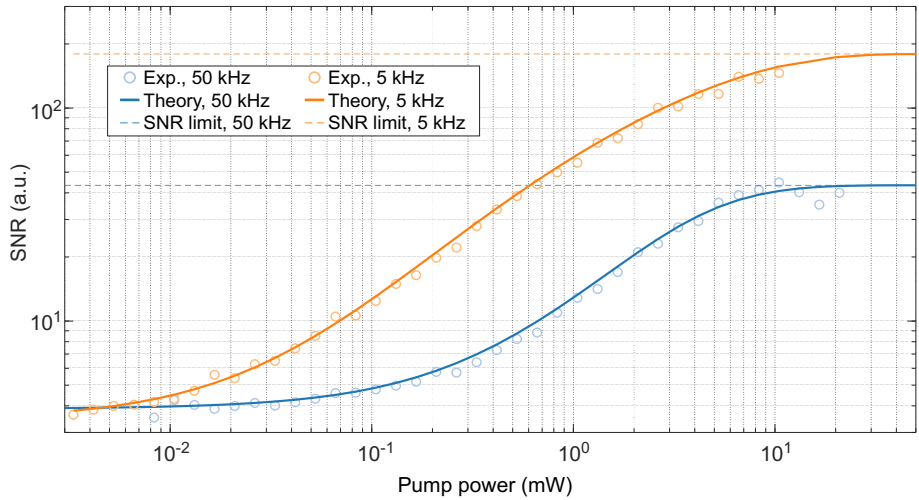

**SI Figure 9: Brillouin signal SNR analysis of a fiber sample using the VIPA spectrometer.** Theoretical and experimental SNR analysis of the fiber's Brillouin signal under different detection bandwidths. The sCMOS camera acquisition mode was set to low gain for all measurements.

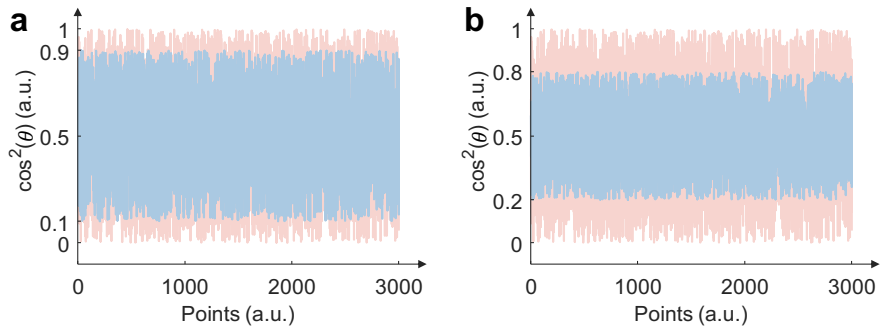

**SI Figure 10: Simulated fluctuation characteristics of $\cos^2(\theta)$ affected by the self-polarization-scrambling effect.** The blue curves illustrate the cases of **a,** $k_{Pol} = 0.8$ and **b,** $k_{Pol} = 0.6$, while the pink curves represent the reference case where $k_{Pol} = 1$.

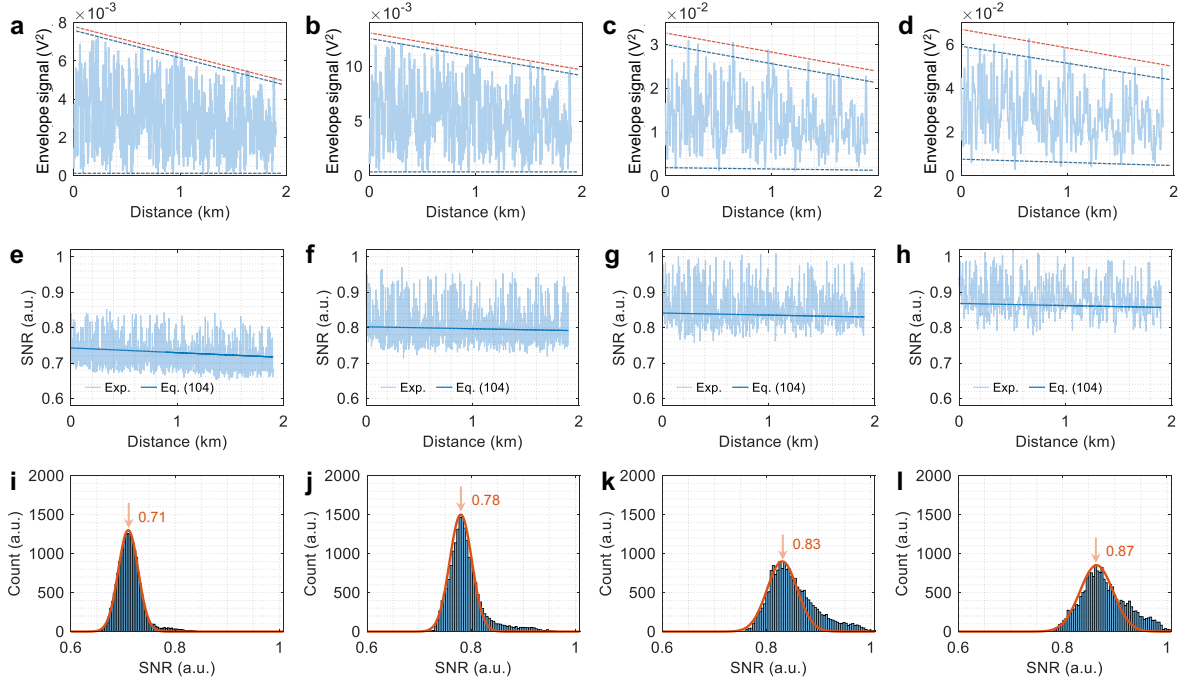

**SI Figure 11: Experimental results of the analysis on the self-polarization-scrambling effect. a-d,** Measured BOTDR envelope signals (1000 trace averaging) along 1.9 km SMF without PSc, illustrating the degree of self-polarization-srambling effect under SRs of 1 m, 2 m, 6 m and 10 m, respectively. The two blue dashed lines indicate the actual upper and lower limits of the envelope signal fluctuation, while the red dashed line indicates the maximum value of the envelope signal in the absence of self-polarization-scrambling effect. **e-h,** Light bule curves show the measured SNR profiles along 1.9 km standard SMF with PSc, under SRs of 1 m, 2 m, 6 m and 10 m, respectively. Dark blue curves show the corresponding SNR values predicted by Eq. (104). **i-l,** Distribution histograms of the SNR under SRs of 1 m, 2 m, 6 m and 10 m, respectively. The red line represents the fitting result for the primary Gaussian-like histogram.

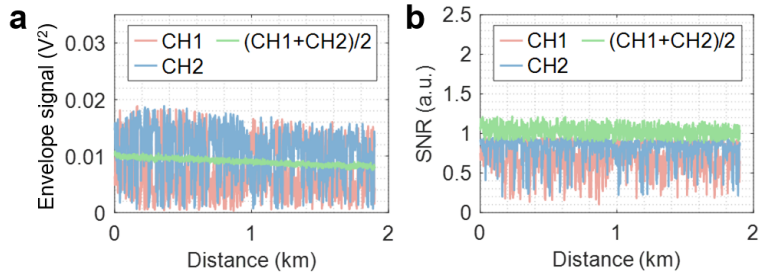

**SI Figure 12: Experimental results of BOTDR based on polarization diversity coherence receiver (PDCR). a,** Envelope signals traces and **b,** SNR traces of PDCR before and after the combination of the two channels for the 2 m SR case.

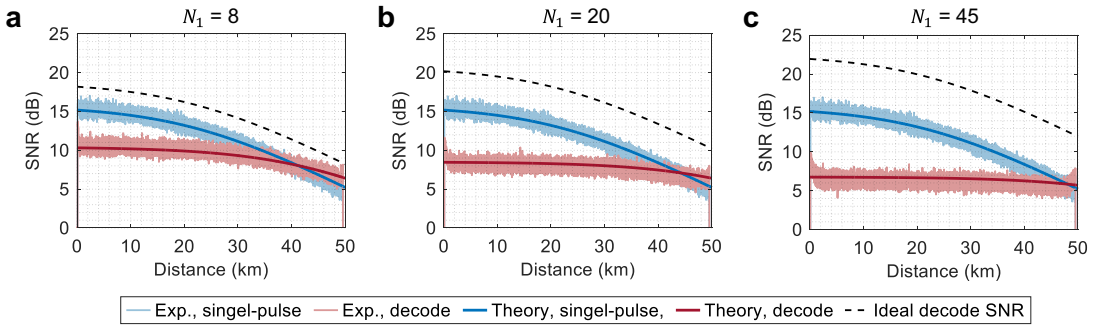

**SI Figure 13 | SNR behaviour of coded-pulse BOTDR along a 50 km standard SMF. a,** $N_1 = 8$, **b,** $N_1 = 20$, and, **c,** $N_1 = 45$. The three subplots share a single-pulse SNR curve as a reference. The duration of each coded pulse is 20 ns, matching that of the single pulse. For different coding bit numbers, the peak power of the first coded pulse is set equal to that of the single pulse, which is optimized to the modulation instability threshold.

# SUPPLEMENTARY NOTES
## SI Note 1: Mathematical characterization for spontaneous Brillouin scattering fluctuations.

This SI note provides detailed derivation process on investigating the physical formation of the intensity fluctuation of spontaneous Brillouin scattering (SpBS) and its stochastic characteristics. The derivation is based on solving standard three-wave couple partial differential equations for SpBS[1–3]:

$$\left(\frac{1}{v_g}\frac{\partial}{\partial t}+\frac{\partial}{\partial z}+\frac{\alpha}{2}\right)E_P(z,t)=0 \quad (1)$$

$$\left(\frac{1}{v_g}\frac{\partial}{\partial t}-\frac{\partial}{\partial z}+\frac{\alpha}{2}\right)E_S(z,t)=j\kappa\rho^*(z,t)E_P(z,t) \quad (2)$$

$$\frac{\partial\rho(z,t)}{\partial t}+\frac{1}{2}\Gamma\rho(z,t)=f(z,t) \quad (3)$$

where $E_P(\cdot)$, $E_S(\cdot)$, and $\rho(\cdot)$ represent the complex amplitudes of the pump light, the Stokes SpBS light and the acoustic wave (density wave), respectively. These parameters depend on both position $z$ and time $t$. The superscript * stands for the complex conjugate. Additionally, $v_g$ represents the velocity of light in the medium, defined as $v_g = c/n_{eff}$, where $c$ is the speed of light in a vacuum, and $n_{eff}$ is the effective group index of the propagating mode. Other parameters include the medium attenuation coefficient $\alpha$, the imaginary unit j, the coupling coefficient $\kappa$, and the phonon (intensity) decay rate $\Gamma$. The Langevin noise source $f$ describes the thermal (or quantum) excitation of acoustic waves in the density of the medium that leads to spontaneous Brillouin scattering (SpBS) process, which is a complex Gaussian random process with zero mean that is white in both space and time.

The methodology is on the foundation of strategically performing discretization operations on certain parameters, enabling both intuitively visualizing the relevant physical insight and later on mathematically characterizing the stochastics of the SpBS intensity fluctuations. First, we discretize the light-matter interacting length of the medium into $N_z$ ($\to \infty$) short segments (**Fig. 1** in main text), each with length of $\Delta z$ ($\to +0$). Thus the position of $k$-th segment is denoted as $z_k = k\Delta z$, where $k \in [1, N_z]$. With each segment, the local thermally activated vibration is represented by a stochastic process of Langevin noise $f(t)$, which is here modelled as the sum of $N_t$ ($\to \infty$) Dirac delta functions $\delta(\cdot)$ with a step delay $\Delta t$ ($\to +0$) (**Fig. 1(i)** in main text):

$$f(z_k,t)=A^f(z_k,t)e^{j\varphi^f(z_k,t)}=\sum_{i=1}^{N_t}A^f(z_k,t_i)\Delta t\cdot\delta(t-t_i)\cdot e^{j\varphi^f(z_k,t_i)} \quad (4)$$

where $A^f(\cdot)$ and $\varphi^f(\cdot)$ stand for the random amplitude and random phase of the complex Gaussian noise, respectively; $t_i = i\Delta t$ means the $i$-th moment of the stochastic process. The activated acoustic wave, due to its much smaller propagation velocity than that of the light, is also considered as a local, time-dependent but non-propagating process. Substituting Eq. (4) into Eq. (3), the local acoustic wave excited within the certain discrete segment $z_k$ satisfies:

$$\frac{\partial\rho(z_k,t)}{\partial t}+\frac{1}{2}\Gamma\rho(z_k,t)=\sum_{i=1}^{N_t}A^f(z_k,t_i)\Delta t\cdot\delta(t-t_i)\cdot e^{j\varphi^f(z_k,t_i)} \quad (5)$$

We take the Laplace transformation[3] of Eq. (5), leading to:

$$s\tilde{\rho}(z_k,s)+\frac{1}{2}\Gamma\tilde{\rho}(z_k,s)=\sum_{i=1}^{N_t}A^f(z_k,t_i)\Delta t\cdot e^{-t_i s}\cdot e^{j\varphi^f(z_k,t_i)} \quad (6)$$

where the ~ hat denotes the Laplace transform; $s$ is the Laplace variable. Rearranging Eq. (6) yields:

$$\tilde{\rho}(z_k,s)=\frac{\sum_{i=1}^{N_t}A^f(z_k,t_i)\Delta t\cdot e^{-t_i s}\cdot e^{j\varphi^f(z_k,t_i)}}{s+\frac{1}{2}\Gamma} \quad (7)$$

Performing the inverse Laplace transformation to Eq. (7), the time-domain solution for the local acoustic wave complex amplitude at $z_k$ is obtained:

$$\rho(z_k,t)=\sum_{i=1}^{N_t}A^f(z_k,t_i)\Delta t\cdot e^{-\frac{1}{2}\Gamma(t-t_i)}u(t-t_i)\cdot e^{j\varphi^f(z_k,t_i)} \quad (8)$$

where $u(\cdot)$ conventionally represents the Heaviside unit step sequence. Incorporating the acoustic carrier wave into Eq. (8) further yields the full time-domain representation of the local acoustic wave:

$$\vec{\rho}(z_k,t)=\rho(z_k,t)e^{j\omega_\rho t}=\sum_{i=1}^{N_t}A^f(z_k,t_i)\Delta t\cdot e^{-\frac{1}{2}\Gamma(t-t_i)}u(t-t_i)\cdot e^{j\varphi^f(z_k,t_i)}e^{j\omega_\rho t} \quad (9)$$

where $\omega_\rho$ is the angular frequency of the acoustic wave. Eq. (9) indicates that $\vec{\rho}(t)$ can be represented on the time scale as the superposition of $N_t$ waves with time delays at step of $\Delta t$ (**Fig. 1a(ii)** in main text): each wave shares the same angular frequency $\omega_\rho$, but featuring random initial amplitude $A^f(\cdot)\Delta t$ and phase $\varphi^f(\cdot)$ that are both dictated to those of corresponding Langevin noise components given by Eq. (4). In addition, the amplitude envelope of each wave exhibits an exponential decay trend, with time constant determined by the acoustic lifetime of a certain material.

Substituting Eq. (9) into Eq. (2), the SpBS light field returning from the light-matter interacting region to the pump-launching input, defined as $\overrightarrow{E_{Sp}}(t)$, is derived as:

$$\overrightarrow{E_{Sp}}(t) \stackrel{\text{def}}{=} \overrightarrow{E_S}(z_1, t) = j\kappa_1 \sum_{k=1}^{N_z} \vec{\rho}^*\left(z_k, t - \frac{z_k}{v_g}\right) \overrightarrow{E_P}\left(z_k, t - \frac{z_k}{v_g}\right) \tag{10}$$

where $z_1$ means the first segment at the start end; $\kappa_1 = \sqrt{P_P/A_{eff}}\kappa$, $P_P$ being the power of incident pump light and $A_{eff}$ is the effective mode area; $E_P(\cdot)$ is the pump light reaching $z_k$. With the substitution that considers a typical rectangular window function of the pump light:

$$\overrightarrow{E_P}(z_k, t) = E_P^0 \Delta t \left[u\left(t - \frac{z_k}{v_g}\right) - u\left(t - D_P - \frac{z_k}{v_g}\right)\right] e^{-j\omega_P t} \tag{11}$$

where $E_P^0$, $D_P$ and $\omega_P$ denote amplitude, duration, and angular frequency of the incident pump light, respectively, Eq. (10) can be rewritten as:

$$\overrightarrow{E_{Sp}}(t) = j\kappa_1 E_P^0 \Delta t \sum_{k=1}^{N_P} \vec{\rho}^*\left(z_k, t - \frac{z_k}{v_g}\right) e^{-j\left[\omega_P\left(t - \frac{z_k}{v_g}\right)\right]} \tag{12}$$

where $N_P = D_P v_g / \Delta z + 1$ is the number of segments corresponding to the light-matter interaction length of pump light. Eq. (12) straightforwardly indicates that $\overrightarrow{E_{Sp}}(t)$ is the windowing and time-delay superposition of $\vec{\rho}(t)$ at different light-matter interaction segments, as intuitively shown by **Fig. 1** in the main text.

Substituting Eq. (9) into Eq. (12) gives rises to the full expression of $\overrightarrow{E_{Sp}}(t)$:

$$\overrightarrow{E_{Sp}}(t) = j\kappa_1 E_P^0 \Delta t \sum_{k=1}^{N_P} \sum_{i=1}^{N_t} A^f(z_k, t_i) \Delta t \cdot e^{-\frac{1}{2}\Gamma\left(t - \frac{z_k}{v_g} - t_i\right)} \cdot u(t - t_i) \cdot e^{-j\varphi^f(z_k, t_i)} \cdot e^{-j\left[\omega_{Sp}\left(t - \frac{z_k}{v_g}\right)\right]} \tag{13}$$

where $\omega_{Sp} = \omega_P - \omega_\rho$ denotes the angular frequency of $\overrightarrow{E_{Sp}}(t)$. Eq. (13) is not physically intuitive but will be utilized later on in **SI** Note 2 for studying the stochastic property of the SpBS intensity.

**SI Note 2: Mathematical SNR investigation on SpBS intensity fluctuations.**
The seminal study correctly attributes SpBS intensity fluctuations to the linear amplification of light scattered by thermally generated phonons, and hence interprets the signal-to-noise ratio (SNR) as unity - mirroring the stochastic equivalence between phonon vibrations and thermal light, which likewise exhibits unity SNR[4,5]. where the latter features a unity SNR. Here, we build on this physical insight by presenting a rigorous mathematical derivation of the SpBS SNR based solely on the analytical expressions in **SI Note 1**, without invoking approximations or presuming any specific intensity distribution. By uniting this time-domain derivation with a complementary frequency-domain analysis, we deliver a comprehensive treatment of the statistical properties underlying SpBS intensity fluctuations.

*1. Analytically validating the conventional assumption of SNR=1*
The derivation on the statistics of $P_{Sp}(t)$ starts from manipulating $\vec{\rho}(z_k, t)$ expressed by Eq. (9), due to its close relation with $\overrightarrow{E_{Sp}}(t)$ as expressed by Eq. (12). First, the negative exponential function in Eq. (9) is Fourier expanded to $n$ terms as[6]:

$$e^{-\frac{1}{2}\Gamma(t-t_i)} = \sum_{x=1}^{n} A_x^e e^{j[\omega_x^e(t-t_i)+\varphi_x^e]} = \sum_{x=1}^{n} A_x^e e^{j(\omega_x^e t - \omega_x^e t_i + \varphi_x^e)} \tag{14}$$

where $A_x^e$, $\varphi_x^e$, and $\omega_x^e$ represent the coefficient, phase, and angular frequency of the $x$-th term after expansion, respectively. Substituting Eq. (14) into Eq. (9), the time-delayed conjugate expression of the acoustic wave can be written as:

$$\vec{\rho}^*\left(z_k, t - \frac{z_k}{v_g}\right) = \sum_{i=1}^{N_t} \left\{ A^f(z_k, t_i) \Delta t e^{-j\varphi^f(z_k, t_i)} e^{-j\omega_\rho\left(t - \frac{z_k}{v_g}\right)} \sum_{x=1}^{n} A_x^e e^{-j\left[\omega_x^e\left(t - \frac{z_k}{v_g}\right) - \omega_x^e t_i + \varphi_x^e\right]} \right\}$$

$$= \sum_{i=1}^{N_t} \left\{ A^f(z_k, t_i) \Delta t \sum_{x=1}^{n} A_x^e e^{-j\left\{(\omega_\rho + \omega_x^e)t + \left[\varphi_x^e + \varphi^f(z_k, t_i) - \omega_x^e t_i - (\omega^\rho + \omega_x^e)\frac{z_k}{v_g}\right]\right\}} \right\} \tag{15}$$

Let $A_{ki}^f = A^f(z_k, t_i)\Delta t$, $\Omega_x = \omega_\rho + \omega_x^e$, and $\phi_{kix} = \varphi_x^e + \varphi^f(z_k, t_i) - \omega_x^e t_i - \Omega_x \frac{z_k}{v_g}$, and by switching the order of summation, Eq. (15) can be rewritten as:

$$\vec{\rho}^*\left(z_k, t - \frac{z_k}{v_g}\right) = \sum_{i=1}^{N_t}\left\{A_{ki}^f \sum_{x=1}^{n} A_x^e e^{-j(\Omega_x t + \phi_{kxi})}\right\} = \sum_{x=1}^{n}\left\{A_x^e \sum_{i=1}^{N_t} A_{ki}^f e^{-j(\Omega_x t + \phi_{kix})}\right\} \quad (16)$$

Let $F_{kx}$ and $\Phi_{kx}$ denote the amplitude and phase variables after the first summation, respectively, as:

$$F_{kx}^2 = \left[A_x^e \Sigma_{i=1}^{N_t} A_{ki}^f \cos(\phi_{kix})\right]^2 + \left[A_x^e \Sigma_{i=1}^{N_t} A_{ki}^f \sin(\phi_{kix})\right]^2 \quad (17)$$

$$\Phi_{kx} = \arctan\left[\frac{A_x^e \Sigma_{i=1}^{N_t} A_{ki}^f \sin(\phi_{kix})}{A_x^e \Sigma_{i=1}^{N_t} A_{ki}^f \cos(\phi_{kix})}\right] \quad (18)$$

leading to a simplified version of Eq. (16),

$$\vec{\rho}^*\left(z_k, t - \frac{z_k}{v_g}\right) = \sum_{x=1}^{n} F_{kx} e^{-j(\Omega_x t + \Phi_{kx})} \xrightarrow{\text{Real}} \sum_{x=1}^{n} F_{kx} \cos(\Omega_x t + \Phi_{kx}) \quad (19)$$

where the indicator 'Real' located above the arrow denotes taking the real part. Note that Eq. (19) is similar to the expression of thermal light[5], but differs in that it incorporates a variable carrier frequency $\Omega_x$, rather than a fix carrier frequency in the case of thermal light.

By substituting Eq. (19) into Eq. (12), and switching the order of summation, $\overrightarrow{E_{Sp}}(t)$ is further stated as:

$$\overrightarrow{E_{Sp}}(t) = \epsilon \sum_{k=1}^{N_P}\left\{\sum_{x=1}^{n} F_{kx} \cos(\Omega_x' t + \Phi_{kx}')\right\} = \epsilon \sum_{x=1}^{n}\left\{\sum_{k=1}^{N_P} F_{kx} \cos(\Omega_x' t + \Phi_{kx}')\right\} \quad (20)$$

where $\Omega_x' = \Omega_x + \omega_P$; $\Phi_{kx}' = \Phi_{kx} - \omega_P \frac{z_k}{v_g}$; $N_P$ is the number of segments within the pump light duration; $\epsilon = j\kappa_1 E_P^0 \Delta t$ is a constant with a given pump light. Similar to the calculation procedure from Eqs. (17)-(18) to Eq. (19), we define $\xi_x$ and $\psi_x$ as the amplitude and phase variables after the first summation of Eq. (20):

$$\xi_x^2 = \left[\Sigma_{k=1}^{N_P} F_{kx} \cos(\Phi_{kx}')\right]^2 + \left[\Sigma_{k=1}^{N_P} F_{kx} \sin(\Phi_{kx}')\right]^2 \quad (21)$$

$$\psi_x = \arctan\left[\frac{\Sigma_{k=1}^{N_P} F_{kx} \sin(\Phi_{kx}')}{\Sigma_{k=1}^{N_P} F_{kx} \cos(\Phi_{kx}')}\right] \quad (22)$$

resulting in:

$$\overrightarrow{E_{Sp}}(t) = \epsilon \sum_{x=1}^{n} \xi_x \cos(\Omega_x' t + \psi_x) \quad (23)$$

Eq. (19) and Eq. (23) show exactly the same formation, indicating that $\overrightarrow{E_{Sp}}(t)$ and $\vec{\rho}(t)$ actually share same statistical properties. In most practical Brillouin systems, the power envelope of $\overrightarrow{E_{Sp}}(t)$, i.e., $P_{Sp}(t)$, is typically analyzed, which is mathematically equivalent to take low-pass filtering on the squared $\overrightarrow{E_{Sp}}(t)$ that can be expressed as:

$$\overrightarrow{E_{Sp}}(t)^2 = \epsilon^2 \left\{\sum_{x=1}^{n} \xi_x \cos(\Omega_x' t + \psi_x)\right\}^2$$

$$= \epsilon^2 \left\{\sum_{x=1}^{n} \xi_x^2 \cos^2(\Omega_x' t + \psi_x) + \sum_{x \neq y}^{\frac{n(n-1)}{2}} 2\xi_x \xi_y \cos(\Omega_x' t + \psi_x)\cos(\Omega_y' t + \psi_y)\right\}$$

$$= \epsilon^2 \left\{\begin{array}{l} \sum_{x=1}^{n} \frac{1}{2}\xi_x^2 [\cos(2\Omega_x' t + 2\psi_x) + 1] \\ + \sum_{x \neq y}^{\frac{n(n-1)}{2}} \xi_x \xi_y \{\cos[(\Omega_x' + \Omega_y')t + (\psi_x + \psi_y)] + \cos[(\Omega_x' - \Omega_y')t - (\psi_x - \psi_y)]\} \end{array}\right\} \quad (24)$$

which includes $n$ squared terms and $\frac{n(n-1)}{2}$ cross terms. The squared and cross terms in Eq. (24) are each simplified by order reduction, allowing for an intuitive distinction between high-frequency and low-frequency components. Then, $P_{Sp}(t)$ can be expressed as:

$$P_{Sp}(t) = \epsilon^2 \text{LPF}\{\overrightarrow{E_{Sp}}(t)^2\} = \epsilon^2 \left\{ \sum_{x=1}^{n} \frac{1}{2}\xi_x^2 + \sum_{x \neq y}^{\frac{n(n-1)}{2}} \xi_x \xi_y \cos[(\Omega_x' - \Omega_y')t - (\psi_x - \psi_y)] \right\} \quad (25)$$

where LPF{·} represents the operator for low-pass filtering.

Observing Eq. (25), it can be found that the statistical characteristics of $P_{Sp}(t)$ essentially depend on statistical properties of $\xi_x$ that has been expressed by Eq. (21). Let $a_x = \Sigma_{k=1}^{N_P} F_{kx} \cos(\Phi_{kx}')$ and $b_x = \Sigma_{k=1}^{N_P} F_{kx} \sin(\Phi_{kx}')$ to rewrite Eq. (21) as $\xi_x^2 = a_x^2 + b_x^2$ for simplicity here. According to the central limit theorem, as $N_P$ is a large number, both $a_x$ and $b_x$ follow the normal distribution. Therefore, the mathematical expectation and variance of $a_x$ can be obtained as:

$$\overline{a_x} = N_P \overline{F_{kx} \cos(\Phi_{kx}')} = 0 \quad (26)$$

$$D\{a_x\} = N_P D\{F_{kx} \cos(\Phi_{kx}')\} = N_P \overline{F_{kx}^2 \cos^2(\Phi_{kx}')} = \frac{1}{2} N_P \overline{F_{kx}^2} \quad (27)$$

where D{·} represents the operator for evaluating the variance. Further, the statistical properties of $a_x^2$ can be obtained as follows:

$$\overline{a_x^2} = D\{a_x\} \quad (28)$$

$$D\{a_x^2\} = \overline{a_x^4} - \overline{a_x^2}^2 = 3D^2\{a_x\} - D^2\{a_x\} = 2D^2\{a_x\} \quad (29)$$

As it can be derived that $\overline{b_x^2} = \overline{a_x^2}$ and $D\{b_x^2\} = D\{a_x^2\}$, the statistical properties of $\xi_x^2$ can be obtained as follows:

$$\overline{\xi_x^2} = \overline{a_x^2} + \overline{b_x^2} = 2D\{a_x\} \quad (30)$$

$$D\{\xi_x^2\} = D\{a_x^2\} + D\{b_x^2\} + 2\text{Cov}\{a_x^2, b_x^2\} = 4D^2\{a_x\} \quad (31)$$

where $\text{Cov}\{a_x^2, b_x^2\}$ means the covariance of $a_x^2$ and $b_x^2$, equal to zero. Moreover, we can obtain:

$$\overline{\xi_x^4} = D\{\xi_x^2\} + \overline{\xi_x^2}^2 = 2\overline{\xi_x^2}^2 = 8D^2\{a_x\} \quad (32)$$

Based on Eqs. (25)-(32), the mathematical expectation of $P_{Sp}(t)$ can be easily obtained as:

$$\overline{P_{Sp}(t)} = \epsilon^2 \sum_{x=1}^{n} \frac{1}{2} \overline{\xi_x^2} = n\epsilon^2 D\{a_x\} \quad (33)$$

And the mathematical expectations of $P_{Sp}(t)^2$ is:

$$\overline{P_{Sp}(t)^2} = \epsilon^4 \sum_{x=1}^{n} \frac{1}{4} \overline{\xi_x^4} + \epsilon^4 \sum_{x \neq y}^{\frac{n(n-1)}{2}} \overline{\frac{1}{2}\xi_x^2 \xi_y^2 + \xi_x^2 \xi_y^2 \cos^2[(\Omega_x' - \Omega_y')t - (\psi_x - \psi_y)]}$$

$$= \epsilon^4 \sum_{x=1}^{n} \frac{1}{4} \overline{\xi_x^4} + \epsilon^4 \sum_{x \neq y}^{\frac{n(n-1)}{2}} \overline{\xi_x^2} \, \overline{\xi_y^2} = 2n^2 \epsilon^2 D^2\{a_x\} \quad (34)$$

Thus, the variance of $P_{Sp}(t)$ can be readily derived as:

$$D\{P_{Sp}(t)\} = \overline{P_{Sp}(t)^2} - \overline{P_{Sp}(t)}^2 = n^2 \epsilon^4 D^2\{a_x\} \quad (35)$$

Finally, the SNR of $P_{Sp}(t)$, denoted as $SNR_{Sp}$, is obtained as the ratio of the mathematical expectation of $P_{Sp}(t)$ to its STD:

$$SNR_{Sp} = \frac{\overline{P_{Sp}(t)}}{\sqrt{D\{P_{Sp}(t)\}}} = 1 \quad (36)$$

which agrees well with the value predicted in seminal work[1].

2. *Updating the SNR to be $B_m$ dependent*
Eq. (36) leads to a deeper understanding on the spectral behavior of $P_{Sp}(t)$ shown in **Fig. 2d** of the main text: the contribution of the DC component (signal contribution) is equal to that of the AC component (noise contribution). This view enables to conveniently analyze the impact of practical measurement bandwidth $B_m$ on the stochastic behavior of $P_{Sp}(t)$. There are essentially two conditions: 1) when $B_m \geq B_{Sp}$, all AC apart is kept so that the noise 3-dB bandwidth is $B_{Sp}$; 2) when $B_m < B_{Sp}$, the AC part is truncated so that the noise 3-dB bandwidth becomes $B_m$.

Considering the AC part typically exhibits a Lorentzian shape with a bandwidth of $B_{Sp}$, expressed as:

$$L_{or1}(f) = \frac{a_{Lor1}(2B_{Sp})^2}{(2B_{Sp})^2 + 4f^2}, \qquad f \geq 0 \tag{37}$$

where $f$ is the frequency variable and $a_{Lor1}$ is the peak value at the center frequency, the noise variances (the area of AC part) under abovementioned two conditions can be calculated as:

$$D\{P_{Sp}(t)\} \propto \begin{cases} \int_0^{B_{Sp}} L_{or1}(f)df = \frac{a_{Lor1}B_{Sp}}{2}\arctan(1), & B_m \geq B_{Sp} \\ \int_0^{B_m} L_{or1}(f)df = \frac{a_{Lor1}B_{Sp}}{2}\arctan\left(\frac{B_m}{B_{Sp}}\right), & B_m < B_{Sp} \end{cases} \tag{38}$$

Taking into account the DC signal ($\overline{P_{Sp}(t)}$) that is irrelevant to $B_m$ and equals to the contribution of AC part with bandwidth of $B_{Sp}$, and based on Eq. (36), the practical $B_m$-dependent SNR of $P_{Sp}(t)$ can be finally expressed as:

$$SNR_{Sp} = \begin{cases} 1, & B_m \geq B_{Sp} \\ \sqrt{\frac{\arctan(1)}{\arctan\left(\frac{B_m}{B_{Sp}}\right)}}, & B_m < B_{Sp} \end{cases} \tag{39}$$

## SI Note 3: Mathematical process modelling the SNR for FFT-based coherent detection.

This note presents the analysis and mathematical derivation of the SNR model for FFT-based coherent detection in the main text. We start from modelling the acquired between signal between the Stokes SpBS and the lower OLO sideband, based on their respective optical fields reaching the input of C2 (**Fig. 3a** in main text):

$$\text{Stokes SpBS: } \overrightarrow{E_{Sp1}}(t) = \hat{x}E_{Sp1}(t)e^{j\varphi_{Sp1}^x(t)}e^{j2\pi f_{Sp1}t} \tag{40}$$

$$\text{lower OLO sideband: } \overrightarrow{E_{Lo1}}(t) = \hat{x}E_{Lo1}e^{j2\pi f_{Lo1}t} \tag{41}$$

where $\hat{x}$ stands for the polarization direction; $\varphi_{Sp1}^x(t)$ is the random phase difference in $\hat{x}$ polarization direction between the SpBS signal and OLO light reaching the input of C2; $E_{Sp1}$ and $E_{Lo1}$ are amplitudes of the Stokes wave and OLO; j is an imaginary unit; $f_{Sp1}$ and $f_{Lo1}$ represent the carrier frequencies of the Stokes wave and OLO.

After passing through the 2 × 2 optical coupler and the subsequent balanced photodetector (BPD), the resulting beating signal $s(t)$ is derived as:

$$s(t) = 2\mathcal{R}_p\sqrt{P_{Lo}P_{Sp}(t)}\cos[2\pi f_c t + \Phi_1(t)] \tag{42}$$

where $\mathcal{R}_p \approx 0.95$ A/W is the responsivity of the photodiode; $P_{Sp}(t) \propto E_{Sp1}(t)^2$ and $P_{Lo} \propto E_{Lo1}^2$ are the optical power of the SpBS light and lower OLO sideband, respectively, where $P_{Sp}(t)$ fluctuates randomly following the statistical property shown by Eq. (39) in **SI Note 2**; $f_c = |f_{Sp1} - f_{Lo1}|$ is the beating frequency of SpBS and OLO; $\Phi_1(t) = \varphi_{Sp1}^x(t) - \pi/2$ is the random phase difference between the SpBS signal and OLO reaching the BPD.

Note that, during the photoelectric conversion process, the photo-detection noise $e(t)$ - characterized as zero-mean Gaussian white noise - is inevitably superimposed on the signal $s(t)$. Typically, $e(t)$ consists of thermal noise and shot noise (mainly contributed by the OLO), and the total noise power spectral density (PSD) $\sigma_e^2$ is given by[7,8]:

$$\sigma_e^2 \approx \sigma_T^2 + 2q\mathcal{R}_p P_{Lo} \tag{43}$$

where $\sigma_T^2$ is the PSD of thermal noise (assumed to be flat), and $q = 1.6 \times 10^{-19}$ C is the electron charge. Thereby the total noise variance is expressed as $\sigma_e^2 B_e$, where $B_e$ is the noise bandwidth.

Taking into account the abovementioned detection noise, the time-domain response of the beating signal, designated here as $r(t)$, is the linear superposition of the signal $s(t)$ and photo-detection noise $e(t)$. With a sampling rate of $f_s$ and a total sampling points number of $N_F$, its discrete-time expression with the variable $n$ is:

$$r(n) = s(n) + e(n) \tag{44}$$

where $n=1, 2, 3, …, N_F$. The key operation of FFT-based coherent detection is to perform a discrete-domain fast Fourier transform (FFT) on $r(n)$, obtaining a normalized PSD, which is then used to analyze the SNR performance at the peak of the Brillouin spectrum. Performing the discrete FFT on $r(n)$ results in the discrete frequency-domain expression with the variable $k$:

$$R(k) = S(k) + E(k) \tag{45}$$

where $k=1, 2, 3, \ldots, N_F$. The PSD of $R(k)$, normalized by the points number $N_F$, is given by:

$$\frac{|R(k)|^2}{N_F} = \frac{|S(k)+E(k)|^2}{N_F} = \frac{|S(k)|^2}{N_F} + \frac{|E(k)|^2}{N_F} + \frac{2S_{Re}(k)E_{Re}(k) + 2S_{Im}(k)E_{Im}(k)}{N_F} \tag{46}$$

where $|\cdot|$ represents the modulus of a complex expression; the subscripts $Re$ and $Im$ respectively denote the real and imaginary parts in the complex domain. Eq. (46) indicates that the normalized PSD of $r(n)$ consists of three components: 1) the normalized PSD of $s(n)$, represented by the first term on the right-hand side of Eq. (46), 2) the normalized PSD of $e(n)$, represented by the second term on the right-hand side of Eq. (46), and 3) the cross term between the normalized PSD of $s(n)$ and $e(n)$, represented by the third term on the right-hand side of Eq. (46). In the following, the statistical properties of each component are sequentially analyzed.

**1) The normalized PSD of $s(n)$.** Given the complexity of directly analyzing the PSD of $s(n)$, here an indirect analytical approach is strategically carried out. We first ignore the random-phase term in $s(n)$, thus simplifying it to a single-frequency cosine signal $s_{sim}(n)$ as:

$$s_{sim}(n) = A\cos\left[\frac{2\pi f_c(n-1)}{f_s}\right] = A\cos\left[\frac{2\pi(n-1)}{a}\right] \tag{47}$$

where $A = 2\mathcal{R}_p\sqrt{P_{Lo}P_{Sp}(t)}$; $f_s = af_c$, $N_F = b\frac{f_s}{f_c} = ab$, $a$ and $b$ being positive integers, to avoid spectral leakage. This operation enables that all the signal energy is concentrated at one single frequency point, whose normalized PSD can be conveniently derived, being. Using this single-point PDS, the actual peak value of the normalized PSD of $s(n)$, is later on derived based on the principle of energy conservation.

The following steps focus on obtaining the single-frequency normalized PSD. First, the discrete fast Fourier transformation of $s_{sim}(n)$ is:

$$S_{sim}(k) = \sum_{n=1}^{N_F} s_{sim}(n) e^{-j\frac{2\pi}{N_F}(n-1)(k-1)} \tag{48}$$

The center frequency $f_c$ corresponds to the $k_c$-th ($k_c \neq 1$) point in the frequency domain, where $k_c = f_c/(f_s/N_F) + 1 = b + 1$, as shown in **SI Fig. 1**. We can expand $S_{sim}(k)$ at $k = k_c$ as:

$$S_{sim}(k_c) = \sum_{n=1}^{N_F} A\cos\left[\frac{2\pi(n-1)}{a}\right]\cos\left[\frac{2\pi(n-1)}{a}\right] + j\sum_{n=1}^{N_F} -A\cos\left[\frac{2\pi(n-1)}{a}\right]\sin\left[\frac{2\pi(n-1)}{a}\right]$$

$$= N_F\frac{A}{2} + \sum_{n=1}^{N_F}\frac{A}{2}\cos\left(\frac{n-1}{a}4\pi\right) + j\sum_{n=1}^{N_F} -\frac{A}{2}\sin\left(\frac{n-1}{a}4\pi\right) \tag{49}$$

As the last two terms equal to 0, the normalized PSD of $s_{sim}(n)$ at $k = k_c$ is expressed as:

$$\frac{|S_{sim}(k_c)|^2}{N_F} = \frac{1}{4}N_F A^2 = N_F \mathcal{R}_p^2 P_{Lo}P_{Sp}(t) \tag{50}$$

We now proceed to derive the corresponding peak expression for the PSD of the actual beating signal $s(n)$. In contrast to the single-frequency case of $s_{sim}(n)$, the spectrum of $s(n)$ exhibits a Lorentzian shape with a FWHM of $B_{Sp}$, as illustrated in **SI Fig. 1**, which is denoted as $L_{or2}(f)$ and can be expressed as:

$$L_{or2}(f) = \frac{|S(k_c)|^2}{N_F}\frac{B_{Sp}^2}{B_{Sp}^2 + 4(f-f_c)^2} \tag{51}$$

where $f$ is the frequency variable; $f_c$ is the center frequency; $\frac{|S(k_c)|^2}{N_F}$ is the peak value at the center frequency of the normalized PSD. According to the law of energy conservation, the area under the single-frequency case is equal to that of the Lorentzian-shaped case, and thus we have:

$$\frac{|S_{sim}(k_c)|^2}{N_F}\delta f = \int_{-\infty}^{+\infty} L_{or2}(f)df = \frac{\pi}{2}\frac{|S(k_c)|^2}{N_F}B_{Sp} \tag{52}$$

Then the peak of the normalized PSD of $S(k)$ can be obtained:

$$\frac{|S(k_c)|^2}{N_F} = \frac{2N_F\delta f}{\pi B_{Sp}}\mathcal{R}_p^2 P_{Lo}P_{Sp}(t) = \frac{2f_s}{\pi B_{Sp}}\mathcal{R}_p^2 P_{Lo}P_{Sp}(t) \tag{53}$$

whose mean value then can be readily derived as:

$$\frac{\overline{|S(k_c)|^2}}{N_F} = \frac{2f_s}{\pi B_{Sp}} \mathcal{R}_p{}^2 P_{Lo} \overline{P_{Sp}} \tag{54}$$

And its variance can be derived as:

$$D\left\{\frac{|S(k_c)|^2}{N_F}\right\} = \frac{4f_s{}^2}{\pi^2 B_{Sp}{}^2} \mathcal{R}_p{}^4 P_{Lo}{}^2 D\{P_{Sp}\} = \frac{4f_s{}^2}{\pi^2 B_{Sp}{}^2} \mathcal{R}_p{}^4 P_{Lo}{}^2 \frac{\overline{P_{Sp}}^2}{SNR_{Sp}{}^2} \tag{55}$$

The value of $\overline{P_{Sp}}$, taking into account the stimulated Brillouin scattering (SBS), can be expressed as [1,2,9]:

$$\overline{P_{Sp}} = \beta_1 \left(e^{\beta_2 P_p} - 1\right) \tag{56}$$

where $P_p$ is the pump power; $\beta_1$ and $\beta_2$ are constant coefficients and can be expressed as:

$$\beta_1 = \frac{8\pi\hbar\omega_S(\overline{n}+1)}{ncA_{eff}\Gamma} \tag{57}$$

$$\beta_2 = \frac{g_0 L}{A_{eff}} \tag{58}$$

where $\hbar$ is the reduced Planck constant; $\omega_S$ is the angular frequency of Stokes light; $\overline{n}$ is the mean number of phonons per mode of the acoustic field; $A_{eff}$ is the effective mode field area of the fiber; $\Gamma$ is the phonon (intensity) decay rate; $g_0$ is the Brillouin gain factor; $L$ is the fiber length.

**2) The normalized PSD of $e(n)$.** The analysis starts from the discrete fast Fourier transformation of $e(n)$ as:

$$E(k) = \sum_{n=1}^{N_F} e(n) e^{-j\frac{2\pi}{N_F}(n-1)(k-1)} \tag{59}$$

And the normalized PSD of $e(n)$ at $k = k_c$ ($k_c \neq 1$) is expressed as:

$$\frac{|E(k_c)|^2}{N_F} = \frac{E_{Re}(k_c)^2 + E_{Im}(k_c)^2}{N_F} \tag{60}$$

where $E_{Re}(k_c)$ and $E_{Im}(k_c)$ stand for the real and imaginary parts of $E(k)$ when $k = k_c$, which can be respectively expressed as:

$$E_{Re}(k_c) = \sum_{n=1}^{N_F} e(n) \cos\left(\frac{n-1}{N_F} K\pi\right) \tag{61}$$

$$E_{Im}(k_c) = \sum_{n=1}^{N_F} -e(n) \sin\left(\frac{n-1}{N_F} K\pi\right) \tag{62}$$

where $K = 2(k_c + 1)$ is a positive even number. Eq. (60) indicates that the statistical properties of $|E(k_c)|^2$ can be derived by separately characterizing the statistical properties of $E_{Re}(k_c)^2$ and $E_{Im}(k_c)^2$, both indeed sharing same statistical characteristics according to Eqs. (61) and (62). Thereby the analysis below only focuses on the real part $E_{Re}(k_c)^2$.

Based on Eq. (61), we first present the expression for $E_{Re}(k_c)^2$, which contains $N_F$ squared terms and $\frac{N_F(N_F-1)}{2}$ cross terms as:

$$E_{Re}(k_c)^2 = \sum_{n=1}^{N_F} e(n)^2 \cos^2\left(\frac{n-1}{N_F} K\pi\right) + \sum_{n \neq m}^{\frac{N_F(N_F-1)}{2}} 2e(n) \cos\left(\frac{n-1}{N_F} K\pi\right) e(m) \cos\left(\frac{m-1}{N_F} K\pi\right) \tag{63}$$

Since the cross terms in Eq. (63) have a mean value of zero, the expected value of $E_{Re}(k_c)^2$ is given by:

$$\overline{E_{Re}(k_c)^2} = \sum_{n=1}^{N_F} \overline{e(n)^2} \cos^2\left(\frac{n-1}{N_F} K\pi\right) = \frac{1}{2} N_F \sigma_e{}^2 B_e \tag{64}$$

The variance of $E_{Re}(k_c)^2$ is then calculated as:

$$D\{E_{Re}(k_c)^2\} = D\left\{\sum_{n=1}^{N_F} e(n)^2 \cos^2\left(\frac{n-1}{N_F}K\pi\right)\right\}$$

$$+D\left\{\sum_{\substack{n\neq m}}^{\frac{N_F(N_F-1)}{2}} 2e(n)\cos\left(\frac{n-1}{N_F}K\pi\right)e(m)\cos\left(\frac{m-1}{N_F}K\pi\right)\right\}$$

$$+2\text{Cov}\left\{\sum_{n=1}^{N_F} e(n)^2 \cos^2\left(\frac{n-1}{N_F}K\pi\right), \sum_{\substack{n\neq m}}^{\frac{N_F(N_F-1)}{2}} 2e(n)\cos\left(\frac{n-1}{N_F}K\pi\right)e(m)\cos\left(\frac{m-1}{N_F}K\pi\right)\right\}$$

$$= \sum_{n=1}^{N_F} \cos^4\left(\frac{n-1}{N_F}K\pi\right)D\{e(n)^2\} + \sum_{\substack{n\neq m}}^{\frac{N_F(N_F-1)}{2}} 4\cos^2\left(\frac{n-1}{N_F}K\pi\right)\cos^2\left(\frac{m-1}{N_F}K\pi\right)D\{e(n)e(m)\}$$

$$= \left(\frac{1}{2}N_F^2 + \frac{1}{4}N_F\right)(\sigma_e^2 B_e)^2 \tag{65}$$

Then, the mathematical expectation and variance of the normalized PSD of $E(k_c)$ can be respectively obtained as:

$$\frac{\overline{|E(k_c)|^2}}{N_F} = \frac{\overline{E_{Re}(k_c)^2}}{N_F} + \frac{\overline{E_{Im}(k_c)^2}}{N_F} = \sigma_e^2 B_e \tag{66}$$

$$D\left\{\frac{|E(k_c)|^2}{N_F}\right\} = D\left\{\frac{E_{Re}(k_c)^2}{N_F}\right\} + D\left\{\frac{E_{Im}(k_c)^2}{N_F}\right\} = \left(1 + \frac{1}{2N_F}\right)(\sigma_e^2 B_e)^2 \approx (\sigma_e^2 B_e)^2 \tag{67}$$

It should be noted that, the above results represent the equivalent PSD, as shown by red square in **SI Fig. 2**, which spreads the original energy distribution to sampling bandwidth of $f_s/2$. So the actual PSD under bandwidth of $B_e$, as shown by green square in **SI Fig. 2**, should be further derived, whose expected value and variance are given by:

$$\frac{\overline{|E(k_c)|^2}}{N_F} = \sigma_e^2 B_e \frac{f_s}{2B_e} = \sigma_e^2 \frac{f_s}{2} \tag{68}$$

$$D\left\{\frac{|E(k_c)|^2}{N_F}\right\} \approx \left(\sigma_e^2 B_e \frac{f_s}{2B_e}\right)^2 = \left(\sigma_e^2 \frac{f_s}{2}\right)^2 \tag{69}$$

**3) The cross term.** The mathematical expectation of the cross term between $s(n)$ and $e(n)$, represented by the third term on the right-hand side of Eq. (46), is given by:

$$\overline{\frac{2}{N_F}[S_{Re}(k_c)E_{Re}(k_c) + S_{Im}(k_c)E_{Im}(k_c)]} = \frac{2}{N_F}\left[\overline{S_{Re}(k_c)}\,\overline{E_{Re}(k_c)} + \overline{S_{Im}(k_c)}\,\overline{E_{Im}(k_c)}\right] = 0 \tag{70}$$

Its variance is further derived as:

$$D\left\{\frac{2S_{Re}(k_c)E_{Re}(k_c) + 2S_{Im}(k_c)E_{Im}(k_c)}{N_F}\right\} = \frac{4}{N_F^2}\overline{[S_{Re}(k_c)E_{Re}(k_c) + S_{Im}(k_c)E_{Im}(k_c)]^2}$$

$$= \frac{4}{N_F^2}\left[\overline{S_{Re}(k_c)^2}\,\overline{E_{Re}(k_c)^2} + \overline{S_{Im}(k_c)^2}\,\overline{E_{Im}(k_c)^2} + \overline{2S_{Re}(\ _c)E_{Re}(k_c)S_{Im}(k_c)E_{Im}(k_c)}\right] \tag{71}$$

Given that $\overline{E_{Re}(k_c)E_{Im}(k_c)} = 0$, Eq. (71) can be simplified as:

$$D\left\{\frac{2S_{Re}(k_c)E_{Re}(k_c) + 2S_{Im}(k_c)E_{Im}(k_c)}{N_F}\right\} = 4\left[\frac{\overline{S_{Re}(k_c)^2}}{N_F}\frac{\overline{E_{Re}(k_c)^2}}{N_F} + \frac{\overline{S_{Im}(k_c)^2}}{N_F}\frac{\overline{E_{Im}(k_c)^2}}{N_F}\right]$$

$$= 2\frac{\overline{|S(k_c)|^2}}{N_F}\frac{\overline{|E(k_c)|^2}}{N_F} = \frac{2f_s^2}{\pi B_{Sp}}\mathcal{R}_p^2 P_{Lo}\overline{P_{Sp}}\sigma_e^2 \tag{72}$$

All above-derived statistic parameters - expectation values and variances on the peak of normalized PSD corresponding to the three terms in Eq. (46) - are summarized in SI Table I to eventually model the SNR of FFT-based coherent detection. As shown in the table, the actual signal term is given by Eq. (54), while the constant noise bias given by Eq. (68) remains trivial and can be practically eliminated in measurements by subtracting the PSD obtained in the absence of the pump light.

**SI Table I**
Expectation values and variances of the three terms in Eq. (46)

| Terms | Expectation values | | Variances | |
|---|---|---|---|---|
| $\dfrac{\|S(k_c)\|^2}{N_F}$ | $\dfrac{2f_s}{\pi B_{Sp}}\mathcal{R}_p{}^2 P_{Lo}\overline{P_{Sp}},$ | (Eq. (54)) | $\dfrac{4f_s{}^2}{\pi^2 B_{Sp}{}^2}\mathcal{R}_p{}^4 P_{Lo}{}^2 \dfrac{\overline{P_{Sp}}^2}{SNR_{Sp}{}^2},$ | (Eq. (55)) |
| $\dfrac{\|E(k_c)\|^2}{N_F}$ | $\sigma_e{}^2 \dfrac{f_s}{2},$ | (Eq. (68)) | $\left(\sigma_e{}^2 \dfrac{f_s}{2}\right)^2,$ | (Eq. (69)) |
| Cross term | $0,$ | (Eq. (70)) | $\dfrac{2f_s{}^2}{\pi B_{Sp}}\mathcal{R}_p{}^2 P_{Lo}\overline{P_{Sp}}\sigma_e{}^2,$ | (Eq. (72)) |

Consequently, the SNR of the response acquired through FFT-based coherent detection can be expressed as:

$$\mathrm{SNR}\{r_{Co}^{FFT}\} = \frac{\dfrac{|S(k_c)|^2}{N_F}}{\sqrt{\mathrm{D}\left\{\dfrac{|S(k_c)|^2}{N_F}\right\} + \mathrm{D}\left\{\dfrac{2S_{Re}(k_c)E_{Re}(k_c)+2S_{Im}(k_c)E_{Im}(k_c)}{N_F}\right\} + \mathrm{D}\left\{\dfrac{|E(k_c)|^2}{N_F}\right\}}}$$

$$= \frac{\dfrac{2f_s}{\pi B_{Sp}}\mathcal{R}_p{}^2 P_{Lo}\overline{P_{Sp}}}{\sqrt{\dfrac{4f_s{}^2}{\pi^2 B_{Sp}{}^2}\mathcal{R}_p{}^4 P_{Lo}{}^2 \dfrac{\overline{P_{Sp}}^2}{SNR_{Sp}{}^2} + \dfrac{2f_s{}^2}{\pi B_{Sp}}\mathcal{R}_p{}^2 P_{Lo}\overline{P_{Sp}}\sigma_e{}^2 + \left(\sigma_e{}^2 \dfrac{f_s}{2}\right)^2}}$$

$$= \frac{\mathcal{R}_p{}^2 P_{Lo}\overline{P_{Sp}}}{\sqrt{\dfrac{1}{SNR_{Sp}{}^2}\mathcal{R}_p{}^4 P_{Lo}{}^2 \overline{P_{Sp}}^2 + \dfrac{\pi B_{Sp}}{2}\mathcal{R}_p{}^2 P_{Lo}\overline{P_{Sp}}\sigma_e{}^2 + \left(\dfrac{\pi B_{Sp}}{4}\right)^2 \sigma_e{}^4}} \qquad (73)$$

Note that, while $SNR_{Sp}$ features a bandwidth dependence as depicted by Eq. (39), here the FFT-based coherent detection corresponds to a constant $SNR_{Sp} = 1$ in Eq. (73), consistent with the scenario where $B_m \geq B_{Sp}$ as illustrated in **SI Fig. 3a**. For comparison, the case of FFT-based coherent detection is shown in **SI Fig. 3b**, indicating that selecting a single point at the center frequency of the Brillouin spectrum is functionally equivalent to applying a narrow BPF with a bandwidth equal to the spectral resolution $\delta f$ of FFT. Both the spectral amplitude at zero frequency and the total envelope spectral area exhibit a reduction factor of $(\delta f/B_{Sp})^2$ relative to **SI Fig. 3a**, preserving $SNR_{Sp}$=1.

Therefore, the FFT-based fundamental detection scheme only considers the case of $SNR_{Sp} = 1$, independent of $B_m$. Although a $B_m$-dependent $SNR_{Sp}$ could be induced via a digital LPF with a cutoff bandwidth below $\delta f$, we advocate employing a more direct physical approach – an envelope detector-based detection scheme (detailed in the **Methods** section of the main text) – to systematically investigate the relationship between $B_m$ and $SNR_{Sp}$.

**SI Note 4: Mathematical process modelling the SNR of single-pulse BOTDR with PMF.**
In this note, we model the SNR of the distance-domain envelope signal[10] at the Brillouin resonance, obtained by a frequency-scanning Brillouin optical time domain reflectometer (BOTDR) with a polarization maintaining fiber (PMF), as illustrated in **Fig. 6a** of the main text. No matter using an off-the-shelf envelope detector (ED) or digital post-processing, the envelope extraction process can be mathematically represented by squaring the beating signal filtered by the bandpass filter (BPF) and then applying a low-pass filtering operation, LPF{·}, to eliminate the high-frequency components.

First, we model the distance-domain photocurrent output from BPF, designated here as $I(z)$, which results from the beating between the SpBS components and the OLO sidebands. Its expression is similar to that in Ref. [9][10], except for the absence of polarization-related terms due to the use of PMF:

$$I(z) = I_s(z) + I_e(z)$$
$$= 2\mathcal{R}_p \left\{ \sqrt{P_{Lo1} P_{Sp1}(z)} \cos[H(z) + \Phi_1(z)] + \sqrt{P_{Lo2} P_{Sp2}(z)} \cos[H(z) + \Phi_2(z)] \right\} + I_e(z) \quad (74)$$

where $I_s(z)$ is the photocurrent of the beating signal output from the BPF; $\mathcal{R}_p \approx 0.95$ A/W is the responsivity of the photodiodes; $f_c(z)$ denotes the frequency difference between the SpBS light and the OLO light; $H(z) = 4\pi f_c(z) n_{eff} z / c$ indicates the carrier phase of the beating signal; $\Phi_1(t)$ (or $\Phi_2(t)$) is the random phase differences between the Sokes (or anti-Stokes) SpBS signal and the lower (or upper) OLO sideband. $P_{Lo1}$ and $P_{Lo2}$ are the power of the lower and upper OLO sidebands, respectively, with $P_{Lo1} = P_{Lo2} = P_{Lo}/2$. $P_{Sp1}(z)$ and $P_{Sp2}(z)$ are the power of the Stokes and anti-Stokes SpBS signals, respectively, which fluctuate randomly due to the intrinsic stochastic nature, both with mean value expressed as:

$$\overline{P_{Sp1}(z)} = \overline{P_{Sp2}(z)} = \frac{\overline{P_{Sp}(z)}}{2} = \frac{k_{Sp} c P_p D_p e^{-2\alpha z}}{2 n_{eff}} \quad (75)$$

where $k_{Sp}$ is the backscattering coefficient of SpBS, $P_p$ and $D_p$ are the peak power and duration of the incident pump pulse, respectively, $n_{eff}$ is the effective group index of the propagating mode in the fiber, and $\alpha$ is the fiber attenuation coefficient. $I_e(z)$ in Eq. (74) is the noise photocurrent at the BPF output, mainly attributed to filtered thermal noise and shot noise. The variance of $I_e(z)$ can be characterized as $\sigma_e^2 B_{BPF}$, where $\sigma_e^2$ is the photo-detection noise PSD as described by Eq. (43) in SI Note 3, and $B_{BPF}$ is the BPF bandwidth, approximately equal to the FWHM of the Brillouin spectrum $B_{Sp}$.

The output of the BPF then undergoes an envelope extraction process with a LPF bandwidth $B_m$ (generally equal to $B_{BPF}$), yielding the power envelope signal:

$$r_{Sg}^{PM}(z) = \text{LPF}\{I(z)^2\} = \text{LPF}\{I_s(z)^2\} + \text{LPF}\{2 I_s(z) I_e(z)\} + \text{LPF}\{I_e(z)^2\} \quad (76)$$

We then derive the statistical parameters - the expectation value and variance - for each term in Eq. (76) at every fiber position $z$, as detailed below.

**1) The term of $\text{LPF}\{I_s(z)^2\}$.** Firstly, squaring $I_s(z)$ in Eq. (74) gives:

$$I_s(z)^2 = 4\mathcal{R}_p^2 \left\{ \begin{array}{c} P_{Lo1} P_{Sp1}(z) \cos^2[H(z) + \Phi_1(z)] \\ + P_{Lo2} P_{Sp2}(z) \cos^2[H(z) + \Phi_2(z)] \\ + 2\sqrt{P_{Lo1} P_{Sp1}(z) P_{Lo2} P_{Sp2}(z)} \cos[H(z) + \Phi_1(z)] \cos[H(z) + \Phi_2(z)] \end{array} \right\} \quad (77)$$

After applying the product-to-sum transformation, the terms containing $\cos[2H(z)]$ signify the high-frequency components (typically a few hundred megahertz and beyond $B_m$), which are filtered out by the low-pass filtering operation, leading to:

$$\text{LPF}\{I_s(z)^2\} = 2\mathcal{R}_p^2 \left\{ P_{Lo1} P_{Sp1}(z) + P_{Lo2} P_{Sp2}(z) + 2\sqrt{P_{Lo1} P_{Sp1}(z) P_{Lo2} P_{Sp2}(z)} \cos[\Delta\Phi(z)] \right\} \quad (78)$$

where $\Delta\Phi(z) = \Phi_1(z) - \Phi_2(z)$. Considering that $\Phi_1(z)$ and $\Phi_2(z)$ follow a uniform distribution over the interval $[-\pi, \pi]$, the expectation value of $\cos[\Delta\Phi(z)]$ is considered zero, i.e., $\overline{\cos[\Delta\Phi(z)]} = 0$. This allows us to express the expectation value of $\text{LPF}\{I_s(z)^2\}$, designated as $\mu_{s^2}(z)$, as:

$$\mu_{s^2}(z) = \overline{\text{LPF}\{I_s^{BPF}(z)^2\}} = 2\mathcal{R}_p^2 \left[ P_{Lo1} \overline{P_{Sp1}(z)} + P_{Lo2} \overline{P_{Sp2}(z)} \right] = \mathcal{R}_p^2 P_{Lo} \overline{P_{Sp}(z)} \quad (79)$$

Then based on Eq. (39) and the definition $D\{P_{Sp}(z)\} = \overline{P_{Sp1}(z)^2} - \overline{P_{Sp1}(z)}^2$, the following relations hold:

$$\overline{P_{Sp1}(z)^2} = \left(1 + \frac{1}{SNR_{Sp}^2}\right) \overline{P_{Sp1}(z)}^2 \quad (80)$$

$$\overline{P_{Sp2}(z)^2} = \left(1 + \frac{1}{SNR_{Sp}^2}\right) \overline{P_{Sp2}(z)}^2 \quad (81)$$

enabling calculating the mean value of $\text{LPF}^2\{I_s(z)^2\}$:

$$\overline{\text{LPF}^2\{I_s(z)^2\}} = 4\mathcal{R}_p^4 \left\{ P_{Lo1}^2 \overline{P_{Sp1}(z)^2} + P_{Lo2}^2 \overline{P_{Sp2}(z)^2} + 4 P_{Lo1} \overline{P_{Sp1}(z)} P_{Lo2} \overline{P_{Sp2}(z)} \right\}$$

$$= 4\mathcal{R}_p^4 \left\{ P_{Lo1}^2 \left(1 + \frac{1}{SNR_{Sp}^2}\right) \overline{P_{Sp1}(z)}^2 + P_{Lo2}^2 \left(1 + \frac{1}{SNR_{Sp}^2}\right) \overline{P_{Sp2}(z)}^2 + 4 P_{Lo1} \overline{P_{Sp1}(z)} P_{Lo2} \overline{P_{Sp2}(z)} \right\}$$

$$= \frac{1}{2}\left(3 + \frac{1}{SNR_{Sp}^2}\right) \mathcal{R}_p^4 P_{Lo}^2 \overline{P_{Sp}(z)}^2 \quad (82)$$

Finally, the variance of $\text{LPF}\{I_s(z)^2\}$, designated as $\sigma_{s^2}(z)^2$, can be computed according to its statistic definition:

$$\sigma_{s^2}(z)^2 = \overline{\text{LPF}^2\{I_s(z)^2\}} - \mu_{s^2}(z)^2 = \frac{1}{2}\left(1 + \frac{1}{SNR_{Sp}^2}\right) \mathcal{R}_p^4 P_{Lo}^2 \overline{P_{Sp}(z)}^2 \quad (83)$$

**2) The term of LPF$\{2I_s(z)I_e(z)\}$.** Since the BPF bandwidth is narrower than the system acquisition bandwidth, $I_e(z)$ can be considered as a narrowband random noise, which can be decomposed into its in-phase and quadrature components[11]:

$$I_e(z) = I_e^I(z)\cos[H(z)] - I_e^Q(z)\sin[H(z)] \tag{84}$$

where $I_e^I(z)$ and $I_e^Q(z)$ stand for the in-phase and quadrature components of $I_e(z)$, respectively. These two independent components both follow zero-mean normal distributions with identical variance to $I_e(z)$. By substituting Eq. (74) and Eq. (84) into the $2I_s(z)I_e(z)$ and removing high-frequency components, we obtain:

$$\text{LPF}\{2I_s(z)I_e(z)\} = 2\mathcal{R}_p \left\{ \begin{array}{l} I_e^I(z)\sqrt{P_{Lo1}P_{Sp1}(z)}\cos[\Phi_1(z)] + I_e^Q(z)\sqrt{P_{Lo1}P_{Sp1}(z)}\sin[\Phi_1(z)] \\ + I_e^I(z)\sqrt{P_{Lo2}P_{Sp2}(z)}\cos[\Phi_2(z)] + I_e^Q(z)\sqrt{P_{Lo2}P_{Sp2}(z)}\sin[\Phi_2(z)] \end{array} \right\} \tag{85}$$

The expectation value of LPF$\{2I_s(z)I_e(z)\}$, designated as $\mu_{2se}(z)$, can be calculated as:

$$\mu_{2se}(z) = 0 \tag{86}$$

The variance of LPF$\{2I_s(z)I_e(z)\}$, designated as $\sigma_{2se}(z)^2$, can then be derived using its statistical definition:

$$\sigma_{2se}(z)^2 = \overline{\text{LPF}^2\{2I_s(z)I_e(z)\}} - \mu_{2se}(z)^2$$
$$= 4\mathcal{R}_p^2 \left[P_{Lo1}\overline{P_{Sp1}(z)} + P_{Lo2}\overline{P_{Sp2}(z)}\right]\sigma_e^2 B_{BPF} = 2\mathcal{R}_p^2 P_{Lo}\overline{P_{Sp}(z)}\sigma_e^2 B_{BPF} \tag{87}$$

**3) The term of LPF$\{I_e(z)^2\}$.** By substituting Eq. (84) into LPF$\{I_e(z)^2\}$, we obtain:

$$\text{LPF}\{I_e(z)^2\} = \frac{1}{2}I_e^I(z)^2 + \frac{1}{2}I_e^Q(z)^2 \tag{88}$$

According to the calculation of multiple moments of a normal distribution[12], it can be obtained that:

$$\overline{I_e^I(z)^2} = \overline{I_e^Q(z)^2} = \sigma_e^2 B_{BPF} \tag{89}$$

and

$$\overline{I_e^I(z)^4} = \overline{I_e^Q(z)^4} = 3\sigma_e^4 B_{BPF}^2 \tag{90}$$

The expectation value and variance of LPF$\{I_e(z)^2\}$, designated as $\mu_{e^2}(z)$ and $\sigma_{e^2}(z)^2$, respectively, can then be derived as:

$$\mu_{e^2}(z) = \overline{\text{LPF}\{I_e(z)^2\}} = \sigma_e^2 B_{BPF} \tag{91}$$

$$\sigma_{e^2}(z)^2 = \text{D}\left\{\frac{1}{2}I_e^I(z)^2\right\} + \text{D}\left\{\frac{1}{2}I_e^Q(z)^2\right\} = (\sigma_e^2 B_{BPF})^2 \tag{92}$$

**SI Table II**
Expectation values and variances of the three terms in Eq. (76)

| Terms | Expectation values | Variances |
|---|---|---|
| LPF$\{I_s(z)^2\}$ | $\mathcal{R}_p^2 P_{Lo}\overline{P_{Sp}(z)}$,  (Eq. (79)) | $\frac{1}{2}\left(1+\frac{1}{SNR_{Sp}^2}\right)\mathcal{R}_p^4 P_{Lo}^2 \overline{P_{Sp}(z)}^2$,  (Eq. (83)) |
| LPF$\{2I_s(z)I_e(z)\}$ | 0,  (Eq. (86)) | $2\mathcal{R}_p^2 P_{Lo}\overline{P_{Sp}(z)}\sigma_e^2 B_{BPF}$,  (Eq. (87)) |
| LPF$\{I_e(z)^2\}$ | $\sigma_e^2 B_{BPF}$,  (Eq. (91)) | $(\sigma_e^2 B_{BPF})^2$,  (Eq. (92)) |

All above-derived statistic parameters (expectation values and variances of the three terms in Eq. (76)) are summarized in SI Table II to eventually model the SNR of the single-pulse BOTDR with PMF. Finally, practically subtracting the bias arising from $\mu_{e^2}(z)$ (i.e., the expectation value of LPF$\{I_e(z)^2\}$) in actual measurement, the SNR at each fiber position $z$ can be expressed as:

$$\text{SNR}\{r_{Sg}^{PM}(z)\} = \frac{\mu_{s^2}(z)}{\sqrt{\sigma_{s^2}(z)^2 + \sigma_{2se}(z)^2 + \sigma_{e^2}(z)^2}}$$

$$= \frac{\mathcal{R}_p^2 P_{Lo}\overline{P_{Sp}(z)}}{\sqrt{\frac{1}{2}\left(1+\frac{1}{SNR_{Sp}^2}\right)\mathcal{R}_p^4 P_{Lo}^2 \overline{P_{Sp}(z)}^2 + 2\mathcal{R}_p^2 P_{Lo}\overline{P_{Sp}(z)}\sigma_e^2 B_{BPF} + \sigma_e^4 B_{BPF}^2}} \tag{93}$$

**SI Note 5: Mathematical process modelling the SNR of single-pulse BOTDR with standard SMF.**

Here the SNR model of a conventional BOTDR sensing system based on standard single model fiber (SMF) is established. The system employs a polarization scrambler (PSc) to alleviate polarization fading (i.e., introducing polarization noise). Compared to the PMF case in **SI Note 4**, this scenario incorporates polarization-related effects[10,13,14]. Consequently, the distance-domain signal photocurrent output from the BPF is expressed as:

$$I_s(z) = 2\mathcal{R}_p \cos[\theta(z)] \left\{ \begin{array}{l} \sqrt{P_{Lo1}P_{Sp1}(z)} \cos[H(z) + \Phi_1(z)] \\ + \sqrt{P_{Lo2}P_{Sp2}(z)} \cos[H(z) + \Phi_2(z)] \end{array} \right\} \quad (94)$$

where the angle $\theta(z)$ denotes the local relative polarization rotation of the SpBS with respect to the OLO, varying within the range $[0, \pi/2]$.

Following a mathematical process similar to that in **SI Note 4**, we model the SNR behavior of BOTDR response (power envelope signal, designated as $r_{Sg}^{SMF}(z)$), incorporating the effects of polarization noise.

**1) The term of LPF$\{I_s(z)^2\}$.** Based on Eq. (94), the term of LPF$\{I_s(z)^2\}$ can be expressed as:

$$\text{LPF}\{I_s(z)^2\} = 2\mathcal{R}_p^2 \cos^2[\theta(z)] \left\{ P_{Lo1}P_{Sp1}(z) + P_{Lo2}P_{Sp2}(z) + 2\sqrt{P_{Lo1}P_{Sp1}(z)P_{Lo2}P_{Sp2}(z)} \cos[\Delta\Phi(z)] \right\} \quad (95)$$

where $\cos^2[\theta(z)]$ varies within the range [0, 1], following a uniform distribution over multiple measurements. For estimating the influence of self-polarization-scrambling effect, a factor $k_{Pol}$ is introduced to adopt the statistical properties of $\cos^2[\theta(z)]$ (as later on discussed in **SI Note 6**), leading to:

$$\overline{\cos^2[\theta(z)]} = \frac{1}{2} \quad (96)$$

$$\text{D}\{\cos^2[\theta(z)]\} = \frac{1}{12}k_{Pol}^2 \quad (97)$$

Then, the expectation value of LPF$\{I_s(z)^2\}$, i.e., $\mu_{s^2}(z)$, can be calculated as:

$$\mu_{s^2}(z) = \overline{\text{LPF}\{I_s(z)^2(z)\}} = \mathcal{R}_p^2 \left[ P_{Lo1}\overline{P_{Sp1}(z)} + P_{Lo2}\overline{P_{Sp2}(z)} \right] = \frac{1}{2}\mathcal{R}_p^2 P_{Lo}\overline{P_{Sp}(z)} \quad (98)$$

Then, according to the relationships of Eq. (80)-(81), and:

$$\overline{\cos^4[\theta(z)]} = \text{D}\{\cos^2[\theta(z)]\} + \overline{\cos^2[\theta(z)]}^2 = \frac{k_{Pol}^2 + 3}{12} \quad (99)$$

we obtain:

$$\overline{\text{LPF}^2\{I_s(z)^2\}} = 4\mathcal{R}_p^4 \frac{k_{Pol}^2 + 3}{12} \left\{ P_{Lo1}^2 \overline{P_{Sp1}(z)^2} + P_{Lo2}^2 \overline{P_{Sp2}(z)^2} + 4P_{Lo1}\overline{P_{Sp1}(z)}P_{Lo2}\overline{P_{Sp2}(z)} \right\}$$

$$= 4\mathcal{R}_p^4 \frac{k_{Pol}^2 + 3}{12} \left\{ \begin{array}{l} P_{Lo1}^2 \left(1 + \frac{1}{SNR_{Sp}^2}\right)\overline{P_{Sp1}(z)}^2 \\ + P_{Lo2}^2 \left(1 + \frac{1}{SNR_{Sp}^2}\right)\overline{P_{Sp2}(z)}^2 \\ + 4P_{Lo1}\overline{P_{Sp1}(z)}P_{Lo2}\overline{P_{Sp2}(z)} \end{array} \right\}$$

$$= \frac{k_{Pol}^2 + 3}{24}\left(3 + \frac{1}{SNR_{Sp}^2}\right)\mathcal{R}_p^4 P_{Lo}^2 \overline{P_{Sp}(z)}^2 \quad (100)$$

Therefore, the variance of LPF$\{I_s(z)^2\}$, i.e., $\sigma_{s^2}(z)^2$ can be calculated according to its statistic definition:

$$\sigma_{s^2}(z)^2 = \overline{\text{LPF}^2\{I_s(z)^2\}} - \mu_{s^2}(z)^2 = \frac{3SNR_{Sp}^2(k_{Pol}^2 + 1) + k_{Pol}^2 + 3}{24SNR_{Sp}^2}\mathcal{R}_p^4 P_{Lo}^2 \overline{P_{Sp}(z)}^2 \quad (101)$$

**2) The term of LPF$\{2I_s(z)I_e(z)\}$.** Similar to the calculation process from Eq. (84) to Eq. (87), the expectation value and variance of LPF$\{2I_s(z)I_e(z)\}$, designated as $\mu_{2se}(z)$ and $\sigma_{2se}(z)^2$, respectively, can be derived based on their respective statistic definition:

$$\mu_{2se}(z) = 0 \quad (102)$$

$$\sigma_{2se}(z)^2 = \overline{\text{LPF}^2\{2I_s(z)I_e(z)\}} - \mu_{2se}(z)^2 = 2\mathcal{R}_p^2 \left[ P_{Lo1}\overline{P_{Sp1}(z)} + P_{Lo2}\overline{P_{Sp2}(z)} \right]\sigma_e^2$$

$$= \mathcal{R}_p^2 P_{Lo}\overline{P_{Sp}(z)}\sigma_e^2 B_{BPF} \quad (103)$$

**3) The term of LPF$\{I_e(z)^2\}$.** Since the statistical parameters of LPF$\{I_e(z)^2\}$ - namely its expectation and variance - are unaffected by polarization, the analysis of LPF$\{I_e(z)^2\}$ in this Note is consistent with that provided in **SI Note 4**.

**SI Table III**
Expectation values and variances of the three terms in an expression similar to Eq. (76) for the SMF case

| Terms | Expectation values | Variances |
|---|---|---|
| $\text{LPF}\{I_s(z)^2\}$ | $\frac{1}{2}\mathcal{R}_p^2 P_{Lo}\overline{P_{Sp}(z)}$,  (Eq. (98)) | $\frac{3SNR_{Sp}^2(k_{Pol}^2+1)+k_{Pol}^2+3}{24SNR_{Sp}^2}\mathcal{R}_p^4 P_{Lo}^2 \overline{P_{Sp}(z)}^2$, (Eq. (101)) |
| $\text{LPF}\{2I_s(z)I_e(z)\}$ | 0,  (Eq. (102)) | $\mathcal{R}_p^2 P_{Lo}\overline{P_{Sp}(z)}\sigma_e^2 B_{BPF}$, (Eq. (103)) |
| $\text{LPF}\{I_e(z)^2\}$ | $\sigma_e^2 B_{BPF}$,  (Eq. (91)) | $\sigma_e^4 B_{BPF}^2$,  (Eq. (92)) |

For the SMF case, the statistical parameters of the three terms in an expression similar to Eq. (76) are summarized in Table III, and the SNR at each fiber position $z$ can be represented as:

$$\text{SNR}\{r_{Sg}^{SMF}(z)\} = \frac{\mu_{s^2}(z)}{\sqrt{\sigma_{s^2}(z)^2 + \sigma_{2se}(z)^2 + \sigma_{e^2}(z)^2}}$$

$$= \frac{\frac{1}{2}\mathcal{R}_p^2 P_{Lo}\overline{P_{Sp}(z)}}{\sqrt{\frac{3SNR_{Sp}^2(k_{Pol}^2+1)+k_{Pol}^2+3}{24SNR_{Sp}^2}\mathcal{R}_p^4 P_{Lo}^2 \overline{P_{Sp}(z)}^2 + \mathcal{R}_p^2 P_{Lo}\overline{P_{Sp}(z)}\sigma_e^2 B_{BPF} + \sigma_e^4 B_{BPF}^2}} \quad (104)$$

**SI Note 6: Analysis of self-polarization-scrambling effect.**
The polarization fading is a typical effect in SMF-based BOTDR, manifesting as the distance-domain signal fluctuation caused by the random state of polarization (SOP) difference between the SpBS light propagating in the SMF and the local oscillator (OLO)[13]. Mathematically, this effect is normally described by a random polarization-related term $\cos^2[\theta(z)]$ (e.g., in Eq. (94)), which follows a uniform distribution between [0,1], as shown by the simulated pink curves in **SI Fig.10**. This causes the envelope signal to fluctuate between 0 (when the SOPs of SpBS and OLO are orthogonal) and its maximum value (when the SOPs of SpBS and OLO are aligned) in a standard SMF.

Here we reveal that the self-polarization-scrambling effect attributed to the non-uniform SOP over the pump pulse duration due to fiber birefringence, which leads to a weak polarization scrambling process within the pulse itself, resulting in an alleviated degree of polarization fading. Mathematically, the self-polarization-scrambling effect reduces variance of $\cos^2[\theta(z)]$ by factor of $k_{Pol}^2$, where $k_{Pol}$ [0, 1] is a SR-dependent variable introduced to quantify the impact of self-polarization-scrambling effect. The wider the spatial resolution (SR), the lager the averaging effect caused by the self-polarization-scrambling, resulting in a smaller $k_{Pol}$, as shown by the simulated blue curves in **SI Fig.10**.

*1. Experimentally quantifying $k_{Pol}$ with different SRs*
We experimentally characterize $k_{Pol}$ under SRs of 1 m, 2 m, 6 m, and 10 m, respectively, by quantifying how much polarization fading is alleviated by these SRs. In detail, we use the standard BOTDR setup[10] with a 1.9 km-long SMF, by intentionally removing the PSc in the setup, to result in a BOTDR trance with polarization fading. With all SRs, the pulse peak power is set to 31 dBm, and each BOTDR trace is averaged by 1000 times, guaranteeing a large enough SNR to neglect the impact of detection noise. The obtained BOTDR traces with distinct SRs are shown in **SI Fig. 11a-d**, respectively, each with two blue dashed lines indicating the upper and lower bounds. In each figure, we also illustrate a red dashed line that is obtained by doubling the windowed mean value of each BOTDR trace, representing the theoretical maximum value in the absence of self-polarization-scrambling effect. The $k_{Pol}$ is calculated as the ratio of peak-to-peak value of blue dash lines to the value of red dashed line. For the examined SRs of 1 m, 2 m, 6 m, and 10 m, the corresponding $k_{Pol}$ values are characterized as 0.88, 0.83, 0.78 and 0.69, respectively.

*2. Layered SNR Behavior with a PSc*
Here we clarify that, Eq. (104) that incorporates $k_{Pol}$ to account for the self-polarization-scrambling effect, enables to deliver an accurate lowest level of the actual SNR, which represents the standard metric to qualify any distributed sensor and are used for **Fig. 6e** of the main text. However, as shown by **SI Fig. 11e-h** that illustrate the SNR profiles of above four SRs with PSc, at some fiber positions the actual SNRs that are higher than this predicted value. Although these higher SNRs are not so meaningful for performance point of view, they are interpreted here for theoretical clarity. This phenomenon actually refers to the cases that the pump pulse right passes by fiber locations with sharp SOP transition due to birefringence, causing additional SOP averaging to that of PSc, resulting in smaller polarization noise[14]. The distribution histograms of the SNR shown by **SI Fig. 11e-h** are illustrated by **SI Fig. 11i-l**, respectively, indicating that the SNR expressed by Eq. (104) is the majority, which follows a standard normal distribution as expected.

**SI Note 7: BOTDR based on polarization diversity coherent receiver.**
In this note, we show that, even employing a polarization diversity coherent receiver (PDCR)[15] to eliminate polarization noise by summing up the respectively obtained BOTDR response of parallel and orthogonal SOPs, the SNR is poorly improved as it is ultimately constrained by the SpBS noise.

Replacing the detection scheme of BOTDR setup used for **SI note 6** to PDCR, with 2 m SR, the BOTDR envelope signal of each PDCR channel is obtained through digital post-processing of the output beating signal, with 2500 trace averaging. Results are shown by pink and blue curves in **SI Fig. 12a**, along with their averaged envelope signals (green curve). The SNRs of these three curves are shown in **SI Fig. 12b**. We focus on observing the green curve that represents the SNR of the PDCR approach, which shows an unobvious difference compared to that in the PSc case in **SI Fig. 11f.**, where the combined influence of polarization noise and SpBS noise dominates. This suggests that while the PDCR scheme effectively eliminates polarization fading, it does not obviously improve BOTDR SNR as SpBS noise still presents and ultimately limits the SNR.

**SI Note 8: Mathematical process modelling the SNR and experimental verification of coded-pulse BOTDR with standard SMF.**
In this note, we show that besides standard single-pulse BOTDR, the proposed framework for SpBS noise can also be utilized to evaluate the performance of more sophisticated BOTDR techniques, such as coded-pulse BOTDR, delivery a performance limit revision on its conventional assumption. We theoretically predict and experimentally demonstrate that, although pulse coding technique has been widely recognised as an approach enhancing SNR over single-pulse methods, in BOTDR scenario the technique actually only improves marginally the performance as impacted by both the SpBS noise and polarization noise.

*1. SNR model of coded-pulse BOTDR*
Here we leverage our SpBS noise framework to establish the SNR model of coded-pulse BOTDR. Since the fact that all '0' elements in the coded sequence do not contribute to the signal carrying Brillouin sensing information, only the contribution of the '1' elements is considered here. For aperiodic codes[16–19], as the fiber attenuation over the duration of the coding sequence is almost negligible, the total optical field of the coded SpBS, including Stokes $\overrightarrow{E_{Sp1}}(t)$ and anti-Stokes $\overrightarrow{E_{Sp2}}(t)$, can be treated as a superposition of the fields contributed by each '1' element:

$$\overrightarrow{E_{Sp1}}(t) \approx \sum_{i=1}^{N_1} \overrightarrow{E_{Sp1}^i}(t) = \sum_{i=1}^{N_1} \left\{ \hat{x} \cos[\theta_i(t)] e^{j\varphi_{Sp1}^{xi}(t)} + \hat{y} \sin[\theta_i(t)] e^{j\varphi_{Sp1}^{yi}(t)} \right\} E_{Sp1}^i(t) e^{j2\pi f_{Sp1}(t)t} \quad (105)$$

$$\overrightarrow{E_{Sp2}}(t) \approx \sum_{i=1}^{N_1} \overrightarrow{E_{Sp2}^i}(t) = \sum_{i=1}^{N_1} \left\{ \hat{x} \cos[\theta_i(t)] e^{j\varphi_{Sp2}^{xi}(t)} + \hat{y} \sin[\theta_i(t)] e^{j\varphi_{Sp2}^{yi}(t)} \right\} E_{Sp2}^i(t) e^{j2\pi f_{Sp2}(t)t} \quad (106)$$

where $\hat{x}$ and $\hat{y}$ stand for the polarization direction of the OLO light and its orthogonal direction, respectively; $N_1$ represents the number of "1"s elements in the coded pulse sequence; $i$ is the index of each '1' element after removing all '0' elements in the coded sequence; $\overrightarrow{E_{Sp1}^i}(t)$ and $\overrightarrow{E_{Sp2}^i}(t)$ are the fields of the Stokes and anti-Stokes components contributed by the $i$-th '1' element, respectively. The local relative polarization rotation ($\theta_i(t)$), the random phase differences ($\varphi_{Sp1}^{xi}(t)$, $\varphi_{Sp1}^{yi}(t)$, $\varphi_{Sp2}^{xi}(t)$, $\varphi_{Sp2}^{yi}(t)$), and the SpBS amplitudes ($E_{Sp1}^i(t)$, $E_{Sp2}^i(t)$) are $i$-dependent.

Following a derivation similar to that in Ref. [9][10], the BPF output signal photocurrent in distance domain can be obtained as:

$$I_s(z) = 2\mathcal{R}_p \sum_{i=1}^{N_1} \cos[\theta_i(z)] \left\{ \sqrt{P_{Lo1} P_{Sp1}^i(z)} \cos[H(z) + \Phi_1^i(z)] + \sqrt{P_{Lo2} P_{Sp2}^i(z)} \cos[H(z) + \Phi_2^i(z)] \right\}$$

$$\stackrel{\text{def}}{=} 2\mathcal{R}_p \sum_{i=1}^{N_1} Y_i(z) \quad (107)$$

where $\Phi_1^i(t) = \varphi_{Sp1}^{xi}(t) - \pi/2$, $\Phi_2^i(t) = \varphi_{Sp2}^{xi}(t) - \pi/2$; $P_{Sp1}^i(z) \propto E_{Sp1}^i(z)^2$ and $P_{Sp2}^i(z) \propto E_{Sp2}^i(z)^2$ represent the Stokes and anti-Stokes SpBS optical powers, respectively, having approximately equal mathematical expectations, that is, $\overline{P_{Sp1}^i(z)} = \overline{P_{Sp2}^i(z)} = \overline{P_{Sp}^i(z)}/2$. The mathematical expectation of the total SpBS power corresponding to each element '1', $\overline{P_{Sp}^i(z)}$, can be expressed by Eq. (75). $Y_i(z)$ is used to denote the term inside the summation for the convenience of the following mathematical derivation:

$$Y_i(z) = \cos[\theta_i(z)]\left\{\sqrt{P_{Lo1}P_{Sp1}^i(z)}\cos[H(z) + \Phi_1^i(z)] + \sqrt{P_{Lo2}P_{Sp2}^i(z)}\cos[H(z) + \Phi_2^i(z)]\right\} \quad (108)$$

For the coded-pulse BOTDR system, we further model the SNR behavior of the decoded response (decoded power envelope signal, designated as $r_{Decode}^{SMF}(z)$) taking into account the polarization noise. We start with the analysis on the SNR of the coded response $r_{Code}^{SMF}(z)$, following a mathematical process similar to that in **SI Note 4** and **5**.

**1) The term of $LPF\{I_s(z)^2\}$.** Squaring Eq. (107), we get:

$$I_s(z)^2 = 4\mathcal{R}_p^2\left\{\sum_{i=1}^{N_1}Y_i(z)\right\}^2 = 4\mathcal{R}_p^2\sum_{i=1}^{N_1}Y_i(z)^2 + 4\mathcal{R}_p^2\sum_{i\neq k}^{N_1(N_1-1)/2}2Y_i(z)Y_k(z) \quad (109)$$

Since both summation and low-pass filtering are linear processes, $LPF\{I_s(z)^2\}$ can be expressed as:

$$LPF\{I_s(z)^2\} = 4\mathcal{R}_p^2 LPF\left\{\sum_{i=1}^{N_1}Y_i(z)^2\right\} + 4\mathcal{R}_p^2 LPF\left\{\sum_{i\neq k}^{\frac{N_1(N_1-1)}{2}}2Y_i(z)Y_k(z)\right\}$$

$$= 4\mathcal{R}_p^2\sum_{i=1}^{N_1}LPF\{Y_i(z)^2\} + 4\mathcal{R}_p^2\sum_{i\neq k}^{N_1(N_1-1)/2}LPF\{2Y_i(z)Y_k(z)\} \quad (110)$$

Considering that the covariance of the last two terms in Eq. (110) is zero, the expectation value and variance of $LPF\{I_s(z)^2\}$, designated as $\mu_{s^2}(z)$ and $\sigma_{s^2}(z)^2$, respectively, can be expressed as:

$$\mu_{s^2}(z) = \overline{LPF\{I_s(z)^2\}} = 4\mathcal{R}_p^2\sum_{i=1}^{N_1}\overline{LPF\{Y_i(z)^2\}} + 4\mathcal{R}_p^2\sum_{i\neq k}^{\frac{N_1(N_1-1)}{2}}\overline{LPF\{2Y_i(z)Y_k(z)\}} \quad (111)$$

$$\sigma_{s^2}(z)^2 = D\{LPF\{I_s(z)^2\}\} = 16\mathcal{R}_p^4\left\{\sum_{i=1}^{N_1}D\{LPF\{Y_i(z)^2\}\} + \sum_{i\neq k}^{\frac{N_1(N_1-1)}{2}}D\{LPF\{2Y_i(z)Y_k(z)\}\}\right\} \quad (112)$$

This necessitates to derive the statistical properties of $LPF\{Y_i(z)^2\}$ and $LPF\{2Y_i(z)Y_k(z)\}$.

Firstly, $Y_i(z)^2$ with a low-pass filtering process can be expressed as:

$$LPF\{Y_i(z)^2\} = \frac{1}{2}\cos^2[\theta_i(z)]\left\{\begin{array}{c}P_{Lo1}P_{Sp1}^i(z) + P_{Lo2}P_{Sp2}^i(z)\\ +2\sqrt{P_{Lo1}P_{Sp1}^i(z)P_{Lo2}P_{Sp2}^i(z)}\cos[\Phi_1^i(z) - \Phi_2^i(z)]\end{array}\right\} \quad (113)$$

where $\cos^2[\theta_i(z)]$ follows the same statistical properties described in Eq. (96), Eq. (97) and Eq. (99). Substituting the mathematical expectations of $P_{Sp1}^i(z)$ and $P_{Sp2}^i(z)$, ($\overline{P_{Sp1}(z)}$ and $\overline{P_{Sp2}(z)}$), yielding:

$$\overline{LPF\{Y_i(z)^2\}} = \frac{1}{4}\left[P_{Lo1}\overline{P_{Sp1}(z)} + P_{Lo2}\overline{P_{Sp2}(z)}\right] = \frac{1}{8}P_{Lo}\overline{P_{Sp}(z)} \quad (114)$$

Similar with Eq. (80) and (81), there is a relationship of $\overline{P_{Sp1}^i(z)^2} = \left(1 + \frac{1}{SNR_{Sp}^2}\right)\overline{P_{Sp1}(z)}^2$ and $\overline{P_{Sp2}^i(z)^2} = \left(1 + \frac{1}{SNR_{Sp}^2}\right)\overline{P_{Sp2}(z)}^2$, thus the mathematical expectation of $LPF^2\{Y_i(z)^2\}$ can be calculated as:

$$\overline{LPF^2\{Y_i(z)^2\}} = \frac{1}{4}\frac{k_{Pol}^2+3}{12}\left[\begin{array}{c}P_{Lo1}^2\left(1 + \frac{1}{SNR_{Sp}^2}\right)\overline{P_{Sp1}(z)}^2 + P_{Lo2}^2\left(1 + \frac{1}{SNR_{Sp}^2}\right)\overline{P_{Sp2}(z)}^2\\ +4P_{Lo1}\overline{P_{Sp1}(z)}P_{Lo2}\overline{P_{Sp2}(z)}\end{array}\right]$$

$$= \frac{k_{Pol}^2+3}{384}\left(3 + \frac{1}{SNR_{Sp}^2}\right)P_{Lo}^2\overline{P_{Sp}(z)}^2 \quad (115)$$

Therefore, the variance of $LPF\{Y_i(z)^2\}$ can be calculated as:

$$D\{\text{LPF}\{Y_i(z)^2\}\} = \overline{\text{LPF}^2\{Y_i(z)^2\}} - \overline{\text{LPF}\{Y_i(z)^2\}}^2 = \left[\frac{k_{Pol}^2+3}{384}\left(3+\frac{1}{SNR_{Sp}^2}\right)-\frac{1}{64}\right]P_{Lo}^2\overline{P_{Sp}(z)}^2 \quad (116)$$

Next, the mathematical expectation and variance of $\text{LPF}\{2Y_i(z)Y_k(z)\}$ are calculated as:

$$\overline{\text{LPF}\{2Y_i(z)Y_k(z)\}} = 0 \quad (117)$$

$$D\{\text{LPF}\{2Y_i(z)Y_k(z)\}\} = \overline{\text{LPF}^2\{2Y_i(z)Y_k(z)\}} - \overline{\text{LPF}\{2Y_i(z)Y_k(z)\}}^2$$
$$= \frac{1}{8}\left[P_{Lo1}\overline{P_{Sp1}(z)} + P_{Lo2}\overline{P_{Sp1}(z)}\right]^2 = \frac{1}{32}P_{Lo}^2\overline{P_{Sp}(z)}^2 \quad (118)$$

Finally, based on Eq. (111) and Eq. (112), $\mu_{s^2}(z)$ and $\sigma_{s^2}(z)^2$ can be obtained as:

$$\mu_{s^2}(z) = 4\mathcal{R}_p^2 \sum_{i=1}^{N_1}\frac{1}{8}P_{Lo}\overline{P_{Sp}(z)} = \frac{1}{2}N_1\mathcal{R}_p^2 P_{Lo}\overline{P_{Sp}(z)} \quad (119)$$

$$\sigma_{s^2}(z)^2 = 16\mathcal{R}_p^4\left\{\sum_{i=1}^{N_1}\left[\frac{k_{Pol}^2+3}{384}\left(3+\frac{1}{SNR_{Sp}^2}\right)-\frac{1}{64}\right]P_{Lo}^2\overline{P_{Sp}(z)}^2 + \sum_{i\neq k}^{\frac{N_1(N_1-1)}{2}}\frac{1}{32}P_{Lo}^2\overline{P_{Sp}(z)}^2\right\}$$

$$= \frac{N_1}{24}\left(3k_{Pol}^2 - 3 + \frac{k_{Pol}^2+3}{SNR_{Sp}^2} + 6N_1\right)\mathcal{R}_p^4 P_{Lo}^2\overline{P_{Sp}(z)}^2 \quad (120)$$

**2) The term of $\text{LPF}\{2I_s(z)I_e(z)\}$.** According to Eq. (84) and Eq. (107), $2I_s(z)I_e(z)$ can be expressed as:

$$2I_s(z)I_e(z) = 4\mathcal{R}_p\sum_{i=1}^{N_1}Y_i(z)I_e^I(z)\cos[H(z)] - 4\mathcal{R}_p\sum_{i=1}^{N_1}Y_i(z)I_e^Q(z)\sin[H(z)] \quad (121)$$

Considering that both summation and low-pass filtering are linear processes, the following equation holds:

$$\text{LPF}\{I_s(z)I_e(z)\} = 4\mathcal{R}_p\text{LPF}\left\{\sum_{i=1}^{N_1}Y_i(z)I_e^I(z)\cos[H(z)]\right\} + 4\mathcal{R}_p\text{LPF}\left\{\sum_{i=1}^{N_1}Y_i(z)I_e^Q(z)\sin[H(z)]\right\}$$

$$= 4\mathcal{R}_p\sum_{i=1}^{N_1}\text{LPF}\{Y_i(z)I_e^I(z)\cos[H(z)]\} + 4\mathcal{R}_p\sum_{i=1}^{N_1}\text{LPF}\{Y_i(z)I_e^Q(z)\sin[H(z)]\} \quad (122)$$

The expectation value and variance of $\text{LPF}\{2I_s(z)I_e(z)\}$, designated as $\mu_{2se}(z)$ and $\sigma_{2se}(z)^2$, respectively, can be expressed as:

$$\mu_{2se}(z) = \overline{\text{LPF}\{I_s(z)I_e(z)\}}$$
$$= 4\mathcal{R}_p\sum_{i=1}^{N_1}\overline{\text{LPF}\{Y_i(z)I_e^I(z)\cos[H(z)]\}} + 4\mathcal{R}_p\sum_{i=1}^{N_1}\overline{\text{LPF}\{Y_i(z)I_e^Q(z)\sin[H(z)]\}} \quad (123)$$

$$\sigma_{2se}(z)^2 = D\{\text{LPF}\{I_s(z)I_e(z)\}\}$$
$$= 16\mathcal{R}_p^2\left\{\sum_{i=1}^{N_1}D\{\text{LPF}\{Y_i(z)I_e^I(z)\cos[H(z)]\}\} + \sum_{i=1}^{N_1}D\{\text{LPF}\{Y_i(z)I_e^Q(z)\sin[H(z)]\}\}\right\} \quad (124)$$

We then proceed to derive the statistics of $\text{LPF}\{Y_i(z)I_e^I(z)\cos[H(z)]\}$ and $\text{LPF}\{Y_i(z)I_e^Q(z)\sin[H(z)]\}$. We obtain:

$$\overline{\text{LPF}\{Y_i(z)I_e^I(z)\cos[H(z)]\}} = 0 \quad (125)$$

and

$$D\{\text{LPF}\{Y_i(z)I_e^I(z)\cos[H(z)]\}\} = \overline{\text{LPF}^2\{Y_i(z)I_e^I(z)\cos[H(z)]\}} - \overline{\text{LPF}\{Y_i(z)I_e^I(z)\cos[H(z)]\}}^2$$
$$= \frac{1}{16}\sigma_e^2 B_{BPF}\left[P_{Lo1}\overline{P_{Sp1}(z)} + P_{Lo2}\overline{P_{Sp2}(z)}\right] = \frac{1}{32}P_{Lo}\overline{P_{Sp}(z)}\sigma_e^2 B_{BPF} \quad (126)$$

By analogy, it can be concluded that $\text{LPF}\{Y_i(z)I_e^Q(z)\sin[H(z)]\}$ possesses the same mathematical expectation and variance as $\text{LPF}\{Y_i(z)I_e^I(z)\cos[H(z)]\}$. Based on Eq (123) and (124), we have:

$$\mu_{2se}(z) = 0 \quad (127)$$
$$\sigma_{2se}(z)^2 = N_1\mathcal{R}_p^2 P_{Lo}\overline{P_{Sp}(z)}\sigma_e^2 B_{BPF} \quad (128)$$

**3) The term of LPF$\{I_e(z)^2\}$.** The statistical parameters of LPF$\{I_e(z)^2\}$ - namely its expectation and variance - are the same as the case of single-pulse BOTDR in **SI Note 4** and **5**.

**SI Table IV**
Expectation values and variances of the three terms in an expression similar to Eq. (76) for the coded-pulse case

| Terms | Expectation values | Variances |
|---|---|---|
| LPF$\{I_s(z)^2\}$ | $\frac{1}{2}N_1 \mathcal{R}_p{}^2 P_{Lo}\overline{P_{Sp}(z)}$, (Eq. (119)) | $\frac{N_1}{24}\left(3k_{Pol}{}^2 - 3 + \frac{k_{Pol}{}^2 + 3}{SNR_{Sp}{}^2} + 6N_1\right)\mathcal{R}_p{}^4 P_{Lo}{}^2 \overline{P_{Sp}(z)}^2$, (Eq. (120)) |
| LPF$\{2I_s(z)I_e(z)\}$ | 0, (Eq. (127)) | $N_1 \mathcal{R}_p{}^2 P_{Lo}\overline{P_{Sp}(z)}\sigma_e{}^2 B_{BPF}$, (Eq. (128)) |
| LPF$\{I_e(z)^2\}$ | $\sigma_e{}^2 B_{BPF}$, (Eq. (91)) | $(\sigma_e{}^2 B_{BPF})^2$, (Eq. (92)) |

The statistical parameters of the SMF-based coded-pulse BOTDR (expectation values and variances) derived above are summarized in SI Table IV. Up to this point, we have derived the SNR of the coded response:

$$\text{SNR}\{r_{Code}^{SMF}(z)\}$$
$$= \frac{\frac{1}{2}N_1 \mathcal{R}_p{}^2 P_{Lo}\overline{P_{Sp}(z)}}{\sqrt{\frac{N_1}{24}\left(3k_{Pol}{}^2 - 3 + \frac{k_{Pol}{}^2 + 3}{SNR_{Sp}{}^2} + 6N_1\right)\mathcal{R}_p{}^4 P_{Lo}{}^2 \overline{P_{Sp}(z)}^2 + N_1 \mathcal{R}_p{}^2 P_{Lo}\overline{P_{Sp}(z)}\sigma_e{}^2 B_{BPF} + \sigma_e{}^4 B_{BPF}{}^2}} \quad (129)$$

The SNR of the decoded response can then be obtained via compressing the noise STD by a factor of $\sqrt{N_1/2}$ [18,19]:

$$\text{SNR}\{r_{Decode}^{SMF}(z)\}$$
$$= \frac{\frac{1}{2}\mathcal{R}_p{}^2 P_{Lo}\overline{P_{Sp}(z)}}{\sqrt{\frac{2}{N_1}}\sqrt{\frac{N_1}{24}\left(3k_{Pol}{}^2 - 3 + \frac{k_{Pol}{}^2 + 3}{SNR_{Sp}{}^2} + 6N_1\right)\mathcal{R}_p{}^4 P_{Lo}{}^2 \overline{P_{Sp}(z)}^2 + N_1 \mathcal{R}_p{}^2 P_{Lo}\overline{P_{Sp}(z)}\sigma_e{}^2 B_{BPF} + \sigma_e{}^4 B_{BPF}{}^2}}$$
$$\leq \sqrt{\frac{3}{\left(3 + \frac{1}{SNR_{Sp}{}^2}\right)k_{Pol}{}^2 - 3 + \frac{3}{SNR_{Sp}{}^2} + 6N_1}} \quad (130)$$

Eq. (130) reveals that, under a given $k_{Pol}$ value, the SNR upper limit of the decoded BOTDR is predicted to decrease as $N_1$ increases, and is always lower than the SNR upper limit of the single-pulse sensing system shown by Eq. (104). This is because the decoded SNR performance of coded-pulse BOTDR is severely constrained by the intrinsic fluctuation of SpBS and polarization noise.

The coding gain, typically defined as the SNR improvement provided by the coding scheme in comparison to the single-pulse scheme, can be expressed as:

$$G_c = \frac{\text{SNR}\{r_{Decode}^{SMF}(z)\}}{\text{SNR}\{r_{Sg}^{SMF}(z)\}}$$
$$= \sqrt{\frac{N_1}{2}}\sqrt{\frac{\frac{(3SNR_{Sp}{}^2 + 1)k_{Pol}{}^2 + 3SNR_{Sp}{}^2 + 3}{24SNR_{Sp}{}^2}\mathcal{R}_p{}^4 P_{Lo}{}^2 \overline{P_{Sp}(z)}^2 + \mathcal{R}_p{}^2 P_{Lo}\overline{P_{Sp}(z)}\sigma_e{}^2 B_{BPF} + \sigma_e{}^4 B_{BPF}{}^2}{\frac{N_1}{24}\left(3k_{Pol}{}^2 - 3 + \frac{k_{Pol}{}^2 + 3}{SNR_{Sp}{}^2} + 6N_1\right)\mathcal{R}_p{}^4 P_{Lo}{}^2 \overline{P_{Sp}(z)}^2 + N_1 \mathcal{R}_p{}^2 P_{Lo}\overline{P_{Sp}(z)}\sigma_e{}^2 B_{BPF} + \sigma_e{}^4 B_{BPF}{}^2}}$$
$$= \begin{cases} \sqrt{\dfrac{(3SNR_{Sp}{}^2 + 1)k_{Pol}{}^2 + 3SNR_{Sp}{}^2 + 3}{2(3SNR_{Sp}{}^2 + 1)k_{Pol}{}^2 + 12N_1 SNR_{Sp}{}^2 - 6SNR_{Sp}{}^2 + 6}}, & \text{for high SpBS signal level} \\ \sqrt{\dfrac{N_1}{2}}, & \text{for low SpBS signal level} \end{cases} \quad (131)$$

It can be concluded that the coding gain behaves differently in the two typical sensing regions:
1) The region where the signal power of SpBS is optimized to high enough that the combined effect of SpBS noise and polarization noise (the first noise term in numerator of Eq. (129)) dominates. In this case, due to joint limitations imposed by SpBS and polarization noise, no positive coding gain can be achieved, and the degradation of SNR grows with the number of $N_1$ under a given $k_{Pol}$ value and a $SNR_{Sp}$ value. This situation generally occurs at the long fiber near-end or entire short fiber.

2) The region where the signal power of SpBS is so low that the photo-detection noise (the third noise term in numerator of Eq. (129)) becomes dominant. In this case, since SpBS noise and polarization noise – both related to the signal magnitude - can be neglected, it is possible to achieve positive coding gain and even ideal coding gain $\sqrt{N_1/2}$. This situation typically occurs at the long fiber far-end, where the SpBS power is significantly reduced due to fiber attenuation, or when the duration / power of pump pulse is impractically set to a quite low level.

*2. Experimental verification*

The impact of the SpBS noise in coded-pulse BOTDR is experimentally investigated based on a genetic-optimized code. The experiment employed a standard coherent detection scheme over a 50 km-long standard SMF. Each pulse had a 20 ns duration and was upsampled to 200 ns before being injecting the fiber, with the first pulse amplified by an EDFA to the modulation instability (MI) threshold. We used a 5 dBm dual-sideband OLO and tested three distinct $N_1$ values: 8, 20, and 45. At the receiver, we implemented a physical envelope detection scheme: The hybrid SpBS-OLO signal was captured by a 400-MHz bandwidth BPD, then sequentially processed through 1) an electrical amplifier, 2) a 50-MHz bandpass filter, and 3) a physical envelope detector, before digitization at 500 MSa/s using an ADC. The beat frequency between the SpBS and OLO was carefully aligned with the centre frequency of the BPF passband. We acquired 20 repeated measurements for each trace, enabling subsequent SNR calculation, with each trace representing an average of 2,048 acquisitions and using a 530 μs sampling window. A single-pulse experiment under the same experimental conditions is performed as a reference. **SI Fig. 13a–c** show SNR measurements for $N_1$ = 8, 20, and 45, closely matching the theoretical predictions. Results highlight that actual SNR improvements from coding are lower than previously expected due to SpBS noise, particularly at large $N_1$. As sensing distance increases, attenuation reduces SpBS and polarization noise effects, leading SNR to approach the ideal photodetection-limited case. These findings underscore the significant impact of SpBS noise in coded BOTDR and the need for refined optimization strategies.

## SUPPLEMENTARY REFERENCES

1. Boyd, R. W., Rząewski, K. & Narum, P. Noise initiation of stimulated Brillouin scattering. *Phys. Rev. A* **42**, 5514–5521 (1990).

2. Boyd, R. W. *Nonlinear Optics*. (Academic Press, Burlington, MA, 2008).

3. Beugnot, J.-C., Tur, M., Mafang, S. F. & Thévenaz, L. Distributed Brillouin sensing with sub-meter spatial resolution: modeling and processing. *Opt. Express* **19**, 7381 (2011).

4. Proakis, J. G. & Manolakis, D. G. DIGITAL SIGNAL PROCESSING. 1033.

5. Goodman, J. W. *Statistical Optics*. (Wiley, Hoboken, New Jersey, 2015).

6. Wang, S., Yang, Z., Soto, M. A. & Thévenaz, L. Study on the signal-to-noise ratio of Brillouin optical-time domain analyzers. *Opt. Express* **28**, 19864–19876 (2020).

7. Saleh, B. E. A. & Teich, M. C. *Fundamentals of Photonics*. (Wiley, Hoboken, NJ, 2019).

8. Thévenaz, L. *Advanced Fiber Optics: Concepts and Technology*. (EPFL press CRC press, Lausanne (Suisse) Boca Raton (États-Unis), 2011).

9. Jin, S., Yang, Z., Hong, X. & Wu, J. Analytical Signal-to-Noise Ratio Model on Frequency-Scanned Brillouin Optical Time-Domain Reflectometry. *J. Light. Technol.* **42**, 5786–5796 (2024).

10. Yariv, A., Yeh, P. & Yariv, A. *Photonics: Optical Electronics in Modern Communications*. (Oxford University Press, New York, 2007).

11. Johnson, R. A., Miller, I. & Freund, J. E. *Miller & Freund's Probability and Statistics for Engineers*. (Pearson, Boston, 2017).

12. van Deventer, M. O. & Boot, A. J. Polarization properties of stimulated Brillouin scattering in single-mode fibers. *J. Light. Technol.* **12**, 585–590 (1994).

13. Gao, X. *et al.* Impact of optical noises on unipolar-coded Brillouin optical time-domain analyzers. *Opt. Express* **29**, 22146–22158 (2021).

14. Jostmeier, T., Marx, B., Buntebarth, C., Rath, A. & Hill, W. Long-Distance BOTDR Interrogator with Polarization- Diverse Coherent Detection and Power Evaluation. in *Optical Fiber Sensors Conference 2020 Special Edition* T3.21 (Optica Publishing Group, Washington, DC, 2021). doi:10.1364/OFS.2020.T3.21.

15. Song, H. Y. & Golomb, S. W. Some new constructions for simplex codes. *IEEE Trans. Inf. Theory* **40**, 504–507 (1994).

16. Golay, M. Complementary series. *IEEE Trans. Inf. Theory* **7**, 82–87 (1961).

17. Yang, Z., Li, Z., Zaslawski, S., Thévenaz, L. & Soto, M. A. Design rules for optimizing unipolar coded Brillouin optical time-domain analyzers. *Opt. Express* **26**, 16505–16523 (2018).

18. Sun, X. *et al.* Genetic-optimised aperiodic code for distributed optical fibre sensors. *Nat. Commun.* **11**, 5774 (2020).